\documentclass[twocolumn,aps,prb,superscriptaddress,floatfix]{revtex4-2}
\usepackage[T1]{fontenc}
\usepackage[utf8]{inputenc}
\usepackage[english]{babel}
\usepackage{color}
\usepackage{float}
\usepackage{braket}
\usepackage{amsmath}
\usepackage{amstext}
\usepackage{amssymb}
\usepackage{graphicx}
\usepackage{bbold}
\usepackage[unicode=true,pdfusetitle,
bookmarks=true,bookmarksnumbered=false,bookmarksopen=false,
breaklinks=false,pdfborder={0 0 1},backref=false,colorlinks=false]
{hyperref}
\hypersetup{
	colorlinks,linkcolor=blue,citecolor=blue,urlcolor=blue}
\usepackage{soul}
\makeatletter
\date{}

\makeatother
\setul{0.5ex}{0.3ex}
\definecolor{Blue}{rgb}{0,0.0,1}
\setulcolor{Blue}
\usepackage{babel}
\newcommand{\angstrom}{\textup{\AA}}

\begin{document} 

\author{Tarik P. Cysne}
\email{tarik.cysne@gmail.com}
\affiliation{Instituto de F\'\i sica, Universidade Federal Fluminense, 24210-346 Niter\'oi RJ, Brazil}

\author{Filipe S. M. Guimar\~{a}es}
\affiliation{J\"ulich Supercomputing Centre, Forschungszentrum J\"{u}lich and JARA, 52425 J\"{u}lich, Germany}

\author{Luis M. Canonico}
\affiliation{Catalan Institute of Nanoscience and Nanotechnology (ICN2), CSIC and BIST, Campus UAB, Bellaterra, 08193 Barcelona, Spain}

\author{Marcio Costa}
\affiliation{Instituto de F\'\i sica, Universidade Federal Fluminense, 24210-346 Niter\'oi RJ, Brazil}

\author{Tatiana G. Rappoport}
\affiliation{Instituto de Telecomunicações, Instituto Superior Tecnico, University of Lisbon, Avenida Rovisco Pais 1, Lisboa, 1049001 Portugal}	
\affiliation{Instituto de F\'\i sica, Universidade Federal do Rio de Janeiro, C.P. 68528, 21941-972 Rio de Janeiro RJ, Brazil}

\author{R. B. Muniz}
\affiliation{Instituto de F\'\i sica, Universidade Federal Fluminense, 24210-346 Niter\'oi RJ, Brazil}

\title{Orbital magnetoelectric effect in nanoribbons of transition metal dichalcogenides}

\begin{abstract}
The orbital magnetoelectric effect (OME) generically refers to the appearance of an orbital magnetization induced by an applied electric field. Here, we show that nanoribbons of transition metal dichalcogenides (TMDs) with zigzag (ZZ) edges may exhibit a sizeable OME activated by an electric field applied along the ribbons' axis. We examine nanoribbons extracted from a monolayer (1L) and a bilayer (2L) of MoS$_2$ in the trigonal (H) structural phase. Transverse profiles of the induced orbital angular momentum accumulations are calculated to first order in the longitudinally applied electric field. Our results show that close to the nanoribbon's edge-state crossings energy, the orbital angular momentum accumulations take place mainly around the ribbons' edges. They have two contributions: one arising from the orbital Hall effect (OHE) and the other consists in the OME. The former is transversely anti-symmetric with respect to the principal axis of the nanoribbon, whereas the latter is symmetric, and hence responsible for the resultant orbital magnetization induced in the system. We found that the orbital accumulation originating from the OHE for the 1L-nanoribbon is approximately half that of a 2L-nanoribbon. Furthermore, while the OME can reach fairly high values in 1L-TMD nanoribbons, it vanishes in the 2L ones that preserve spatial inversion symmetry.The microscopic features that justify our findings are also discussed.

\end{abstract}
\maketitle
\section{Introduction}
The possibilities of utilizing the electronic orbital angular momentum (OAM) degrees of freedom to transmit, process, and store information in solids has increased in recent years \cite{DGo-EPL-2021-review, Bernevig-Hughes-Zhang-PhysRevLett.95.066601, Han-HyunWooLeePhysRevLett.128.176601, Go-Jo-Kim-Lee-PhysRevLett.121.086602,Mokrousov-Go-Experiment-PhysRevLett.128.067201, Chen-Zeng-NatCommun2018}. Two relevant phenomena contribute to make this feasible: the orbital Hall effect (OHE) and the orbital magneto-electric effect (OME). 

The OHE consists in the generation of a transverse OAM current induced by a longitudinally applied electric field. It provides a way of producing electrically controllable OAM currents that can be injected into a variety of materials and eventually used to drive magnetization dynamics in spin-orbit coupled systems \cite{Go1PRR2.033401,Arnab-Orbital-Torque-Exp,Go2PRR2.013177,Go3_arxiv.2106.07928,Go4_arxiv.2202.13896}. The OME describes the advent of an electrically induced orbital magnetization. Over the years it has received different names such as orbital Edelstein effect, kinetic magnetoelectric effect, orbital gyrotropic magnetoelectric effect (just to mention a few) that distinguish the mechanisms involved and characteristics of the systems where it occurs \cite{Bhowal&Satpathy3, Murakami1, Osumi-Zhang-Murakami-CommunPhys}. It has potential application for the development of data-storage orbitronic devices, and has been investigated in diverse materials
\cite{Johansson-Merting-PhysRevResearch.3.013275,Salemi-Oppeneer-PRMaterials-2021, Osumi-Zhang-Murakami-CommunPhys, Yoda-Yokoyama-Murakami-NanoLett-2018, Massarelli-Wu-Paramekanti-PhysRevB.100.075136, Hayami-Kusunose-PhysRevB.98.165110, Hayami-JPCM-2016, He-Law-PhysRevResearch.2.012073, Furukawa-Itou-PhysRevResearch.3.023111}. It is noteworthy that both the OHE and the OME do not require the presence of spin-orbit interaction to take place and, therefore, broaden the spectrum of materials that can be useful for spin-orbitronic applications. Two-dimensional (2D) materials provide a fertile ground for prospecting elements with such characteristics \cite{Mele-PRL-PhysRevLett.123.236403-2019, Us1-PRB, Us2-PRBR, Us3-PRL, Cysne-Bhowal-Vignale-Rappoport, Bhowal-Satpathy-PhysRevB.101.121112, Jian-Zhou-NPJCompMat-2021, Bhowal-Vignale-PhysRevB.103.195309, He-2020-tbg-kome-NatCommun, Schaefer-Katja-PhysRevB.103.224426}. Several of them have multi-orbital band structures that enable the appearance of interesting OAM phenomena. In particular, transition metal dichalcogenides (TMDs) comprise prospective candidates for applications in orbitronics. These materials exhibit orbital textures that underlie the OHE even in its insulating phase \cite{Schuler-Pincelli-Beaulieu-PhysRevX.12.011019, Beaulieu-Schusser-2020-PhysRevLett.125.216404, orbitaltexture-microscopicstudy-Seungyun, Us2-PRBR, Us3-PRL, Cysne-Bhowal-Vignale-Rappoport}, where non-trivial topology associated with the OAM is beginning to be unveiled \cite{Quian-Liu-Liu-PhysRevB.105.045417, Jiang-Xie-PhysRevB.104.L161108, Us_Plus_FazzioPeople-Arxiv, Us3-PRL}.

Up to linear order in the applied electric field the magnetoelectric effect may be described by the magnetoelectric susceptibility tensor $\hat{\alpha}$ defined by: 
\begin{equation}
    M^i = \sum_j \alpha^{ij} \mathcal{E}^j.
\label{MES}
\end{equation}
Here, $i$ and $j$ denote the Cartesian directions $(x,y,z)$, $M^i$ and $\mathcal{E}^j$ represent the components of the induced magnetization and applied electric field, respectively, and $\alpha^{ij}$ symbolize the matrix elements of $\hat{\alpha}$.
There are two contributions to $M^i$: one extrinsic (Boltzmann-like) involving intra-band electronic scattering only, induced by the disorder; The other is intrinsic (Kubo-like) and is entirely determined by inter-band transitions between stationary electronic states. Parity and time reversal symmetries play a fundamental role in both contributions. For instance, a necessary condition for the appearance of the extrinsic contribution is parity symmetry ($\mathcal{P}$) breaking, while the intrinsic contribution requires time reversal symmetry ($\mathcal{T}$) to be also broken. In addition, when both $\mathcal{P}$ and $\mathcal{T}$ are violated but $\mathcal{PT}$ is preserved, only the intrinsic contribution survives \cite{Xiao-Niu-PhysRevB.103.045401, Hayami-Kusunose-PhysRevB.98.165110}. 
The elements $\alpha^{ij}$ are also constrained by crystalline symmetries that eventually determine the general form of $\hat{\alpha}$. The lack of inversion symmetry, for instance, is a necessary, but not sufficient condition for the existence of the extrinsic contribution. 

Here, we will work with non-magnetic TMDs that preserve $\mathcal{T}$. In this case, the OME is activated by the intra-band (extrinsic) contribution only, and corresponds to an electric current-induced phenomenon. A detailed list of the crystalline point groups that allow finite magneto electric effect is given in Refs. \citenum{Furukawa-Itou-PhysRevResearch.3.023111, He-Law-PhysRevResearch.2.012073}. An unsupported 2D monolayer of MoS$_2$ in the H structural phase belongs to the $D_{3h}$ point group symmetry, for which all elements $\alpha^{ij}=0$ and hence no OME is expected to take place---even though inversion symmetry is broken~\cite{Furukawa-Itou-PhysRevResearch.3.023111, He-Law-PhysRevResearch.2.012073}. In order to enable the appearance of the magnetoelectric effect, it is necessary to reduce its crystalline symmetry. One way of achieving this is by straining the material, as reported in references  \cite{Son-Kim-2019-PhysRevLett.123.036806,Bhowal-Satpahy-kOME-PhysRevB.102.201403, Lee-2017-NatMaterials-887-9, Faria_Junior_2022}. Another is to couple the TMD layer to a suitable substrate that changes the system's symmetry \cite{CF-PhysRevB.98.045407, C2v-Suplatt-PhysRevB.92.075419} to allow the magnetoelectric effect to occur. Here, however, we do it geometrically by considering nanoribbons with zigzag edges, which belong to the $C_{2v}$ symmetry point group that allows nonzero values of $\alpha^{zx}$.

To explore the OME in these stripes we have calculated the transverse profiles of the OAM accumulations induced by a longitudinally applied electric field for unsupported ZZ-nanoribbons of H-TMDs. We chose MoS$_2$ as an archetype of this family and examine nanoribbons extracted from both monolayer (1L) and bilayer (2L) of this material. We start with a simplified three-band model that provides a reasonable description of ZZ-nanoribbons' edge-states electronic structure~\cite{Xiao-three-band}. Subsequently, we employ a more comprehensive approach based on density functional theory (DFT) calculations for the electronic states, and verify that the main features of the orbital responses obtained by the two methods are in good agreement. We show that monolayer nanoribbons (1L-MoS$_2$) exhibit relatively large orbital Hall and magnetoelectric effects. In unsupported bilayer nanoribbons (2L-MoS$_2$) the OME is absent due to the existence of spatial-inversion symmetry, and the intensity of the OHE is approximately twice the value obtained for the 1L-MoS$_2$ nanoribbon. 

The paper is organized as follows: Sec.~\ref{sec2} is devoted to the orbital response of unsupported nanoribbons using the three-band model. In Sec.~\ref{sec3}, we perform an analogous study using DFT calculations that confirm the main predictions obtained with the three-band model. In Sec.~\ref{sec4}, the physical aspects behind the orbital responses are discussed based on the equilibrium charge distribution and orbital textures. Finally, in Sec.~\ref{sec5} we draw our main conclusions. Technical details about the model and methods are included in the appendices.

\section{Three-band model calculations \label{sec2}}

\begin{figure}[h!]
	\includegraphics[width=1.0\linewidth,clip]{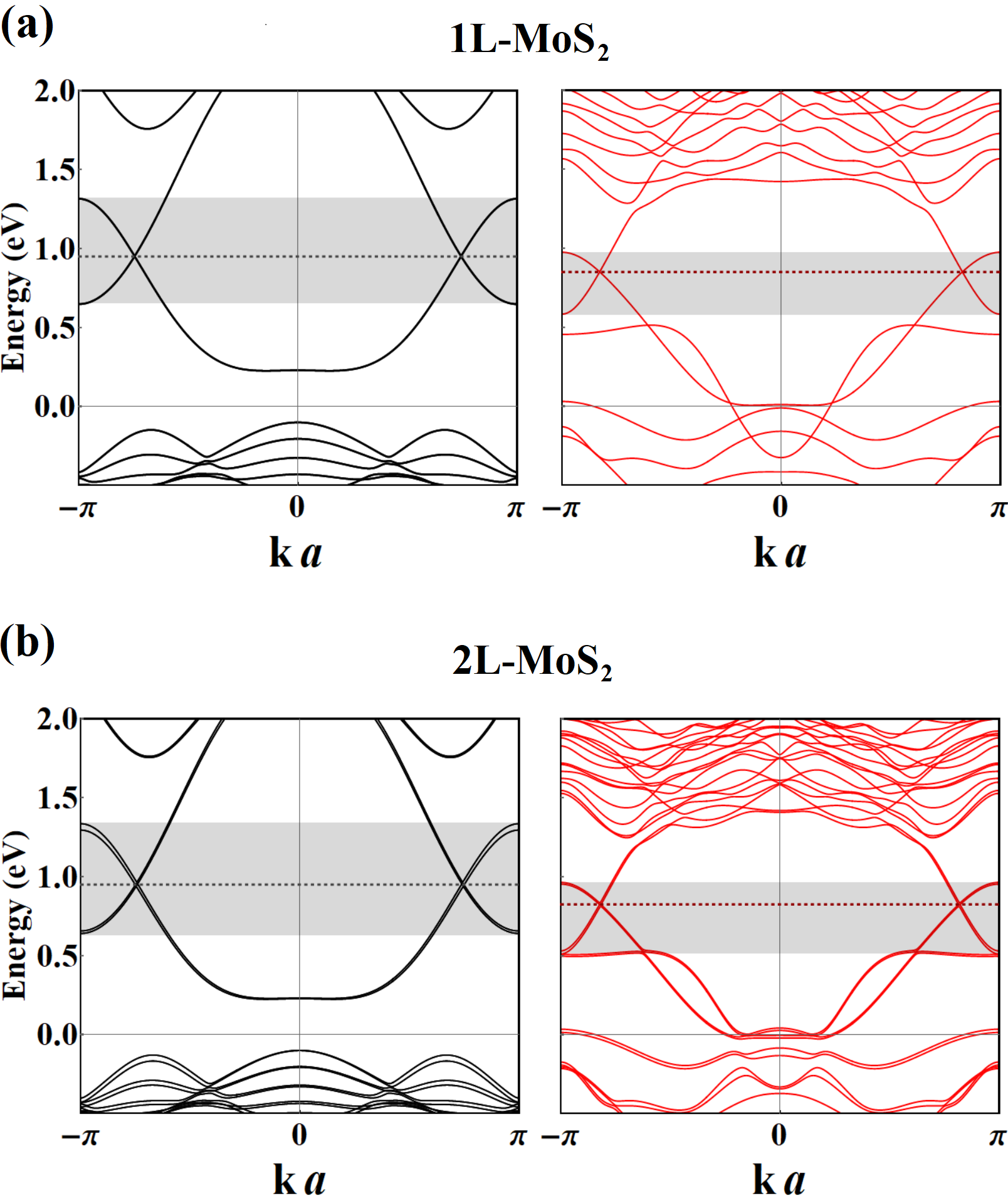}
	\caption{Electronic band structures of ZZ-nanoribbons extracted from 1L-MoS$_2$ (a) and 2L-MoS$_2$ (b). Black curves (left) represent the results obtained with the simplified three-band model [App.~\ref{App-3B-model}] and the red curves (right) results from DFT calculations [App.~\ref{App-complete-TB}]. The horizontal dashed lines designate the energy where the edge-states cross. The shaded regions illustrate the energy ranges that we explored with each method.}
 	\label{fig:figA1}
\end{figure}

\begin{figure}[h!]
	\includegraphics[width=1.0\linewidth,clip]{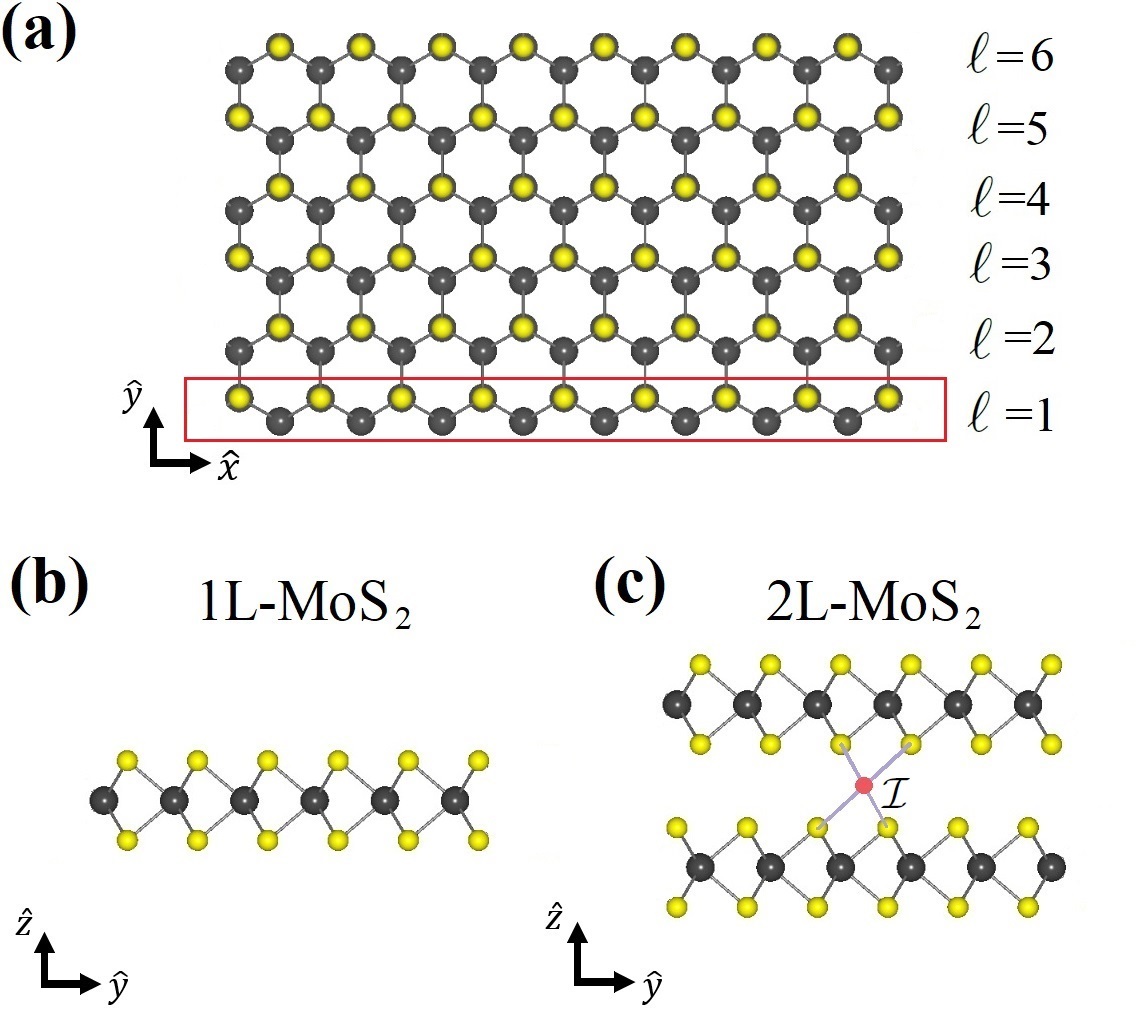}
	\caption{Schematic representation of MoS$_2$ nanoribbons. (a) Top view of ZZ-nanoribbon with $N=6$ lines in breadth. The electric field is applied along the $\hat{x}$-direction, where the nanoribbon is assumed periodic. (b) Frontal view of 1L-MoS$_2$ ZZ-nanoribbon. (c) Frontal view of 2L-MoS$_2$ ZZ-nanoribbon.The red dot between the layers is the spatial inversion symmetry center $\mathcal{I}$ of the bilayer system. }
	\label{fig:figA2} 
\end{figure}

Here, we shall explore the orbital responses of the unsupported nanoribbons using a simplified electronic structure model for MoS$_2$ \cite{Xiao-three-band} that takes into account three $d$-orbitals of the Mo atoms only, namely $d_{x^2-y^2}, d_{xy}$ and, $d_{z^2}$. The effect of the S atoms is introduced via perturbation theory. This concise model describes reasonably well the electronic spectra of ZZ-nanoribbons near the crossing energy of the edge states (see Fig. \ref{fig:figA1}), and has the advantage of being computationally inexpensive and easy to handle. The Hamiltonian of the three-band model was deduced in Ref.~\cite{Xiao-three-band} and is briefly reviewed in Appendix~\ref{APP-A}. The lack of the influence of substrate can be achieved experimentally properly choosing materials that do not interact significantly with MoS$_2$~\cite{Wu-2019-NatComn-10.1038/s41467-019-08629-9} [see also Sec. \ref{sec4} for further discussion]. Here, we neglect the spin-orbit coupling of MoS$_2$ due to the weakness of the spin response compared to the orbital one, as detailed in Appendix \ref{APP-D}. The nanoribbons comprise $N$ lines that are labelled by $\ell=1,2,...,N$, as illustrated in Fig.~\ref{fig:figA2}. The OAM accumulations induced at each line $\ell$ ($\delta \langle L^z_\ell\rangle$) by an electric field applied along the ribbon's axis are calculated using linear response theory as described in Appendix \ref{APP-B}. We use the intra-atomic approximation to describe the OAM operator. This should be a reasonable approximation for MoS$_2$ within this energy range, as discussed in Appendix \ref{APP-D}. The results are depicted in Fig.~\ref{FigMain1} for 1L-MoS$_2$ (a) and 2L-MoS$_2$ (b) ZZ-nanoribbons with $N=15$ lines in breadth for two different values of Fermi-energies ($E_{\text{F}}=1.0\text{eV}$ and $1.3\text{eV}$). 

\begin{figure}[h!]
	\includegraphics[width=1.0\linewidth,clip]{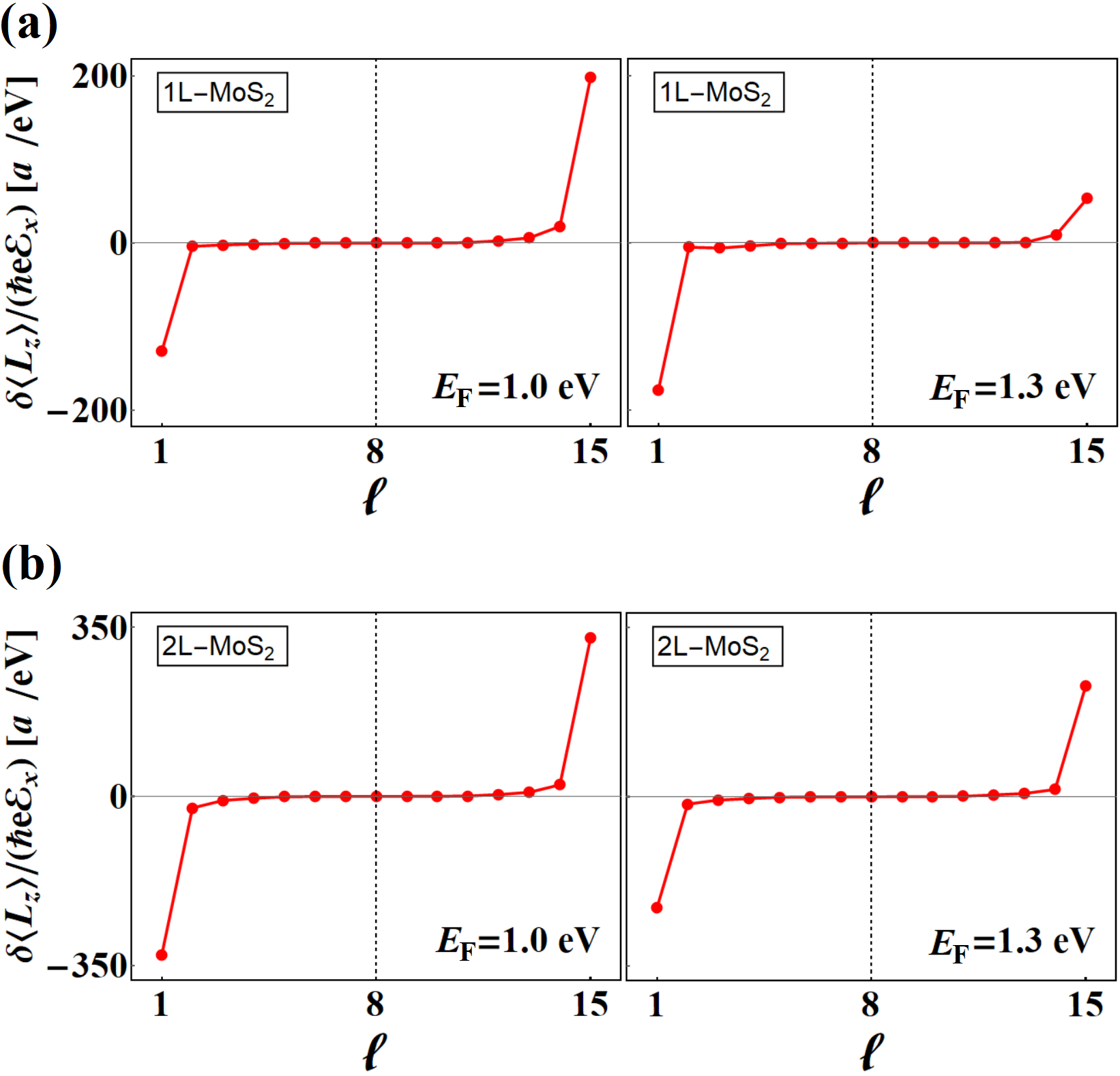}
	\caption{Profiles of the OAM accumulations induced by a longitudinally applied electric field calculated for ZZ nanoribbons of MoS$_2$ with 15 lines in breadth. The electronic structure is described by the three-band model \cite{Xiao-three-band} discussed in appendix \ref{App-3B-model}. Panels (a) and (b) display results for ZZ nanoribbons extracted from 1L-MoS$_2$ and 2L-MoS$_2$, respectively, both calculated for two different values of $E_{\text{F}}$ around the energy where the edge-states cross. All calculations were performed with $\Gamma=1 \text{meV}$.}
	\label{FigMain1} 
\end{figure}
 
We notice that the OAM accumulation profiles for these values of $E_\text{F}$ are mainly concentrated at the nanoribbons' edges. For the unsupported 2L-MoS$_2$ ZZ-nanoribbon they are clearly anti-symmetric with respect to the central line ($\ell=8$), as one would expect from a finite OHE that takes place in bulk 2L-MoS$_2$ \cite{Us3-PRL,Cysne-Bhowal-Vignale-Rappoport}. However, for the 1L-MoS$_2$ nanoribbon, the profiles of OAM accumulation are asymmetric, which indicates the presence of an additional symmetric contribution to the OAM accumulation profile coming from the OME, as discussed in Ref. \cite{Us4-PRB}. A necessary condition for the appearance of this current-induced orbital magnetization is spatial inversion symmetry breaking, which occurs in zigzag stripes extracted from monolayers, but not from bilayers. 
We reiterate that this is a necessary but not a sufficient condition. Crystalline symmetries bring additional constraints to the occurrence of the magneto electric effect. As mentioned in the introduction, a pristine unsupported 2D monolayer of MoS$_2$ belongs to the $D_{3h}$ point group symmetry that forbids the manifestation of the OME \cite{Furukawa-Itou-PhysRevResearch.3.023111, He-Law-PhysRevResearch.2.012073}. However, when it is cut into a ZZ nanoribbon, the point group symmetry reduces to $C_{2v}$ allowing the OME to take place. 

\begin{figure}[h!]
	\includegraphics[width=1.0\linewidth,clip]{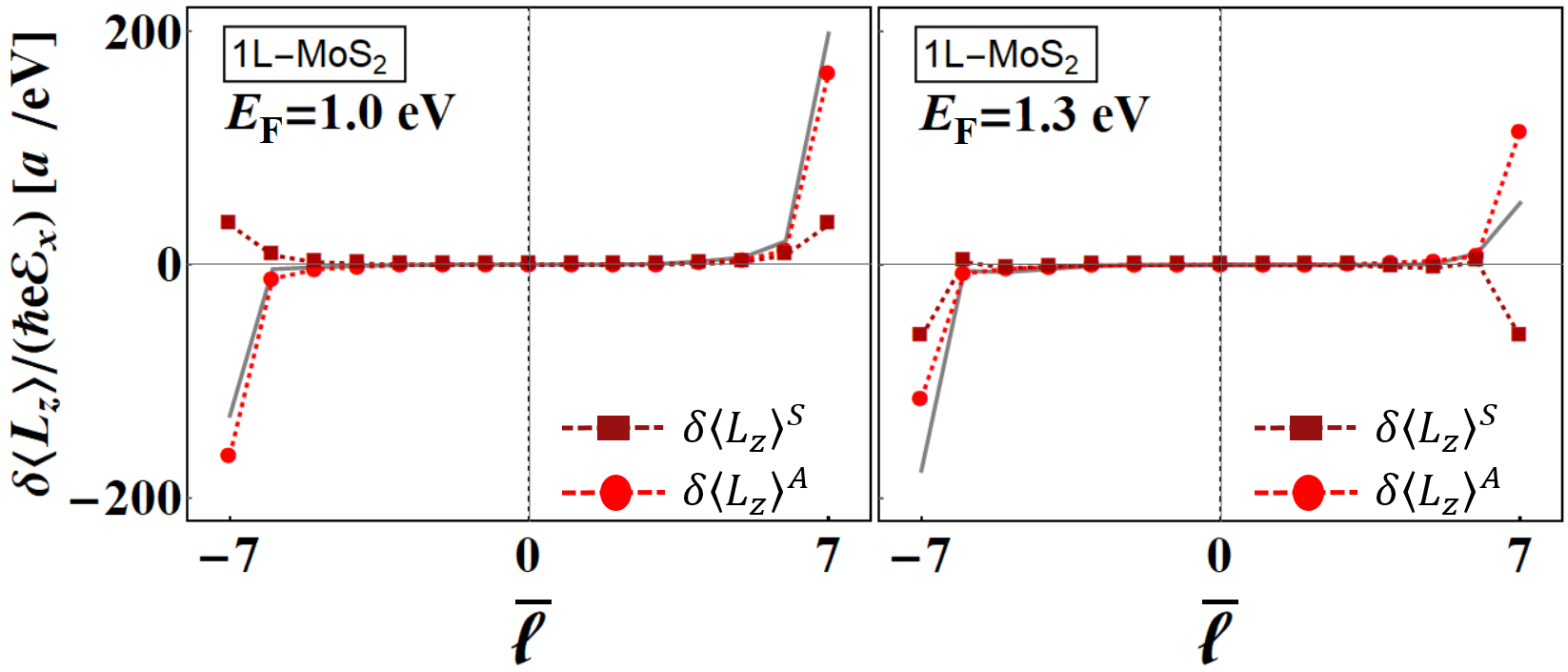}
	\caption{Decomposition into symmetric ($\delta \langle L^z \rangle^S$) and antisymmetric ($\delta \langle L^z \rangle^A$) components of the induced OAM profiles depicted in Fig. \ref{FigMain1} (a).}
	\label{FigMain2} 
\end{figure}

To separate the OHE and the OME, it is convenient to relabel the nanoribbons'  lines from $\ell=1,...,15$ to $\bar{\ell}=-7,...,7$.  We then decompose the OAM accumulation profile into symmetric (S) and anti-symmetric (A) components given by
 \begin{eqnarray}
\delta \langle L_{ \bar{\ell}}^{z}\rangle^{S/A}=\frac{1}{2}\left(\delta \langle L_{\bar{\ell}}^{z}\rangle \pm \delta \langle L_{-\bar{\ell}}^{z}\rangle \right). \label{Decomposition}
\end{eqnarray}
Fig. \ref{FigMain2} illustrates the decomposition of the profiles depicted in Fig. \ref{FigMain1} (a), evincing the relative intensities of the two contributions.

We identify the anti-symmetric component $\delta \langle L_{ \bar{\ell}}^{z}\rangle^A$ with the OHE contribution to the induced OAM profile, and the symmetric one $\delta \langle L_{ \bar{\ell}}^{z}\rangle^S$, which leads to a finite orbital magnetization when summed over all lines, is attributed to the OME. This identification follows the conventional literature on spintronics in nonmagnetic Rashba systems in which spin Hall effects and Edelstein effects are associated with antisymmetric and symmetric profiles of the spin response, respectively \cite{SHE-Review-RevModPhys.87.1213, 2DEG-Rashba-Acc-PhysRevLett.95.046601, 2DEG-Rashba-Acc-PhysRevB.72.245330}.
We note that the symmetric contribution changes sign when the value of $E_{\text{F}}$ varies from 1.0 to 1.3eV, indicating the possibility of manipulating the direction of the induced orbital magnetization with the use of gate voltages.  

In order to examine how both effects vary as functions of $E_{\text{F}}$ it is useful to introduce the quantities $M^z(E_{\text{F}})$, which represents the current-induced orbital magnetization associated with the symmetric component of the OAM accumulation profile, and $\Sigma^{z}_{\text{OH}}(E_{\text{F}})$ that measures the anti-symmetric accumulation of orbital angular momentum on each side of the nanoribbon due to the OHE. They are mathematically defined by
\begin{eqnarray}
M^z(E_{\text{F}})=-2\frac{\mu_B}{\hbar} \sum_{\bar{\ell}} \delta \langle L_{ \bar{\ell}}^{z}\rangle^{S},
\label{MZ}
\end{eqnarray}
where $\mu_B$ is the Bohr magneton and the factor 2 comes from the spin degeneracy. 

\begin{eqnarray}
\Sigma^{z}_{\text{OH}}(E_{\text{F}})= 2\frac{\mu_B}{\hbar} \sum_{\bar{\ell}} \text{sgn}\left(\bar{\ell} \right) \delta \langle L_{ \bar{\ell}}^{z}\rangle^{A},
\label{SigmaOHz}
\end{eqnarray}
where, $\text{sgn}\left(\bar\ell\right)$ represents the usual sign function.
It is noteworthy that both quantities saturate for sufficiently wide ribbons, as shown in Appendix \ref{App-NumberLines}.

\begin{figure}[h!]
	\includegraphics[width=0.8\linewidth,clip]{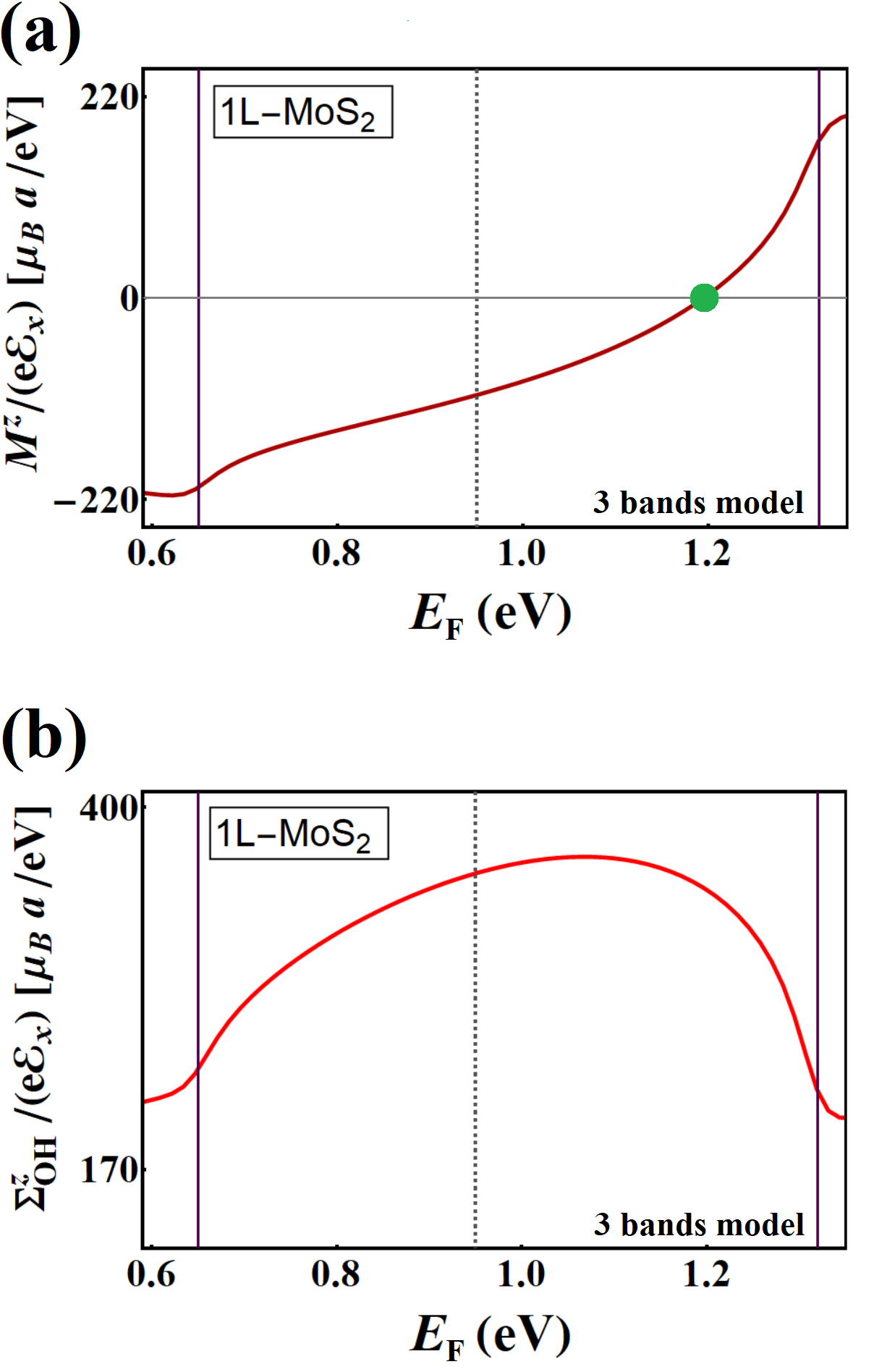}
	\caption{Current-induced orbital magnetization $M^z$ (a) and orbital Hall accumulation (OHA) $\Sigma^z_{\text{OH}}$ (b) per unit cell calculated as functions of $E_\text{F}$ for a ZZ nanoribbon of 1L-MoS$_2$ with 15 lines in breath. The vertical dashed line identifies the edge-states' crossover energy, and the purple solid lines delimit the energy range around it that we are interested in (see Fig. \ref{fig:figA1}).} 
	\label{FigMain3} 
\end{figure}
Fig. \ref{FigMain3} shows $M^z$ and $\Sigma^{z}_{\text{OH}}$ calculated as functions of $E_{\text{F}}$ for a ZZ-nanoribbon of 1L-MoS$_2$ 15  lines wide. The vertical dashed line specifies the energy where the two edge states cross. The vertical purple solid lines delimit the energy range within which the three-band model provides a reasonable description of the electronic edge-states (see Appendix \ref{APP-A}). We note in Fig. \ref{FigMain3} (a) that $M^z$ switches sign within this energy range, as previously pointed out. 

In our calculations we have used a value of $\Gamma=1 \text{meV}$ that correspond to a momentum relaxation time $\tau \approx 0.17 \text{ps}$. For 1L-MoS$_2$, the order of magnitude of current-induced magnetization per unity cell $M^z/(e \mathcal{E}_x)$ is approximately $100 \mu_B a/\text{eV}$. The lattice parameter of MoS$_2$ $a\approx 3\text{\AA}$. Thus, for an electric field intensity $\mathcal{E}=1\times 10^5 \text{V}/\text{m}$, we estimate $M^z\approx 3 \times 10^{-3} \mu_B$ per unit cell. This is an order of magnitude larger than the current-induced magnetic moment for Au(111), slightly larger than values obtained for Bi/Ag(111) and $\alpha$-Sn(001) surface \cite{Mertig-PhysRevB.97.085417}, and of same order of magnitude as the value estimated for NbS$_2$ in Ref. \cite{Bhowal-Satpahy-kOME-PhysRevB.102.201403} assuming a larger value of $\tau \mathcal{E}_x$. It is also instructive to estimate the longitudinal charge current $I_{\text{c}}$ that flows through the 1L-MoS$_2$ ZZ-nanoribbon using the same set of parameters. Within the energy range depicted in Fig. \ref{FigMain3}, we found that $I_{\text{c}}$, calculated for a nanoribbon with 15 lines in breadth, varies between $5-15 \mu$ A, which is compatible with experimental values for wider nanoribbons \cite{ChargeCurrent-MoS2, McClellan-doi:10.1021/acsnano.0c09078}.

\begin{figure}[h!]
	\includegraphics[width=0.8\linewidth,clip]{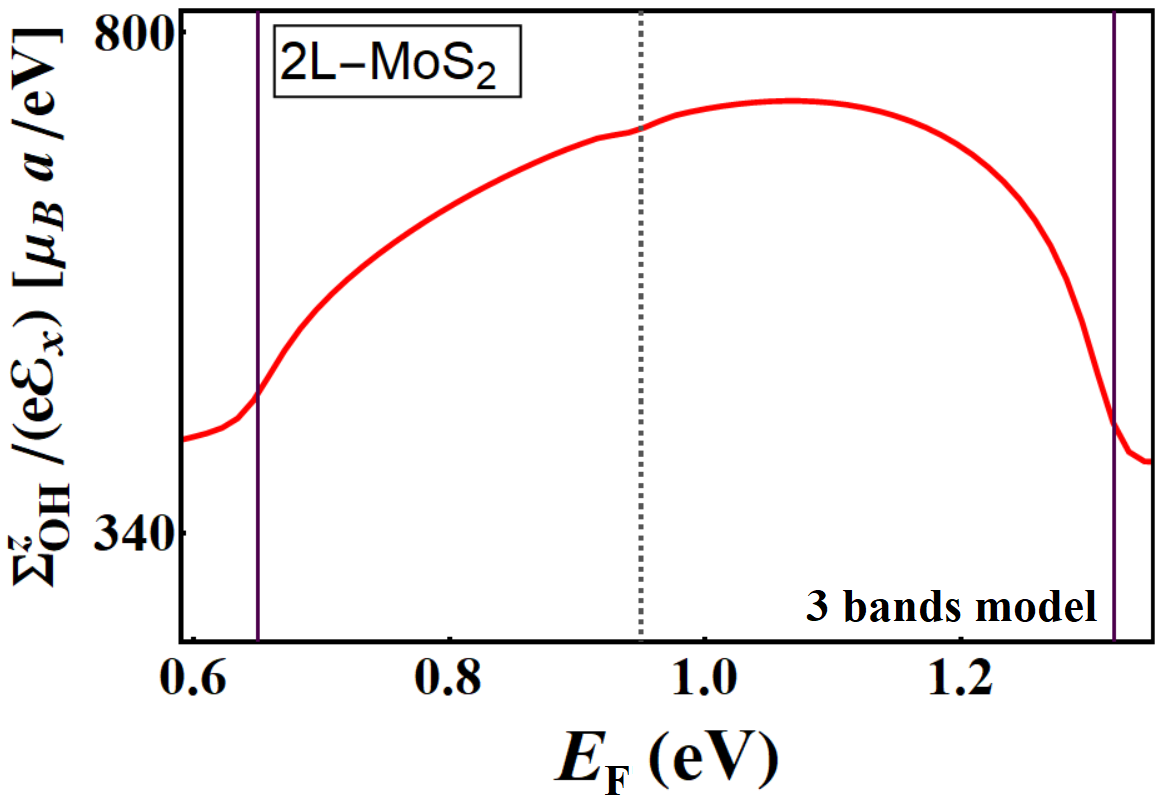}
	\caption{Orbital Hall accumulation $\Sigma^z_{\text{OH}}$ per unit-cell calculated as a function of $E_{\text{F}}$ for a ZZ nanoribbon of 2L-MoS$_2$ with 15 lines in breath. The vertical dashed line identifies the edge-states' crossover energy, and the purple solid lines delimit the energy range around it that we are interested in (see Fig. \ref{fig:figA1}).} 
	\label{FigMain4} 
\end{figure}

Fig. \ref{FigMain4} shows the orbital Hall accumulation $\Sigma^{z}_{\text{OH}}$ calculated as a function of $E_{\text{F}}$ for a 2L-MoS$_2$ ZZ nanoribbon with 15 lines in breadth. It is noteworthy that it exhibits approximately twice the value obtained for the 1L-MoS$_2$ ZZ nanoribbon, which is consistent with previous results of orbital Hall conductivities for mono- and bi-layers of 2H-MoS$_2$,reported in Ref. \onlinecite{Us3-PRL}. However, $M^z=0$ for the 2L-MoS$_2$ ZZ nanoribbon, as expected, because it preserves spatial-inversion symmetry.     

\section{DFT-based calculations} \label{sec3}

\begin{figure}[h!]
	\includegraphics[width=0.8\linewidth,clip]{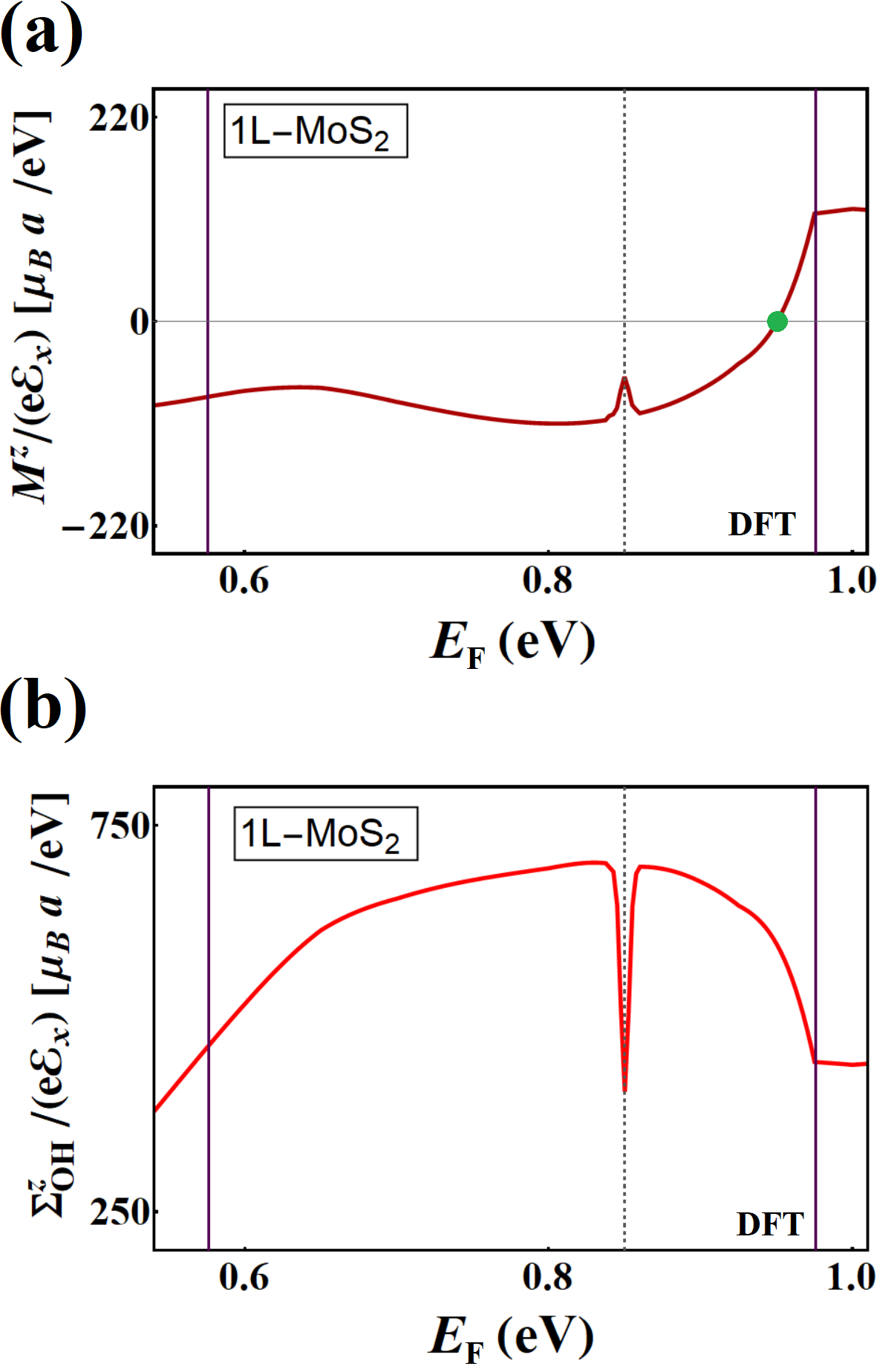}
	\caption{$M^z$ (a) and $\Sigma^z_{\text{OH}}$ (b) per unit cell calculated as functions of $E_\text{F}$ for a ZZ nanoribbon of 1L-MoS$_2$ with 8 lines in breath, using $\mathcal{H}_{\text{PAO}}(k)$. The vertical dashed line identifies the energy where the edge states would cross, and the purple solid lines delimit the energy range that we are interested in (see right panel of Fig.~\ref{fig:figA1}).}
	\label{FigMain5} 
\end{figure}

\begin{figure}[h!]
	\includegraphics[width=0.8\linewidth,clip]{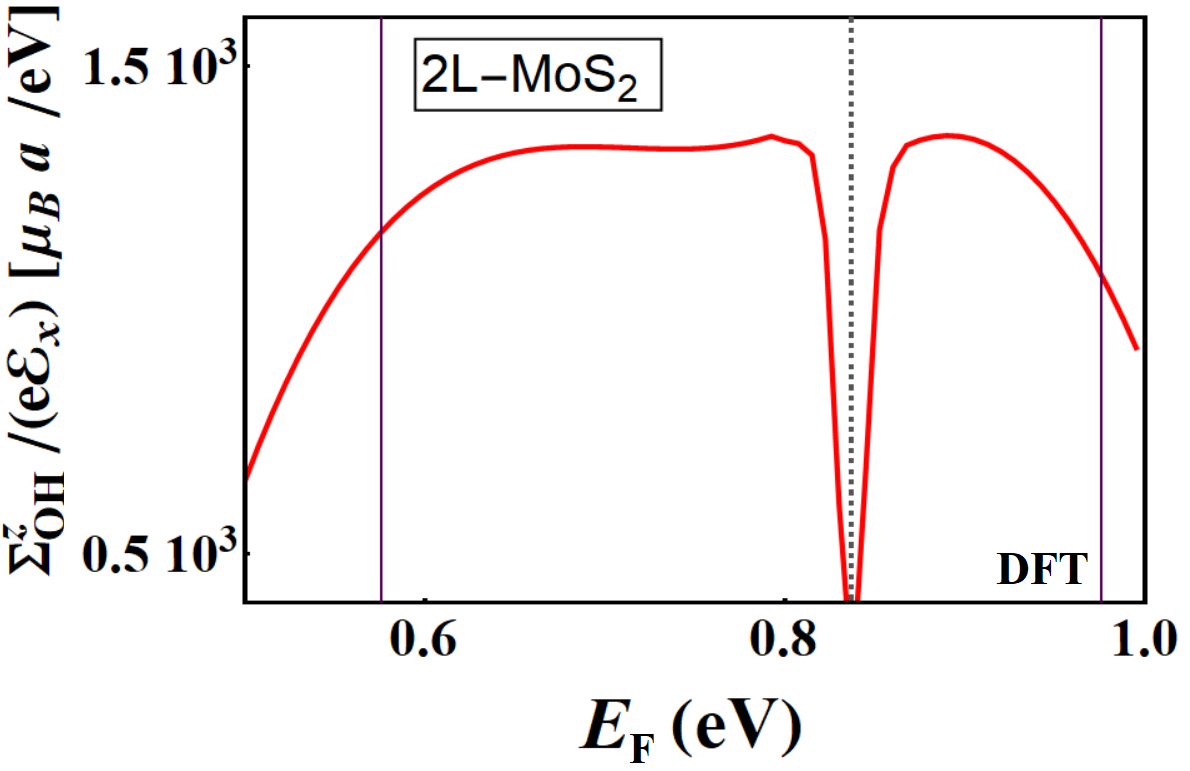}
	\caption{$\Sigma^z_{\text{OH}}$ per unit-cell, calculated as a function of $E_{\text{F}}$, for a ZZ nanoribbon of 2L-MoS$_2$ with 6 lines in breadth, using $\mathcal{H}_{\text{PAO}}(k)$. The vertical dashed line identifies the energy where the edge states would cross, and the purple solid lines delimit the energy range that we are interested in (see right panel of Fig.~\ref{fig:figA1}).}
	\label{FigMain6} 
\end{figure}

\begin{figure}[h!]
	\includegraphics[width=0.8\linewidth,clip]{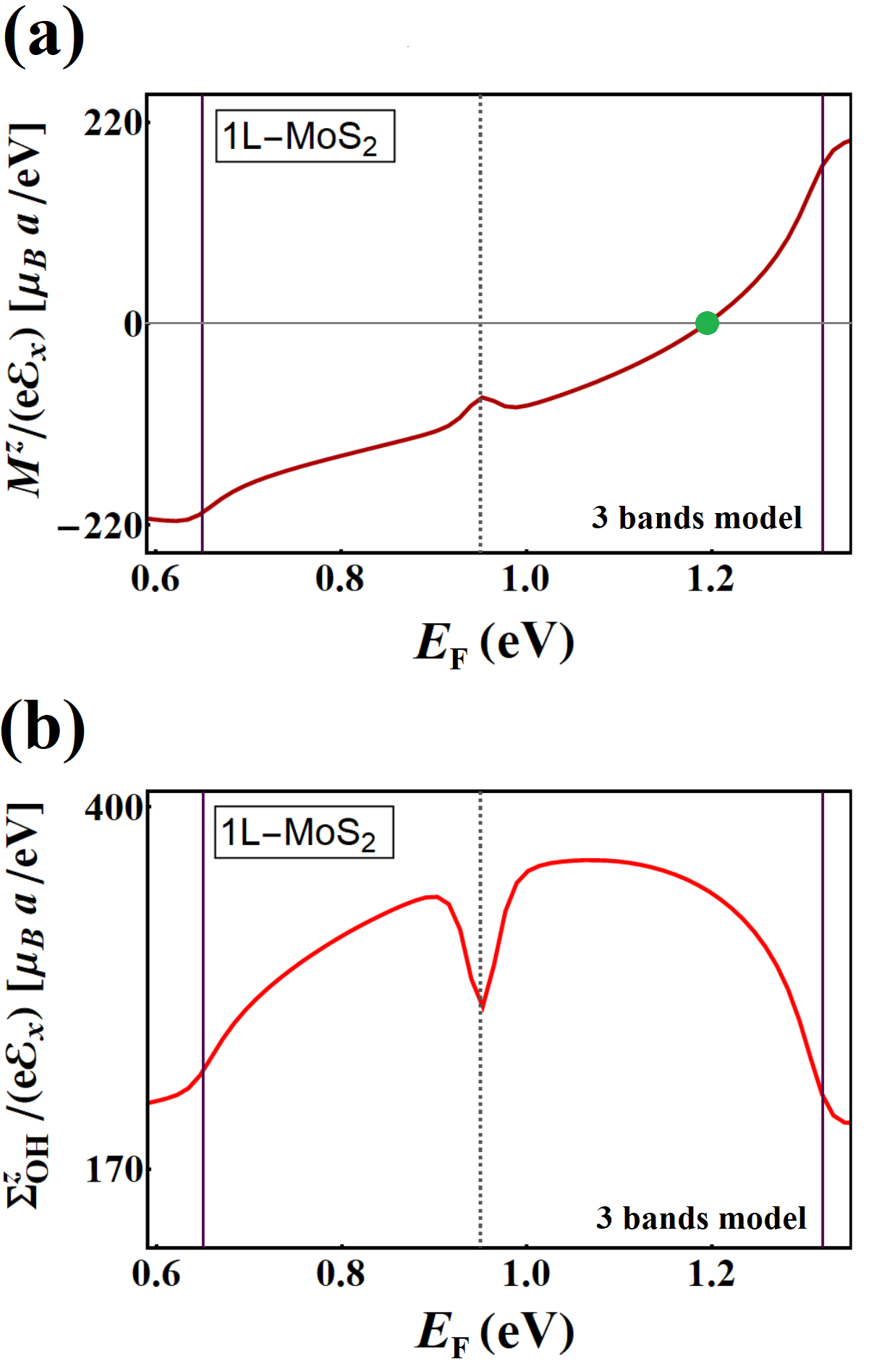}
	\caption{Current-induced orbital magnetization $M^z$ (a) and orbital Hall accumulation (OHA) $\Sigma^z_{\text{OH}}$ (b) per unit cell calculated as functions of $E_\text{F}$ for a ZZ nanoribbon of 1L-MoS$_2$ with 8 lines in breath, using the three-band model. The vertical dashed line identifies the edge-states' crossover energy, and the purple solid lines delimit the energy range around it that we are interested in (see left panel of Fig. \ref{fig:figA1}).} 
	\label{FigMain3b-8l} 
\end{figure}

\begin{figure}[h!]
	\includegraphics[width=0.8\linewidth,clip]{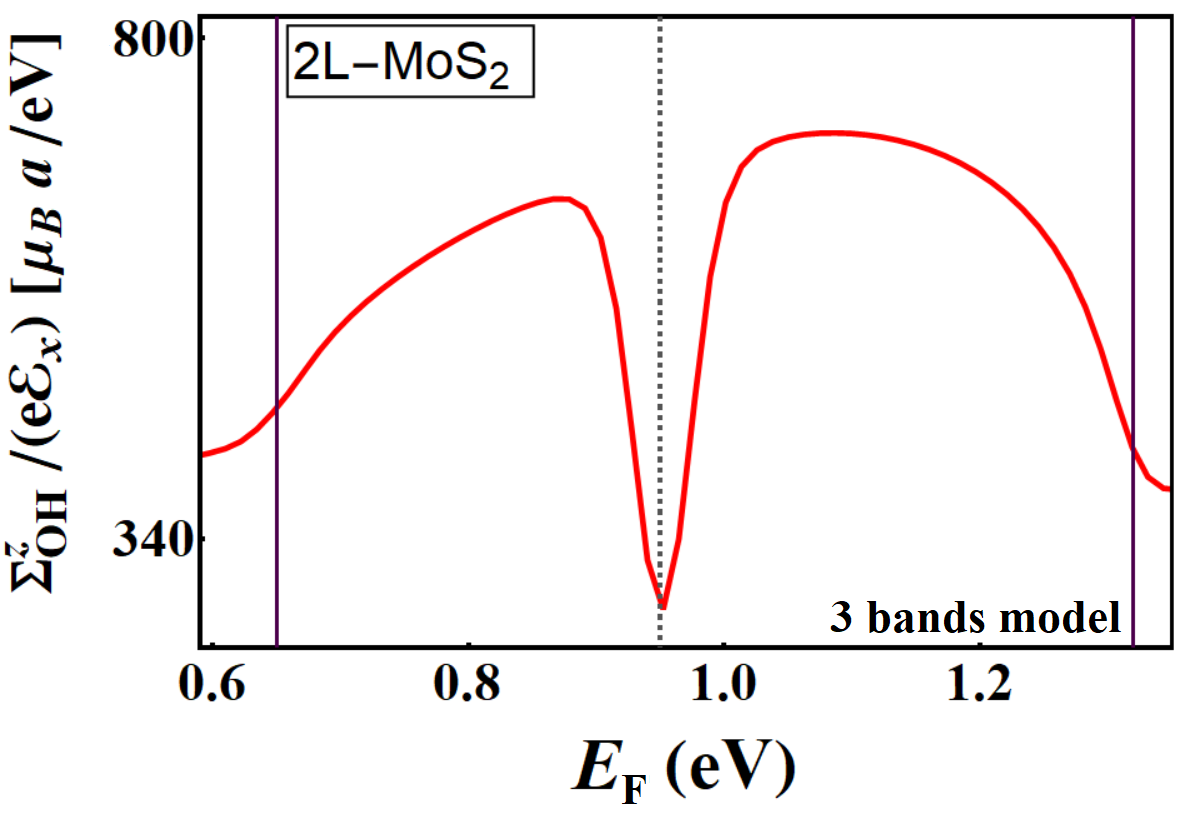}
	\caption{Orbital Hall accumulation $\Sigma^z_{\text{OH}}$ per unit-cell calculated as a function of $E_{\text{F}}$ for a ZZ nanoribbon of 2L-MoS$_2$ with 6 lines in breath, using the three-band model. The vertical dashed line identifies the edge-states' crossover energy, and the purple solid lines delimit the energy range around it that we are interested in (see left panel of Fig. \ref{fig:figA1}).} 
	\label{FigMain43b-6l} 
\end{figure}

To verify the quality of the results obtained with the three-band model, it is instructive to compare them with those of more elaborate methods. For this purpose, we recalculate $M^z$ and $\Sigma^z_{\text{OH}}$ as functions of $E_{\text{F}}$ employing a DFT-based approach. 
In our DFT calculations we use the pseudo atomic orbitals (PAO) projection method \cite{PAO1,PAO6} to construct an effective Hamiltonian $\mathcal{H}_{\text{PAO}}(k)$ with a basis that includes $sspd$ and $sp$ orbitals for the Mo and S atoms, respectively (more details are given in Appendix \ref{APP-A}). The OAM accumulation profiles are obtained by the method described in the Appendix \ref{APP-B}, and Eqs.~\ref{MZ} and~\ref{SigmaOHz} are used to compute $M^z$ and $\Sigma^z_{\text{OH}}$. However, due to the relatively high computational costs of DFT, we limited our calculations to narrower nanoribbons: $N=8$ for 1L-MoS$_2$ and $N=6$ for 2L-MoS$_2$ ZZ nanoribbons. The results are depicted in Figs. \ref{FigMain5} and \ref{FigMain6}, respectively. It is noteworthy that a very small energy band gap appears in the electronic spectra of these narrow ribbons due to lateral confinement of the electronic wave functions \cite{PKim-PhysRevLett.98.206805}, producing a protuberance in $M^z$ and a depression in $\Sigma^{z}_{\text{OH}}$ in the vicinity of the energy where the edge-states would otherwise cross. The same peculiarities happen when the three-band model is used to describe the electronic structure of these ribbons, as Figs.~\ref{FigMain3b-8l} and \ref{FigMain43b-6l} illustrate. Nevertheless, this small band gap diminishes rapidly as the ribbon width increases, vanishing for nanoribbons over 14 lines wide, as we shown Appendix \ref{App-NumberLines}. 

The qualitative agreement between the two approaches shows that the three-band model captures the main features of the electrically induced orbital angular accumulations in these systems. 

\section{Discussions \label{sec4}}

It is instructive to examine the microscopic features involved in the orbital responses of these nanoribbons. We start with the OHE, which is associated with \emph{equilibrium} in-plane orbital texture in momentum space \cite{Go-Jo-Kim-Lee-PhysRevLett.121.086602}. This texture is revealed by evaluating the expectation values of the OAM operator components for each eigenstate, i.e., $\langle L^{\mu} \rangle_{nk}=\langle nk | L^{\mu} | nk \rangle$, where $\mu=x,y,z$. It is worth mentioning that within the three-band tight-binding model we are restricted to a sector of the $L=2$ angular momentum vector space spanned by the eigenstates of $L^z$ associated to $m_L = 0,\pm 2\hbar$ only. Within this sector, the matrix representation of the operator $L^z$ is given by
\begin{eqnarray}
L^{z}= \hbar\begin{bmatrix}
0 & 0 & 0\\
0 & 0 & 2i\\
0 & -2i & 0
\end{bmatrix}. 
\label{lz}
\end{eqnarray}
It is useful to introduce a pseudo-angular momentum algebra in this sector where the in-plane components of the OAM operator are obtained from $\hat{L}^z$ by enforcing the commutation relations $[\frac{1}{2}L^i,\frac{1}{2}L^j]=-2i\hbar\epsilon_{ijk}\frac{1}{2}L^k$, as discussed in Ref.~\onlinecite{Us2-PRBR}. Here, 
$i,j,k$ denote the Cartesian directions $x,y,z$, respectively, and the Einstein’s summation convention is used for the indices. We then find 
\begin{eqnarray}
L^x=\hbar\begin{bmatrix}
  0  &  0  & 0  \\
  0  &  2 &  0 \\
  0 &  0 & -2\end{bmatrix}, 
  \label{lx}
\end{eqnarray}
and
\begin{eqnarray}
L^y=\hbar\begin{bmatrix}
  0  &  0  & 0  \\
  0  &  0 &  2 \\
  0 &  2 & 0\end{bmatrix}.  
  \label{LxLy3B}
\end{eqnarray}

\begin{figure}[h!]
	\includegraphics[width=1.0\linewidth,clip]{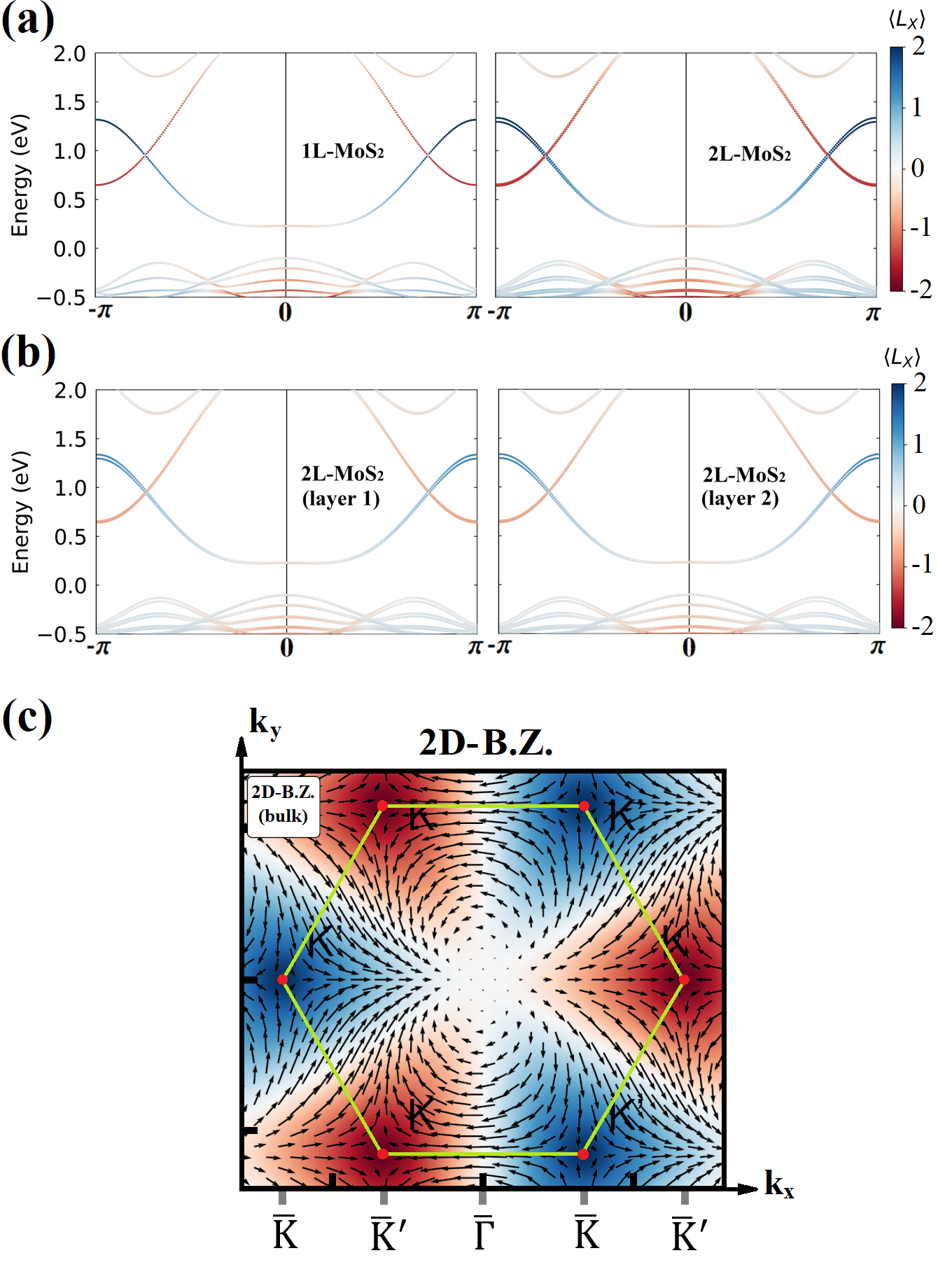}
	\caption{(a) Expectation values of the $x$-component of the OAM operator $\langle L^x \rangle_{nk}$ calculated for each eigenstate for 1L-MoS$_2$ and 2L-MoS$_2$ ZZ nanoribbons with 15 lines in breadth. (b) Layer-projected values of $\langle L^x \rangle_{nk}$ for the 2L-MoS$_2$ ZZ nanoribbon. (c) Orbital texture calculated for the valence band of the MoS$_2$ within the 2D Brillouin zone, and its projection along the ZZ nanoribon's wavevector $k\equiv k_x$.}
	\label{Fig_Lx} 
\end{figure}

\begin{figure}[h!]
	\includegraphics[width=1.0\linewidth,clip]{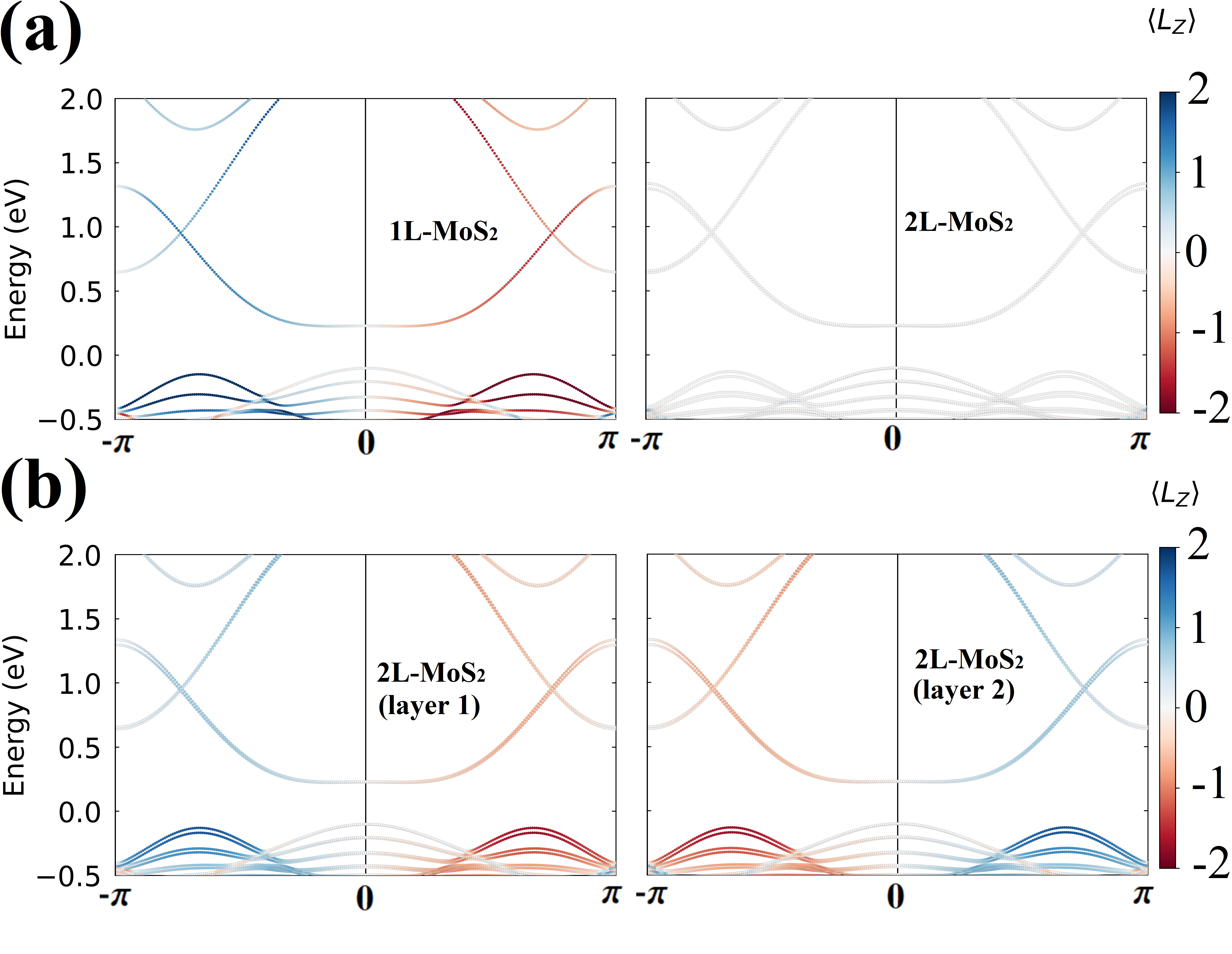}
	\caption{(a) Expectation values of the $z$-component of the OAM operator $\langle L^z \rangle_{nk}$ calculated for each eigenstate associated with band $n$, as functions of the wave vector $k$, for 1L-MoS$_2$ and 2L-MoS$_2$ ZZ nanoribbons with 15 lines in breadth. (b) layer-projected values of $\langle L^z \rangle_{nk}$ for 2L-MoS$_2$ ZZ nanoribbon.}
	\label{Fig_Lz} 
\end{figure}

Figs.~\ref{Fig_Lx} and~\ref{Fig_Lz} show the results of the expectation values of the OAM components $\langle L^x \rangle_{nk}$ and $\langle L^z \rangle_{nk}$, respectively. They are calculated, as functions of the wave vector $k$, for each energy band $n$ of ZZ nanoribbons with 15 lines in breadth, extracted from a monolayer and from a bilayer of MoS$_2$. $\langle L^y \rangle_{nk}=0$ for both cases. In Fig.~\ref{Fig_Lx}~(a) we note that the spectrum of $\langle L^x \rangle_{nk}$ for the monolayer is similar to the bilayer one, although the latter contributes with twice the number of bands. The values of $\langle L^x \rangle_{nk}$ projected on each layer of the bilayer system are shown in Fig.~\ref{Fig_Lx}~(b). They are identical and justify why $\Sigma^z_{\text{OH}}$ for the bilayer is approximately twice that for the monolayer. 

Two-dimensional MoS$_2$ displays in-plane orbital textures that lead to OHE \cite{orbitaltexture-microscopicstudy-Seungyun, Us2-PRBR, Us3-PRL, Cysne-Bhowal-Vignale-Rappoport}. These textures have been observed in some TMDs with the use of optical probes \cite{Schuler-Pincelli-Beaulieu-PhysRevX.12.011019, Beaulieu-Schusser-2020-PhysRevLett.125.216404}.
Fig.~\ref{Fig_Lx}~(c) illustrates the orbital texture calculated for the valence band of a monolayer of MoS$_2$ within the 2D Brillouin zone (BZ), together with its projection along the ZZ nanoribon's wavevector $k \equiv k_x$. Inspection of the OAM texture depicted in Fig. \ref{Fig_Lx}~(c) shows that $\langle L^y \rangle_{k_x,k_y}=-\langle L^y \rangle_{k_x,-k_y}$,  which leads to $\langle L^y \rangle_{nk}=0$ when projected onto $k$.

Fig.~\ref{Fig_Lz} shows the expectation values of the $z$-component of the OAM operator $\langle L^z \rangle_{nk}$ calculated for each eigenstate associated with band $n$, as functions of the wave vector $k$, for 1L-MoS$_2$ and 2L-MoS$_2$ ZZ nanoribbons with 15 lines in breadth. We note in Fig.~\ref{Fig_Lz} (a) that $\langle L^z \rangle_{nk}$ switches sign when $k \rightarrow -k$ for the ZZ nanoribbon extracted from a monolayer of MoS$_2$, and it vanishes for the one taken from the bilayer. Fig.~\ref{Fig_Lz} (b) shows the values of $\langle L^z \rangle_{nk}$ in each layer of the 2L-MoS$_2$ ZZ nanoribbon. They reveal that the $\langle L^z \rangle_{nk}$ spectra in each layer have opposite signs that cancel in the bilayer. 

The z-component of angular momentum expectation value in a 2D monolayer of MoS$_2$ is related to its topological features. MoS$_2$ is topologically trivial with respect to $\mathbb{Z}_2$-index, and its nanoribbon edge-states are not protected by Kramer's degeneracy \cite{Emilia-Caio-PhysRevB.95.035430}. Notwithstanding, MoS$_2$ has a non-trivial topology related to OAM \cite{Quian-Liu-Liu-PhysRevB.105.045417, Jiang-Xie-PhysRevB.104.L161108, Us_Plus_FazzioPeople-Arxiv}, which may be categorized by an orbital Chern number \cite{Us2-PRBR, Us3-PRL, Cysne-Bhowal-Vignale-Rappoport} that indicates the presence of orbital-polarized edge states, as those shown in Fig. \ref{Fig_Lz}. 

\begin{figure}[h!]
	\includegraphics[width=0.7\linewidth,clip]{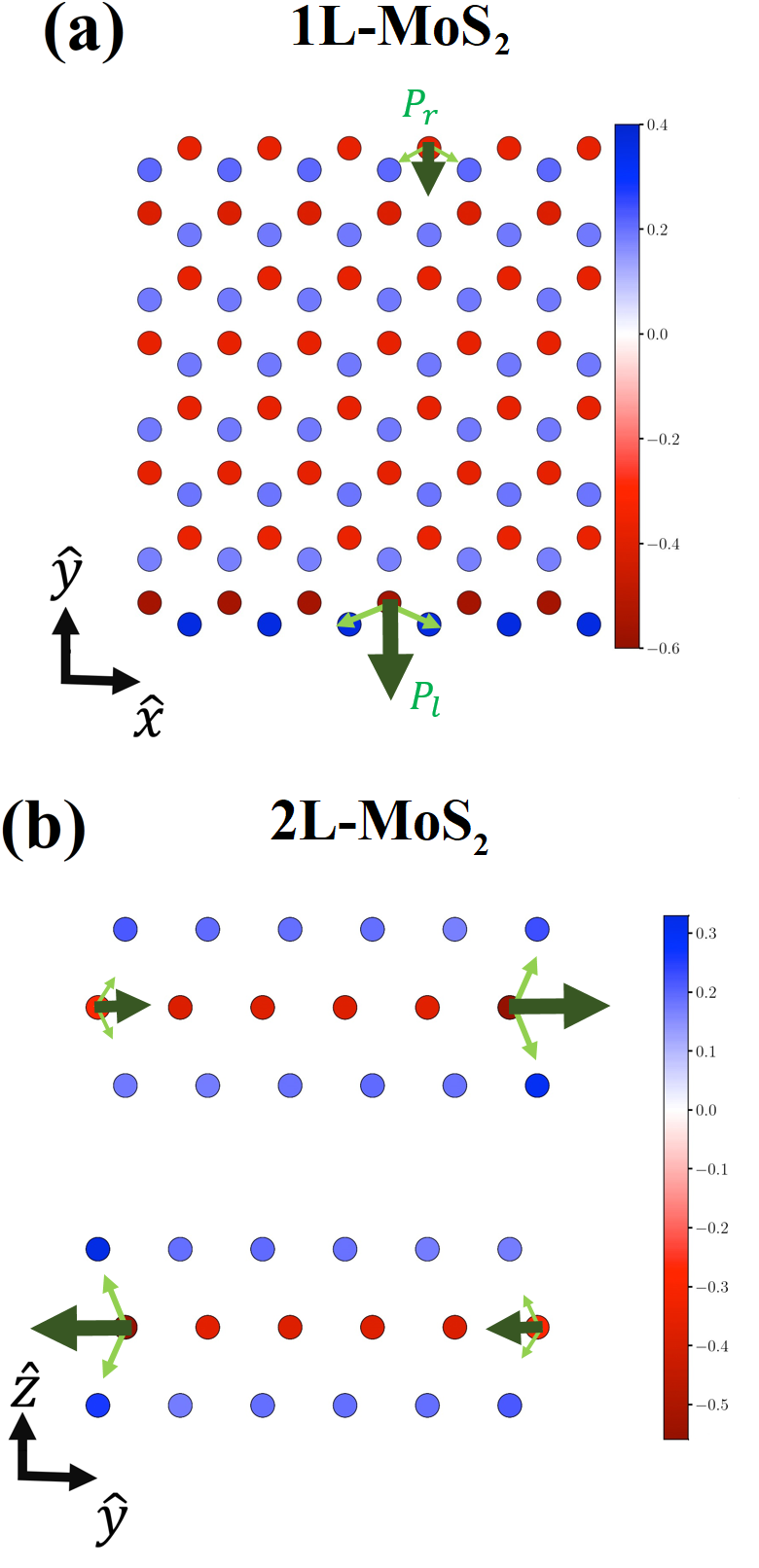}
	\caption{The equilibrium charge density of ZZ-nanoribbon obtained from DFT calculations. (a) Top view of the 1L-MoS$_2$. (b) Frontal view of the 2L-MoS$_2$. The green arrows show the dipole moment formed near one of the edges.}
	\label{Charge-density} 
\end{figure}

\begin{figure}[h!]
	\includegraphics[width=1.0\linewidth,clip]{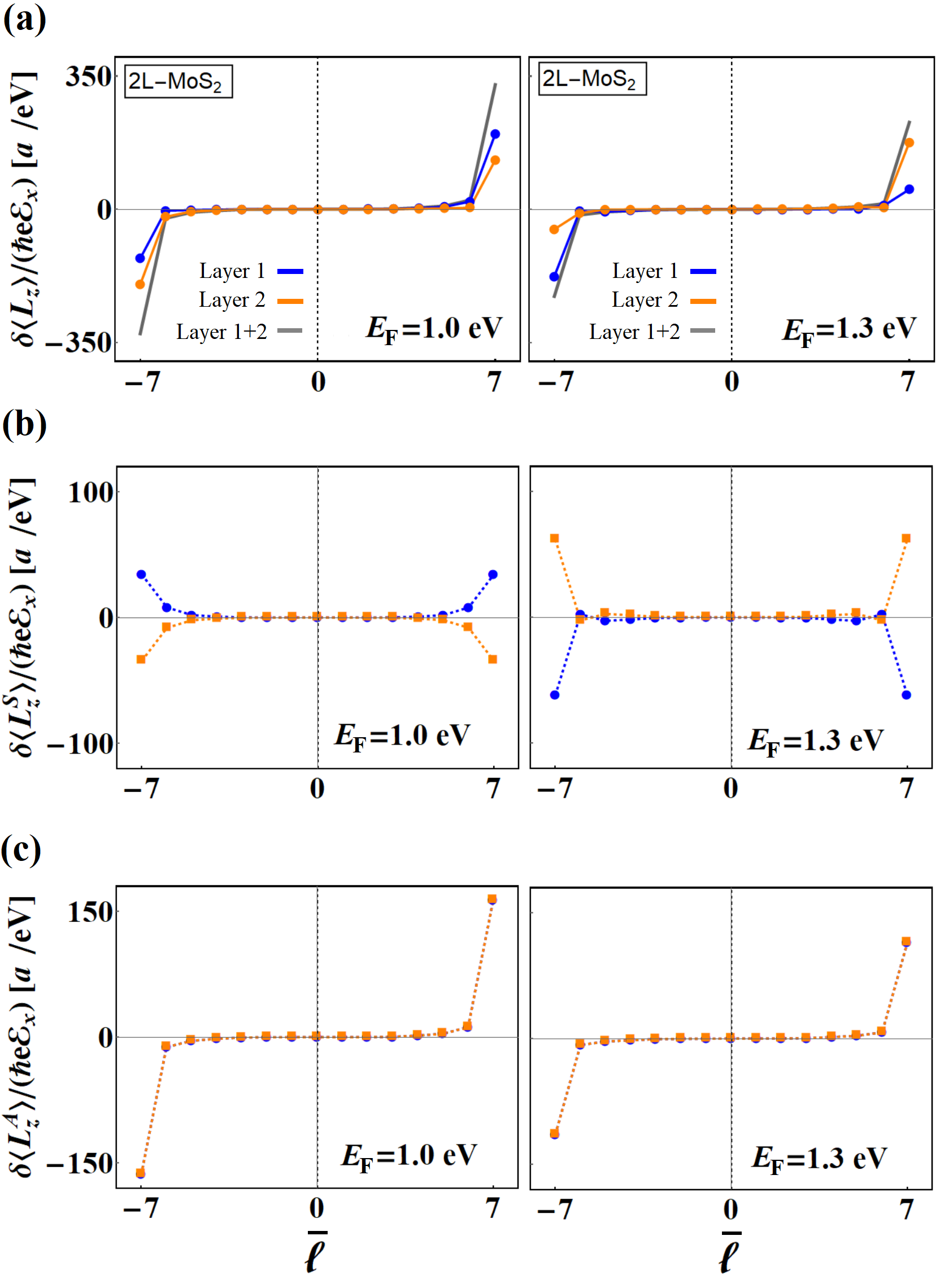}
	\caption{Profiles of the OAM accumulations induced by a longitudinally applied electric field calculated for ZZ nanoribbons with 15 lines in breadth, extracted from a bilayer of MoS$_2$. The electronic structure is described by the three-band model. All calculations were performed with $\Gamma=1 \text{meV}$ for $E_{\text{F}}= 1.0$, and $1.3$eV. Layer-resolved profiles projected onto layer 1 (blue) and 2 (orange) are shown in panel(a). The gray line illustrates the total OAM profile. The symmetric and anti-symmetric components are depicted in panels (b) and (c), respectively, for each layer.}
	\label{FigMain2L-NoGate} 
\end{figure}

Let us now address the appearance of the OME in these nanoribbons. We recall that in the formation of a MoS$_2$ molecule a small amount of electronic charge is transferred from the Mo to the S atoms, leading to the appearance of finite electric dipoles. 
For an unsupported 2D monolayer of MoS$_2$ in the H structural phase the molecular dipoles compensate each other and the net electric polarization of the system vanishes. However, when a nanoribbon with zigzag edges is extracted from the MoS$_2$ monolayer, an overall in-plane polarization $\vec{P}$ emerges along the transverse $\hat{y}$ direction, as schematically illustrated in Fig.\ref{Charge-density} (a). Therefore, an electric field $\vec{\mathcal{E}}$ applied along the $\hat{x}$ direction, will exert torque in system and induce a net orbital magnetization $M^z \propto P^y \mathcal{E}_x$ \cite{NatComn-10.1038-Salemi2019, Us4-PRB}. For ZZ nanoribbons taken from the 2H-MoS$_2$ bilayer the OME vanishes because inversion symmetry is restored. Nevertheless, each layer separately exhibits a net in-plane polarization of the same intensity but with opposite signs \cite{Layer-resolved-polarization}, as Fig.\ref{Charge-density} (b) illustrates. Consequently, under a longitudinally applied electric field, each layer would display OME with induced orbital magnetization pointing at opposite directions, which leads to no net OME. This microscopic feature is illustrated in Fig.~\ref{FigMain2L-NoGate}. It shows the layer-resolved profiles of the OAM accumulations induced by a longitudinally applied electric field calculated for ZZ nanoribbons with 15 lines in breadth, extracted from a bilayer of MoS$_2$. The electronic structure is described by the three-band model for two different values of $E_{\text{F}}$. The layers are numbered by 1 and 2, respectively. Panel (a) the shows the induced total OAM profile (gray solid line) and  layer-resolved ones associated with layer 1 (blue) and 2 (orange), respectively. The pure anti-symmetric character of the gray solid line demonstrates that the OME vanishes in this system. However, the asymmetric aspects of both the blue and orange layer-resolved profiles clearly indicate the presence of OME in each layer separately. In panel (b) we depict the symmetric components of the layer-resolved profiles to show that the induced orbital magnetization in each layer do indeed appear with the same intensity but point at opposite directions. Panel (c) shows that anti-symmetric components of the layer resolved profiles are identical and equally contribute to the OHE.     

\begin{figure}[h!]
	\includegraphics[width=1.0\linewidth,clip]{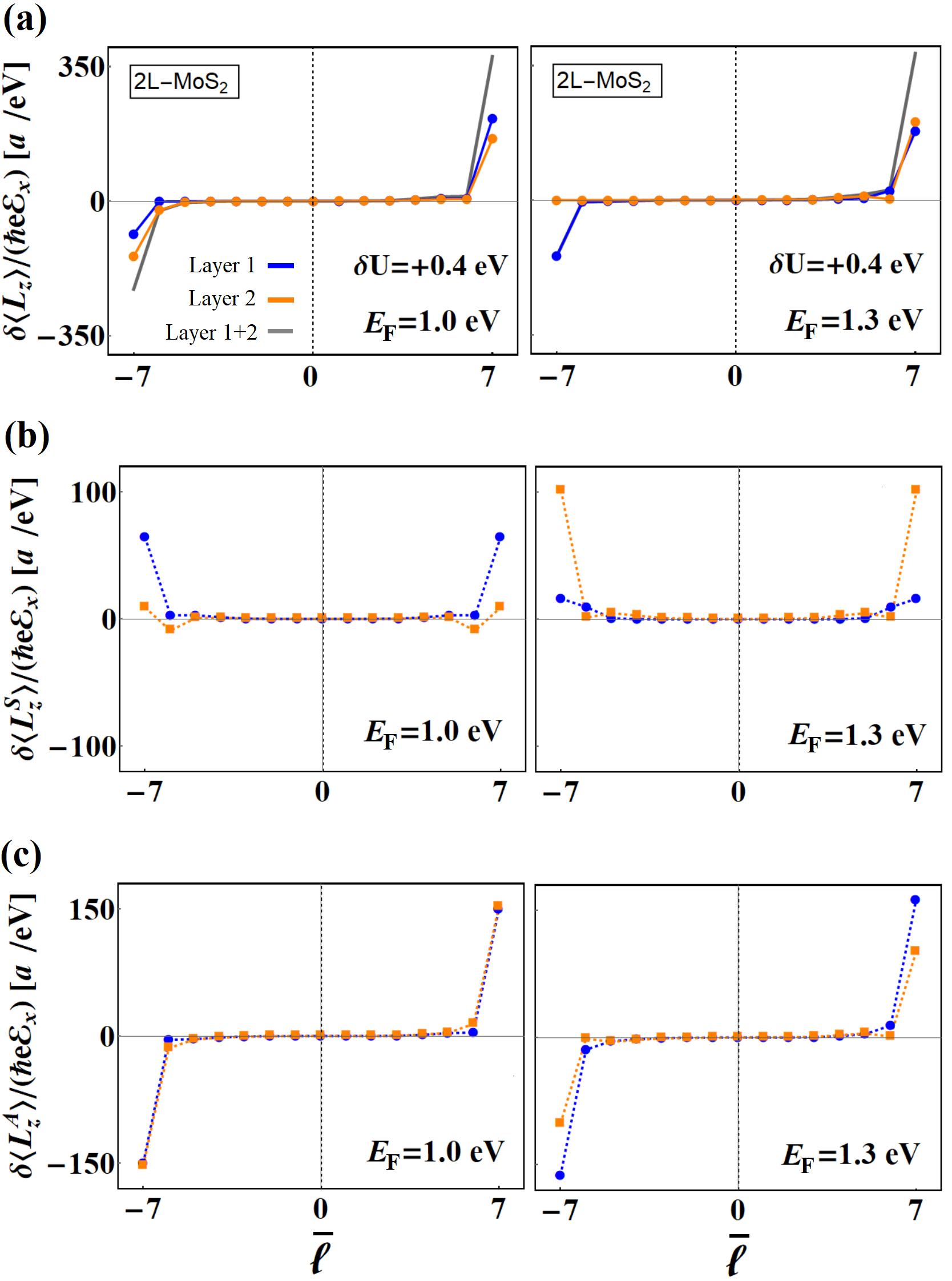}
	\caption{Layer-resolved OAM response to an applied electric-field for biased ($\delta U=U_2-U_1=+0.4$ eV) 2L-MoS$_2$ ZZ-nanoribbon. The left (right) panels shows the results for $E_\text{F}=1.0 (1.3)$ eV. Panel (a) shows the OAM response profile projected in layers 1 (blue) and 2 (orange). The gray line shows the total orbital response profile. In panel (b) we show the symmetric component of the profile for each layer. In panel (c) we show the anti-symmetric component. Here, we used $\Gamma=1 \text{meV}$.}
	\label{FigMain2L-WithGate} 
\end{figure}

\begin{figure}[h!]
	\includegraphics[width=0.8\linewidth,clip]{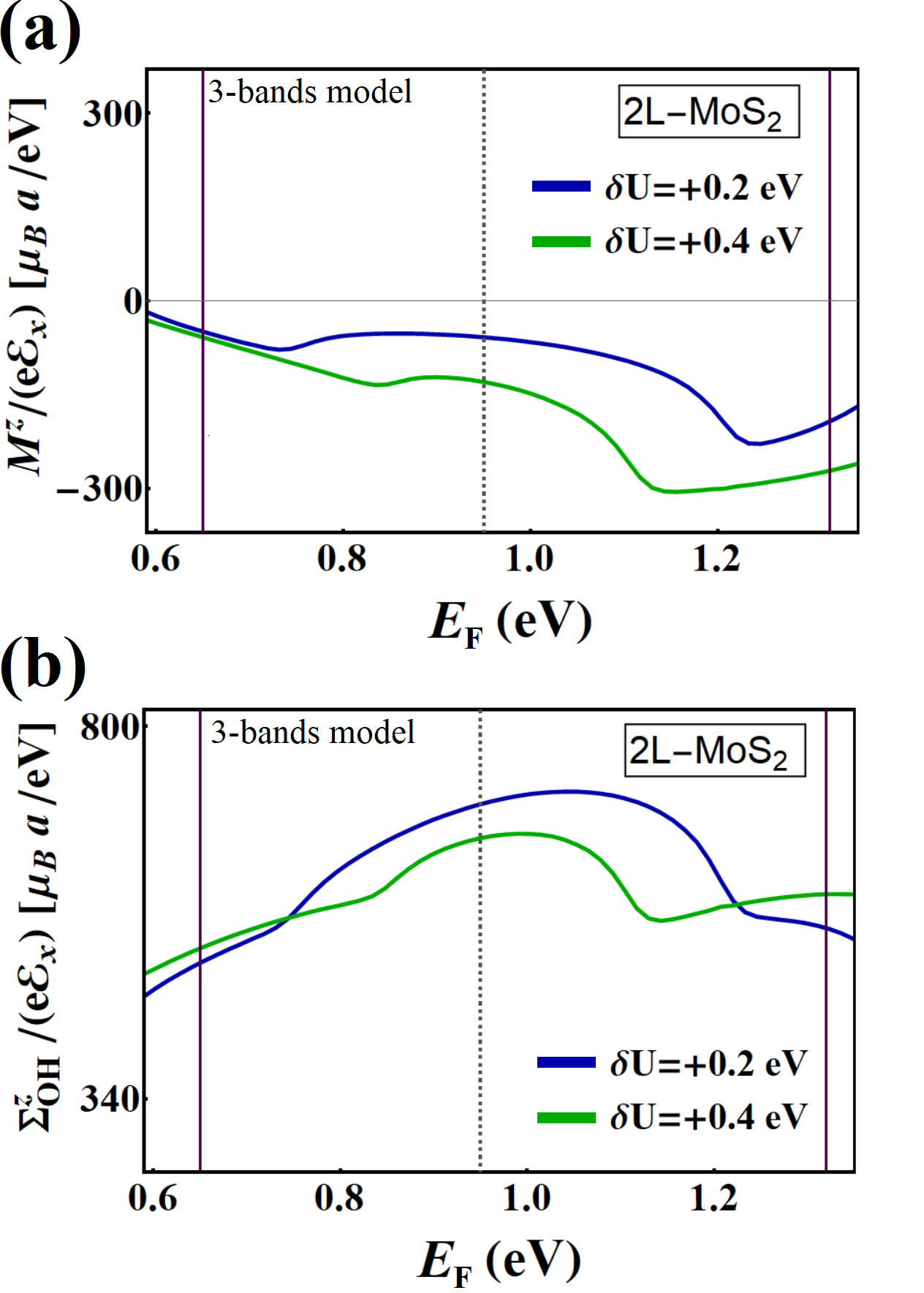}
	\caption{Current-induced orbital magnetization $M^z$ (a) and orbital Hall accumulation (OHA) $\Sigma^z_{\text{OH}}$ (b) per unit cell calculated as functions of $E_\text{F}$ for a biased ZZ nanoribbon of 2L-MoS$_2$ with 15 lines in breath. We used two distinct values of gate bias $\delta U=+0.2$ and $+0.4$ eV and set $\Gamma=1$ meV. The vertical dashed line identifies the energy where the edge-states cross, and the purple solid lines delimit the energy range around it that we are interested in (see Fig. \ref{fig:figA1}).}
	\label{FigMain2L-MagSigma-WithGate} 
\end{figure}

It is instructive to enquire into the effects that spatial inversion symmetry breaking may have on the orbital magnetoelectric response of zigzag edged nanoribbons extracted from MoS$_2$ bilayers. One may simulate it, in a simplistic way, by introducing a layer-dependent on-site potential $U_l$ to mimic the presence of an external electric field (gate bias) applied perpendicularly to the nanoribbon's layers. We choose $U_l=\mp0.2$eV, with $l=1,2$ for layers $1$ (top) and $2$ (bottom), respectively, and recalculate the induced OAM profiles in the presence of this perpendicular field. The results are shown in Fig. \ref{FigMain2L-WithGate}.
The spatial symmetry breaking clearly brings a symmetric component to the induced total OAM accumulation profile, indicating that one may continuously tune the appearance of OME in these systems with the use of a gate bias. One may compute the current-induced orbital magnetization and orbital Hall accumulation in the gated 2L-MoS$_2$ ZZ-nanoribbon using Eqs. (\ref{MZ}, \ref{SigmaOHz}). The results are shown in Fig. \ref{FigMain2L-MagSigma-WithGate} for two values of $U_l$. Finite values of $\delta U=U_2-U_1$ slightly distorts the profile of $\Sigma^z_{\text{OH}}$ but does not modify its order of magnitude. On the other hand, $\delta U$ also leads to finite values of $M^z$ in 2L-MoS$_2$ nanoribbon. One may thus tune the current induced orbital magnetization of a 2L-MoS$_2$ ZZ-nanoribbon from zero to a finite value with the use of an electric-field applied perpendicularly to the nanoribbon plane. The OME in 2L-MoS$_2$ nanoribbon also switches sign when the direction of the perpendicularly applied electric field is inverted, providing a versatile tool for controlling the current-induced orbital magnetization. The layer-dependent potential $U_l$ simulates the effect of a substrate that interacts with the 2L-MoS$_2$ nanoribbon. It also mimics the effect of gate voltage used in field effect transistors to vary the density of electrons in experiments with layered materials.  The effect of the perpendicular electric field in the noncentrosymmetric 1L-MoS$_2$ ZZ-nanoribbon would consist solely in the change of the Fermi-energy due to electronic doping. Hence, the results for the 1L-MoS$_2$ nanoribbon reported through the work would not change.

To go from the charge-neutrality point ($E_{F}=0$ eV) to the edge-states crossings region ($E_{F}\approx 1$ eV), it is necessary to occupy $N_b\approx 2$ bands associated with edge-states, as illustrated in the right panels of Fig. \ref{fig:figA1}. This number does not change as the number of lines in the nanoribbon increases, because they are edge-state bands that cross the bulk gap region and are associated with the MoS$_2$ topology. The area of the unit cell of a MoS$_2$ ZZ-nanoribbon with $N$ lines is $A_u=N a^2 \sqrt{3} /2\approx N (8.76\angstrom^2)$. Thus, the change in the electronic density necessary to reach the energy where the edge-states cross scales with $N^{-1}$ and is given by $\delta n_e=N_b/A_u\approx (0.23/N) 10^{16} \text{cm}^{-2}$. For the very narrow nanoribbons considered here, such electron densities would be quite high. However, for wider nanoribbons with $N=500$ lines in breadth (width $W \approx 0.1 \mu$m) the electronic densities are compatible with the ones achieved in experiments \cite{ElectronicDensity-baugher2013intrinsic}.

\begin{figure}[h!]
	\includegraphics[width=0.85\linewidth,clip]{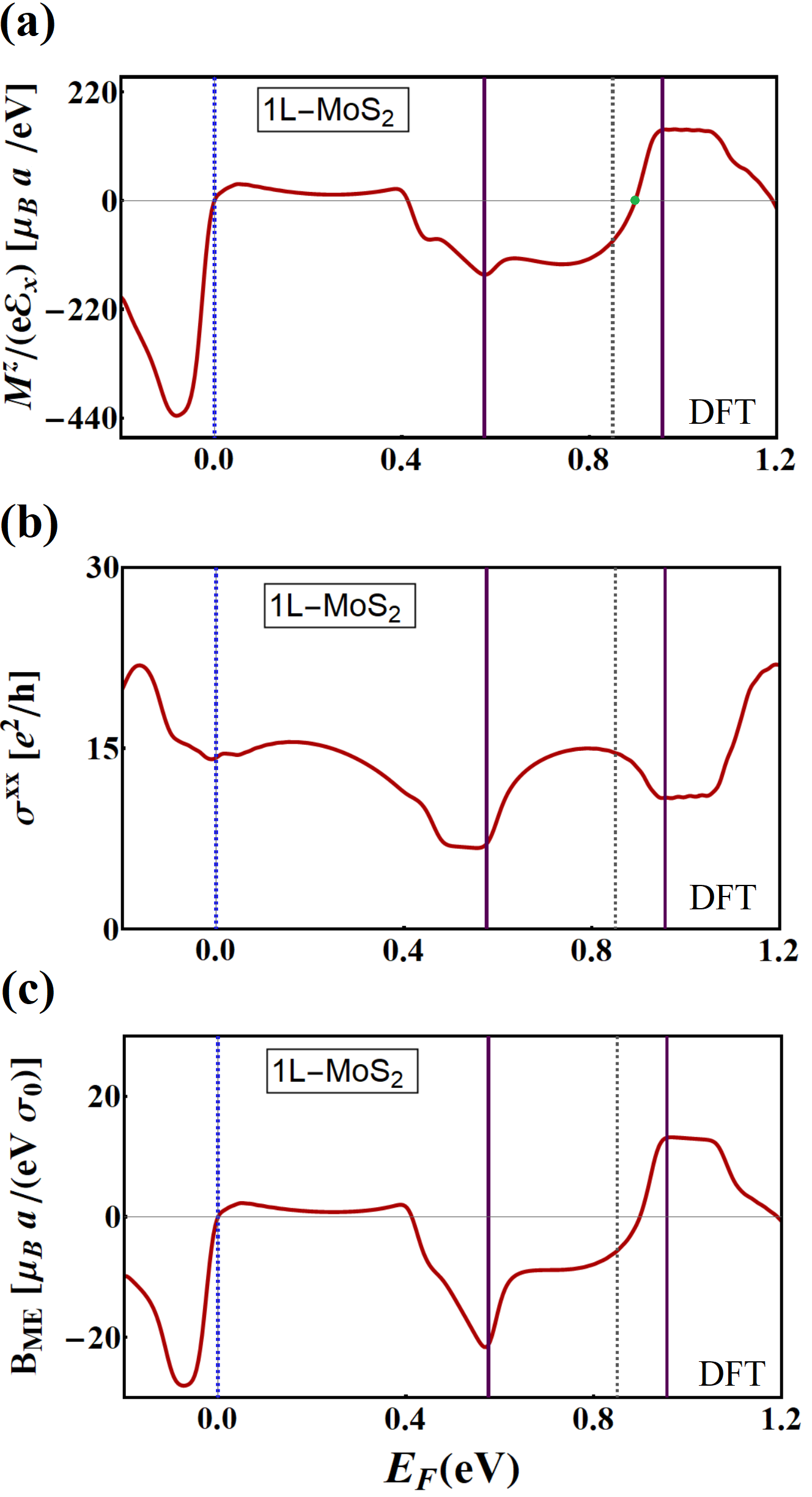}
	\caption{Current-induced orbital magnetization $M^z$ (a), longitudinal charge conductivity $\sigma^{xx}$ (b) and magnetoelectric coefficient $B_{\text{ME}}=\beta^{zx}/e$ (c), calculated as functions of $E_{\text{F}}$ for 1L-MoS$_2$ ZZ-nanoribbon using DFT. The vertical blue dashed line at $E_F=0$ eV indicates the charge neutrality point of the system. The vertical thick purple lines delimit the energy range where the 3-bands model provides a reasonable description of the ZZ-nanoribbon electronic structure [see Fig. \ref{fig:figA1}]. In our calculations we have used $\Gamma=1$meV. We defined $\sigma_0=e^2/h$.}
	\label{FigFullRange} 
\end{figure}

So far, we have focused on the electron-doped regime in which the Fermi-energy ($E_{\text{F}}$) lies around $\approx 1$ eV, where the edge states dominate the orbital response of the MoS$_2$ ZZ-nanoribbon [see Fig. \ref{fig:figA1}]. This allows a direct comparison between results calculated using DFT with those obtained by means of the simplified three-band model [see Sec. \ref{sec3}]. It is noteworthy, however, that the appearance of the OME effect is not restricted to such energy region. In Fig. \ref{FigFullRange} (a), we show DFT calculations of the current-induced orbital magnetization for 1L-MoS$_2$ ZZ-nanoribbons for a much wider range of $E_{\text{F}}$ values. The vertical blue dashed line indicates the charge-neutrality point at $E_{\text{F}}=0$ eV for the 1L-MoS$_2$ ZZ-nanoribbon \cite{Pezo2019-FermiLevel-NR, FermiLevel-NR-PhysRevB.96.165436}.  It is worth noting that for negative values of $E_{\text{F}}$ close to $E_{\text{F}}=0$ the intensity of the OME effect reaches values approximately three times larger than in the vicinity of $E_{\text{F}}=1$eV. 

The OME studied here is an electric current-induced phenomenon mediated by states near $E_{\text{F}}$. Hence, it is useful to rewrite Eq. (\ref{MES}) as $M^i=\sum_j \beta^{ij} \mathcal{J}^j$, where $\mathcal{J}^j$ is the electric current density, and $\beta^{ij}=\alpha^{ij}/ \sigma^{jj}$. Here, $\sigma^{jj}$ represents the longitudinal charge conductivity along the $j$-direction. Since both $\alpha^{ij}$ and $\sigma^{jj}$ are proportional to momentum relaxation time $\tau$, the coefficient $\beta^{ij}$ does not depend upon $\tau$. Figs. \ref{FigFullRange} (b) and (c) also show the longitudinal conductivity $\sigma^{xx}$ and the orbital magnetoelectric coefficient $\beta^{zx}$ calculated as a function of $E_{\text{F}}$ for the 1L-MoS$_2$ ZZ-nanoribbon, respectively. 

Earlier works have studied magnetoelectric effects in chiral structures such as elemental tellurium \cite{Ivchenko-Pikus-Chiral-JETP, Vorobev-JETP}, which have a well-defined helicity that determines the direction of the current-induced magnetization. Later it was shown that spin magnetoelectric effects may occur in layered 2D systems with broken inversion symmetry in the direction perpendicular to the atomic planes and with relatively large spin-orbit coupling \cite{Edelstein1990}. The latter belong to a so-called polar symmetry group, in which the direction of the induced magnetization is basically determined by the direction of the applied current. Of all non-centrosymmetric point groups that allow the occurrence of current-induced magnetoelectric effect, eleven are chiral, and ten are polar \cite{Furukawa-Itou-PhysRevResearch.3.023111}. A nanoribbon with zigzag edges extracted from 1L-MoS$_2$ belongs to the polar point group $C_{2v}$. Thus, for an electric current flowing in plane along the ribbon's axis, the induced magnetization is perpendicular to the ribbon's plane.

Finally, it is noteworthy that edge disorder may significantly affect the current-induced orbital magnetization, especially within the energy range where current flows through edge states. Inquiries into this are certainly relevant and most welcome. From a theoretical point of view, its quantification may require large-scale simulations in real space, which are beyond the scope of this work. However, relatively large samples of high quality nanoribbons have already been obtained \cite{MOS2Nanoribbons-Syntetization-NatMaterials} and we imagine that the effects described in our work may be observed in such samples.

\section{Conclusions \label{sec5}}

We have presented a detailed study of the accumulations of orbital angular momentum induced by a longitudinal electric field in nanoribbons with zig-zag edges extracted from mono and bi-layers of MoS$_2$.  We have shown that nanoribbons with zig-zag edges of MoS$_2$ monolayers exhibit a significant magnetoelectric effect, which together with the orbital Hall effect give rise to asymmetric orbital angular momentum accumulation profiles in these ribbons. For zig-zag edged nanoribbons extracted from bilayers that preserve spatial inversion symmetry we found no net orbital magnetoelectric effect,  but the orbital angular momentum accumulations due to the orbital Hall effect in these systems are twice as large as those obtained for nanoribbons of monolayers. We also show that the absence of OME in the bilayer nanoribbons comes from a cancellation between the orbital magnetic moments induced in each layer separately that have opposite directions. By applying a perpendicular electric field that breaks spatial-inversion symmetry, the orbital magneto-electric effect emerges in the bilayer nanoribbons. We unveiled the underlying physics of both orbital response phenomena involved by analyzing the equilibrium orbital texture and charge dipole distribution on nanoribbons.

\begin{acknowledgments}
	We acknowledge CNPq/Brazil, CAPES/Brazil, FAPERJ/Brazil and INCT Nanocarbono for financial support. FSMG gratefully acknowledge the computing time granted through JARA on the supercomputer JURECA \cite{jureca} at Forschungszentrum Jülich. MC acknowledge the National Laboratory for Scientific Computing (LNCC/MCTI, Brazil) for providing HPC resources of the SDumont supercomputer. TGR acknowledges funding from Fundação para a Ciência e a Tecnologia and Instituto de Telecomunicações - grant number UID/50008/2020 in the framework of the project Sym-Break. She thankfully acknowledges the computer resources at MareNostrum and the technical support provided by Barcelona Supercomputing Center (FI-2020-2-0033). L.M.C. acknowledges funding from The Army Research Office under Grant No. W911NF-21-1-0004. ICN2 is funded by the CERCA Programme/Generalitat de Catalunya and supported by the Severo Ochoa Centres of Excellence program, funded by the Spanish Research Agency (Grant No. SEV-2017-0706).
\end{acknowledgments}	

\appendix

\section{Hamiltonian of ZZ nanoribbons of TMDs \label{APP-A}}

In the main text, we used two different descriptions of the ZZ-nanoribbon of MoS$_2$. The first is a simplified three-band model that provides a qualitative description of the orbital response of nanoribbon. This simplified model has the advantage of being easy to implement. The second description is based on DFT calculations and contains the complete information on the orbital structure of MoS$_2$ nanoribbons. In this appendix, we give details on the Hamiltonians used in both descriptions and compare their electronic spectra. 

\subsection{Three-band model \label{App-3B-model}}

The three-band model \cite{Xiao-three-band} takes into account only the orbitals $d_{z^2}, d_{x^2-y^2}$ and $d_{xy}$ of the transition metal (Mo). They reasonably describe the top of the valence and bottom of the conduction bands in the 2D-bulk system \cite{Faria_Junior_2022}. The effect of chalcogen (S) is introduced via perturbation theory.

\subsubsection*{Monolayer (1L) of MoS$_2$}

Following Ref. \cite{Xiao-three-band}, we define the basis: $\beta_{1L}=\{ \big|d_{z^2}\big>, \big|d_{xy}\big>, \big|d_{x^2-y^2}\big>\} \otimes \{\big|\ell \big>\}$, where $\big| \ell \big>$ indexes the line of the nanoribbon [see Fig. \ref{fig:figA2} (a)]. For a nanoribbon with $N$ lines in breadth we have, $\ell=1, 2, ..., N$. The Hamiltonian of nanoribbon on $k$-space presents a tridiagonal block form, 

\begin{widetext}
\begin{eqnarray}
H^{1L}_{\text{z.z.}}(k)&=&
\begin{bmatrix}
    h_1(k)                  &     h^{\dagger}_2(k)              &                 &             &     \\
    h_2(k) &      h_1(k)              &   h^{\dagger}_2(k)    &             &     \\
                               & h_2(k)  & h_1(k)      & \ddots   &     \\
                               &                            &   \ddots     &  \ddots  & h^{\dagger}_2(k)    \\
                               &                            &                 &  h_2(k) & h_1(k) 
\end{bmatrix}, \label{Hmonolayer}
\end{eqnarray}
where, 
\begin{eqnarray}
h_1(k)&=&
\begin{bmatrix}
  \epsilon_1+2t_0\cos(k a)  &  2i t_1\sin (k a)                    & 2t_2\cos(k a)       \\
   -2it_1\sin(k a)                &  \epsilon_2+2t_{11}\cos(k a)  & 2it_{12}\sin(k a)   \\
     2t_2\cos(k a)               &  -2it_{12}\sin(k a)                & \epsilon_2+2t_{22}\cos(k a)
\end{bmatrix}, \label{h13B}
\end{eqnarray}
and,
\begin{eqnarray}
h_2(k)&=&
\begin{bmatrix}
 2t_0\cos\big(\frac{k a}{2}\big)                        &  i(t_1-\sqrt{3}t_2)\sin\big(\frac{k a}{2}\big)                                                              & -(\sqrt{3}t_1+t_2)\cos\big(\frac{k a}{2}\big)  \\
 -i(t_1+\sqrt{3}t_2) \sin\big(\frac{k a}{2}\big)   & (\frac{1}{2})(t_{11}+3t_{22})\cos\big(\frac{k a}{2}\big)                                            & -i(\frac{\sqrt{3}}{2}t_{11}+2t_{12}-\frac{\sqrt{3}}{2}t_{22})\sin\big(\frac{k a}{2}\big)  \\
  (\sqrt{3}t_1-t_2)\cos\big(\frac{k a}{2}\big)     &-i(\frac{\sqrt{3}}{2}t_{11}-2t_{12}-\frac{\sqrt{3}}{2}t_{22})\sin\big(\frac{k a}{2}\big) & (\frac{1}{2})(3t_{11}+t_{22})\cos\big(\frac{k a}{2}\big)
\end{bmatrix}. \label{h23B}
\end{eqnarray}
\end{widetext}

For MoS$_2$, the parameters involved in this model \cite{Xiao-three-band}: $\epsilon_1=1.046 \text{eV}$, $\epsilon_2=2.104 \text{eV}$, $t_0=-0.184 \text{eV}$, $t_1=0.401 \text{eV}$, $t_2=0.507 \text{eV}$, $t_{11}=0.218 \text{eV}$, $t_{12}=0.338 \text{eV}$, $t_{22}=0.057 \text{eV}$. Here, we shall neglect the relatively weak spin-orbit interaction in MoS$_2$. The intra-atomic OAM operator representation in the basis $\beta_{1L}$ is given by Eq. (\ref{lz}) of the main text \cite{vanderbiltBook}. 

\subsubsection*{Bilayer (2L) of MoS$_2$}

We construct the model for the bilayer from Eq. (\ref{Hmonolayer}), assuming that it preserves spatial inversion symmetry ($\mathcal{P}$). The inversion symmetry point $\mathcal{I}$ of the system is located in the space between the layers as illustrated in the Fig. \ref{fig:figA2} (c). The action of $\mathcal{P}$ in $d$-orbitals is trivial: $\mathcal{P}\big|d_{z^2}, d_{xy}, d_{x^2-y^2} \rangle= \big|d_{z^2}, d_{xy}, d_{x^2-y^2} \rangle$. On the other hand, the symmetry inverts the line numbers and the electronic crystal moments:
\begin{eqnarray}
\mathcal{P}&&\big|\ell\big>=\big|\ell'\big>, \label{Invell} \\
\mathcal{P}&&: k \rightarrow -k, \label{Invk}
 \end{eqnarray}
where, $\ell'= (N+1)-\ell$. To write the Hamiltonian of 2L-MoS$_2$ nanoribbon we define the basis $\beta_{2L}=\beta_{1L} \otimes \{\big| t\rangle, \big| b\rangle \}$ where $\big|t (b) \rangle$ refers to top (bottom) layer and $\beta_{1L}$ where defined in the previous subsection. The Hamiltonian of nanoribbon reads
\begin{eqnarray}
H^{2L}_{\text{z.z.}}(k)&=&
\begin{bmatrix}
    H^{1L}_{\text{z.z.}}(k)       &     T      \\
    T^{\dagger}                  &     \tilde{H}^{1L}_{\text{z.z.}}(k)
\end{bmatrix}.\label{Hbilayer}
\end{eqnarray}
where, $\tilde{H}^{1L}_{\text{z.z.}}(k)=\mathcal{P} H^{1L}_{\text{z.z.}}(k) \mathcal{P}^{\dagger}$. In a first approximation, the hybridization between orbitals $d_{z^2}$ of different layers can be neglected. With this, the interlayer coupling matrix reads
\begin{eqnarray}
T=
\sum^{N}_{\ell=1}
\left( \frac{t_{\perp}}{2}\right)\begin{bmatrix}
  0  &  0  & 0  \\
  0  &  1 &  i \\
  0 &  -i &  -1\end{bmatrix} \otimes \big|t\big>\big<b\big|\otimes\big|\ell\big>\big<\ell\big|,  \nonumber
\end{eqnarray}
where $t_{\perp}=0.043 \text{eV}$ for 2L-MoS$_2$ \cite{Gong_2013}.

\subsection{\textit{Ab-Initio} Simulations \label{App-complete-TB}}

We perform density functional theory calculations (DFT)~\cite{DFT1,DFT2} using the plane-wave-based code \textsc{Quantum Espresso}~\cite{QE-2017}. The exchange and correlation potential is treated within the generalised gradient approximation (GGA)~\cite{PBE}. The ionic cores were described with fully relativistic projected augmented wave (PAW) potentials~\cite{PAW}. We used a cutoff energy of 63 Ry for the wavefunctions and a value 10 times larger for the charge density. In order to reproduce the interlayer distance of the MoS$_{2}$ bilayer we have used the DFT-D3~\cite{DFT-D3} method to describe the dispersion forces. The reciprocal space sampling was 10$\times$1$\times$1 $k$-points. To avoid spurious interaction due to periodic boundary conditions we insert a vacuum spacing of 15\AA. We constructed an effective tight-binding Hamiltonian from our DFT calculations using the pseudo atomic orbital projection (PAO) method~\cite{PAO1,PAO3} as implemented in the \textsc{PAOFLOW} code~\cite{PAO5,PAO6}. The PAO method consists of projecting the DFT Kohn-Sham orbitals into the compact subspace spanned by the pseudo atomic orbitals which are naturally built-in into the PAW potentials. The PAW potentials used for the Mo and S were constructed with a $sspd$ and $sp$ basis, respectively. Once the PAO Hamiltonian $\mathcal{H}_{\text{PAO}}(k)$ is constructed we can calculate the orbital responses to an applied electric field using linear$-$response theory. This method have been used to investigate several other systems, ranging from topological to time-dependent properties \cite{Costa2019,Costa2018,fegete,hoti}.

\subsection{Electronic spectra of both models \label{App-Spectra}}

In the Fig. \ref{fig:figA1} of the main text, we show the electronic spectra of ZZ-nanoribbons of 1L-MoS$_2$ (a) and bilayer 2L-MoS$_2$ (b) systems. The black curves represent the spectra of the three-band model, and the red curves are the spectra from DFT calculations. The dashed lines in the Fig. \ref{fig:figA1} depict the ZZ-nanoribbon edge-states crossings energy within the description of each model. The shaded region on the left panels of the Fig. \ref{fig:figA1} delimit the range of energy that the three-band model aims to describe $[E_{min}^{3b}-E_{max}^{3b}]$. Similarly, the shaded region in the right panels of the Fig. \ref{fig:figA1} delimit the correspondent energy range in DFT calculations [$E^{dft}_{min}-E^{dft}_{max}$].

\section{Linear response theory \label{APP-B}}

In the main text, we used two linear response methods to compute the orbital response in the ZZ-nanoribbons of TMDs. The first method follows the direct calculation of eigenstates and eigenvalues of nanoribbon Hamiltonians \cite{Go1PRR2.033401,Go2PRR2.013177}. The second method express the orbital response in terms of generalized susceptibilities in static limit \cite{PhysRevB.92.220410, Guimaraes2017}. Both methods are equivalent in the long scattering-time (dilute) regime \cite{Bobien-Manchon-PhysRevB.102.085113, Us4-PRB}.

The first method was used to obtain results of sec. \ref{sec2} of the main text and express the orbital response $\delta \langle L^z_{\ell}\rangle$ to an electric-field in the lines of ribbon as a sum of two contributions
\begin{eqnarray}
\delta \langle L^z_{\ell}\rangle^\text{Intra}&=&-\frac{e\hbar \mathcal{E}_x}{2\Gamma} \sum_{n,k}\frac{\partial f}{\partial E} \bigg|_{E=E_{nk}} \nonumber \\
&& \times \big<nk\big|L^z_{\ell} \big|nk\big>\big<nk\big|v(k)\big|nk\big>, \label{Ointra}
\end{eqnarray}
and,
\begin{eqnarray}
\delta \langle L^z_{\ell} \rangle^\text{Inter}&=&e\hbar \mathcal{E}_x \sum_{n,m,k} (f_{nk}-f_{mk}) \nonumber \\
&& \times \text{Im}\Bigg[\frac{\big<nk\big|L^z_{\ell}\big|mk\big>\big<mk\big|v(k)\big|nk\big>}{(E_{nk}-E_{mk}+i\eta)^2}\Bigg].
\label{Ointer}
\end{eqnarray}
Here, $L^z_{\ell}$ is the OAM operator projected on line $\ell$ of ZZ-nanoribbon. We follow Refs. \cite{Go1PRR2.033401,Go2PRR2.013177} and use intra-atomic approximation to the OAM operator \cite{vanderbiltBook}. $E_{nk}$ are the eigenvalues and $\ket{nk}$ the corresponding eigenvectors of the nanoribbon Hamiltonian evaluated in the reciprocal space; $n$ denotes the energy band index, $k$ is the wave vector, and $f_{nk}=f(E_{nk})$ symbolizes the Fermi-Dirac distribution function $f(E)=\big(\text{exp} [(E-E_{\text{F}})/k_{B}T] +1\big)^{-1}$ associated with the state $\ket{nk}$. In all numerical calculations performed in this work, we set $k_{B}T=1$meV. $\hat{v}(k)=\hbar^{-1}(\partial \hat{H}(k)/\partial k)$ is the velocity operator, $e$ is the modulus of the electronic charge, and $\mathcal{E}_x$ denotes the intensity of the applied electric field. $\Gamma = \hbar/(2\tau)$, where $\tau$ is the momentum relaxation time, is treated here as a phenomenological parameter that simulates effects of disorder in the transport properties of the nanoribbons within the constant relaxation-time approximation \cite{Go1PRR2.033401,Go2PRR2.013177, Bobien-Manchon-PhysRevB.102.085113}; $\eta$ is a small positive quantity arising from a conventional artifice to ensure that the external perturbation is turned on adiabatically.

The contributions to OAM responses that come from Eqs. (\ref{Ointra}, \ref{Ointer}) have distinct physical origins \cite{Go1PRR2.033401,Go2PRR2.013177}. Eq. (\ref{Ointra}) captures the contributions from states near the Fermi-surface (Intraband) and is associated with the electronic scattering processes. The contributions of Eqs. (\ref{Ointer}) are associated with interband electronic transitions.  Eqs. (\ref{Ointra}, \ref{Ointer}) transform differently under spatial-inversion ($\mathcal{P}$) and time-reversal ($\mathcal{T}$) symmetry. To the occurrence of finite orbital magnetization coming \emph{purely} from Eq. (\ref{Ointra}), it is necessary that, at equilibrium, the system preserves time-reversal $\mathcal{T}$ and breaks spatial inversion $\mathcal{P}$. To the occurrence of a finite orbital magnetization originating \emph{purely} from Eq. (\ref{Ointer}) is necessary that equilibrium Hamiltonian breaks both $\mathcal{P}$ and $\mathcal{T}$ symmetries, but their product $\mathcal{P}.\mathcal{T}$ must be conserved. These constraints are derived in Refs. \cite{Hayami-Kusunose-PhysRevB.98.165110, Xiao-Niu-PhysRevB.103.045401, Hikaru-Youichi-PhysRevB.96.064432}, and are summarized in table \ref{table1}. Here we deal with systems that preserve time-reversal symmetry at equilibrium. In these cases, the electrically induced magnetization comes entirely from the intra-band term that determines both the magnetoelectric and the Hall responses \cite{PhysRevB.95.014403, Salemi-Oppeneer-PRMaterials-2021}. Eqs. (\ref{Ointra}) and (\ref{Ointer}) are rather general and may be easily extended to calculate, for example, orbital torque in complex heterostructures based on TMDs \cite{Spin-Orbit-Torques-MoS2-nanolett}.

\begin{table}[h!]
	\centering
	\begin{tabular}{||c c c c c c c  ||} 
		\hline
		Pure contribution & & $\mathcal{P}$ & &  $\mathcal{T}$ & &  $\mathcal{P}.\mathcal{T}$ \\ [0.5ex] 
		\hline
		\hline
		Intra-band [Eq. (\ref{Ointra})] & & $\times$ & & $\bigcirc$ & & $\times$ \\
		Inter-band [Eq. (\ref{Ointer})] & & $\times$ & & $\times$ & & $\bigcirc$ \\ [0.5ex] 
		\hline
	\end{tabular}
	\caption{Symmetry constraints for contribuitions on linear response formulas for OME (finite $M^z(E_\text{F})$) [Eqs. (\ref{Ointra}) and (\ref{Ointer})] with respect to spatial inversion ($\mathcal{P}$), time-reversal ($\mathcal{T}$), and the product of both symmetries ($\mathcal{P} \mathcal{T}$). Symbols: $\bigcirc$ the symmetry is preserved by the equilibrium Hamiltonian,  $\times$ the symmetry is broken by the equilibrium Hamiltonian. The derivation of this table from symmetry operations in the linear-response formula is detailed in references \cite{Hayami-Kusunose-PhysRevB.98.165110, Xiao-Niu-PhysRevB.103.045401, Hikaru-Youichi-PhysRevB.96.064432}.}
	\label{table1}
\end{table}

The second method was used in sec. \ref{sec3}. It is based on generalized susceptibilities calculated in the static limit. More details about it are described in Refs. \cite{PhysRevB.92.220410, Guimaraes2017}. A spatially uniform and time dependent harmonic electric field with small amplitude $\mathcal{E}_x$ is applied to the system, and the change in the expectation value of the physical observable is calculated within linear response theory, providing the local angular momentum disturbances per atom in each line $\ell$ given by
\begin{equation}
\delta\langle L^z_{\ell}(t)\rangle =-\lim_{\omega\rightarrow0}\frac{e\mathcal{E}_x}{\hbar\omega}\operatorname{Im}\left\{e^{-i\omega t}\mathcal{D}_{\ell}(\omega)\right\}\ , \label{StaticSucep1}
\end{equation}
 where
\begin{equation}
\mathcal{D}_{\ell}(\omega) = \sum_{\substack{k}}\sum_{\substack{\ell_1\ell_2\\\mu\nu\gamma\xi}} \big( L_{\ell}^z\big)_{\mu \nu} \chi^{\mu\nu\gamma\xi}
_{\ell\ell\ell_1\ell_2}(k,\omega)~\frac{\partial t_{\ell_2\ell_1}^{\gamma\xi}(k)}{\partial k}\ . \label{StaticSucep2}
\end{equation}
Here, $\mu$, $\nu$, $\gamma$, and $\xi$ denote the atomic orbitals, and $\ell$, $\ell_1$, and $\ell_2$ label the atomic lines. In our case, since we are neglecting electronic Coulomb interaction, $\chi(k,\omega)$ represent generalized non-interacting spin susceptibilities. $t_{\ell_2\ell_1}^{\gamma\xi}(k)$ are the elements of hopping matrix in reciprocal space. We used this method in the results of sec. \ref{sec3} because it is easier to integrate with our codes used to obtain the PAO Hamiltonian ($\mathcal{H}_{\text{PAO}}(k)$). 

\section{Dependence of $M^z$ and $\Sigma^z_{\text{OH}}$ with the number of the lines \label{App-NumberLines}}

\begin{figure}[h!]
	\includegraphics[width=0.8\linewidth,clip]{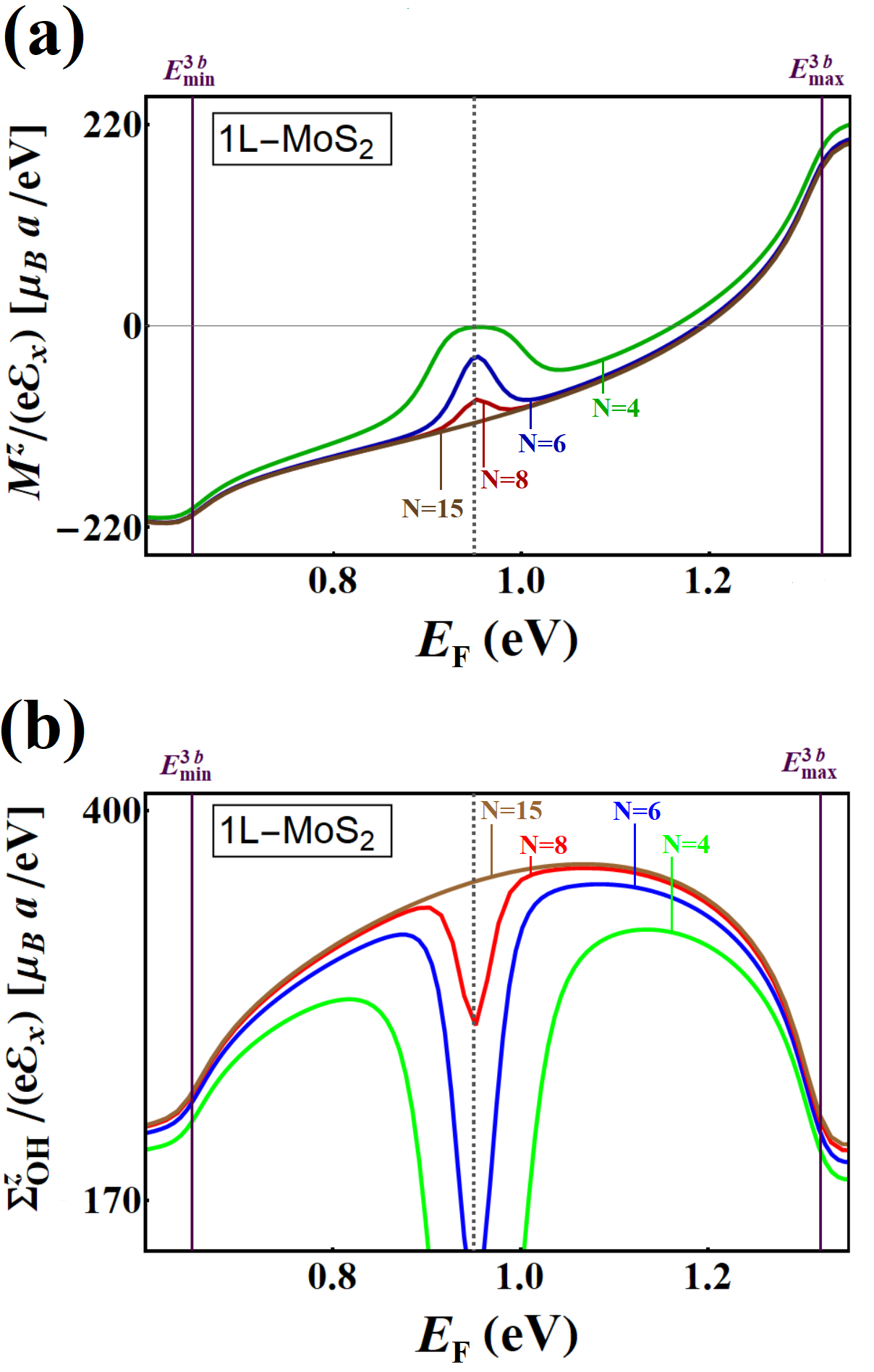}
	\caption{The current induced orbital magnetization (a) and the orbital Hall accumulation (b) for 1L-MoS$_2$ ZZ-nanoribbons with $N=4,6,8,15$ lines. Here we used the three-band model \cite{Xiao-three-band} of TMDs discussed in the appendix \ref{App-3B-model} and set $\Gamma=1 \text{meV}$. The dip near the edge-states crossings energy [vertical dashed line] of ZZ-nanoribbon reduces by increasing the number of lines. This behavior also occurs in the nanoribbons of 2L-MoS$_2$ and in the ZZ-nanoribbons described by the complete PAO Hamiltonian of appendix \ref{App-complete-TB}.}
	\label{fig:figA3} 
\end{figure}

As mentioned in the main text, for narrow ZZ-nanoribbons, a small energy gap appears near the edge-state crossings [represented by dashed lines in Fig. \ref{fig:figA1}]. This energy gap is associated with the lateral confining of electrons in nanoribbons with a small number of lines \cite{PKim-PhysRevLett.98.206805}. The gap generated by electronic confinement causes a reduction of $\Sigma^z_{\text{OH}}$ and $M^z$ near the edge-states crossings energy (vertical dashed lines) in Figs. \ref{FigMain5} and \ref{FigMain6} of main text. By increasing the number of lines of nanoribbon, the lateral confining effects become unimportant, and the energy gap rapidly disappears. Consequently, the dips near dashed lines in Figs. \ref{FigMain5} and \ref{FigMain6} also disappear for wider ZZ-nanoribbons. Here, we illustrate this effect for 1L-MoS$_2$ ZZ-nanoribbon described by the three-band effective model. In Fig. \ref{fig:figA3}, we show the current-induced orbital magnetization and the orbital Hall accumulation for 1L-MoS$_2$ ZZ-nanoribbons described by Hamiltonian of Eqs. (\ref{Hmonolayer}) for nanoribbons with $N=4, 6, 8, 15$ lines in breadth. For ZZ-nanoribbon with 4 and 6 lines, the electronic confinement is important, causing a strong reduction of $\Sigma^z_{\text{OH}}$ and $M^z$ near the edge-states crossings energy. In 1L-MoS$_2$ ZZ-nanoribbons with 8 lines in breadth, the effects of electronic confinement start to disappear, smoothing the dip near the edge-states crossings energy. For wider nanoribbons, the effect of the lateral electronic confinement totally disappears (see the case $N=15$ lines in Fig. \ref{fig:figA3}) and, the energy gap at edge-state crossings vanishes. This behavior of $\Sigma^z_{\text{OH}}$ and $M^z$ at edge-states crossings energy with the number of lines, detailed here for 1L-MoS$_2$ ZZ-nanoribbons described within the three-band model, also occurs in ZZ-nanoribbons described by the complete PAO Hamiltonian and in 2L-MoS$_2$ ZZ-nanoribbons.  

\section{Spin-orbit coupling and intersite OAM contributions \label{APP-D}}

Here, we briefly discuss two features that we have neglected in our calculations: I- the spin-orbit coupling. II- The intersite contribution to the OAM. In what follows we shall use the three-band model to argue that these ingredients should not significantly impact the results reported in the main text.

\subsection{Spin-orbit coupling contribution to the angular momentum accumulation}

\begin{figure}[h]
	\centering
	 \includegraphics[width=0.995\linewidth,clip]{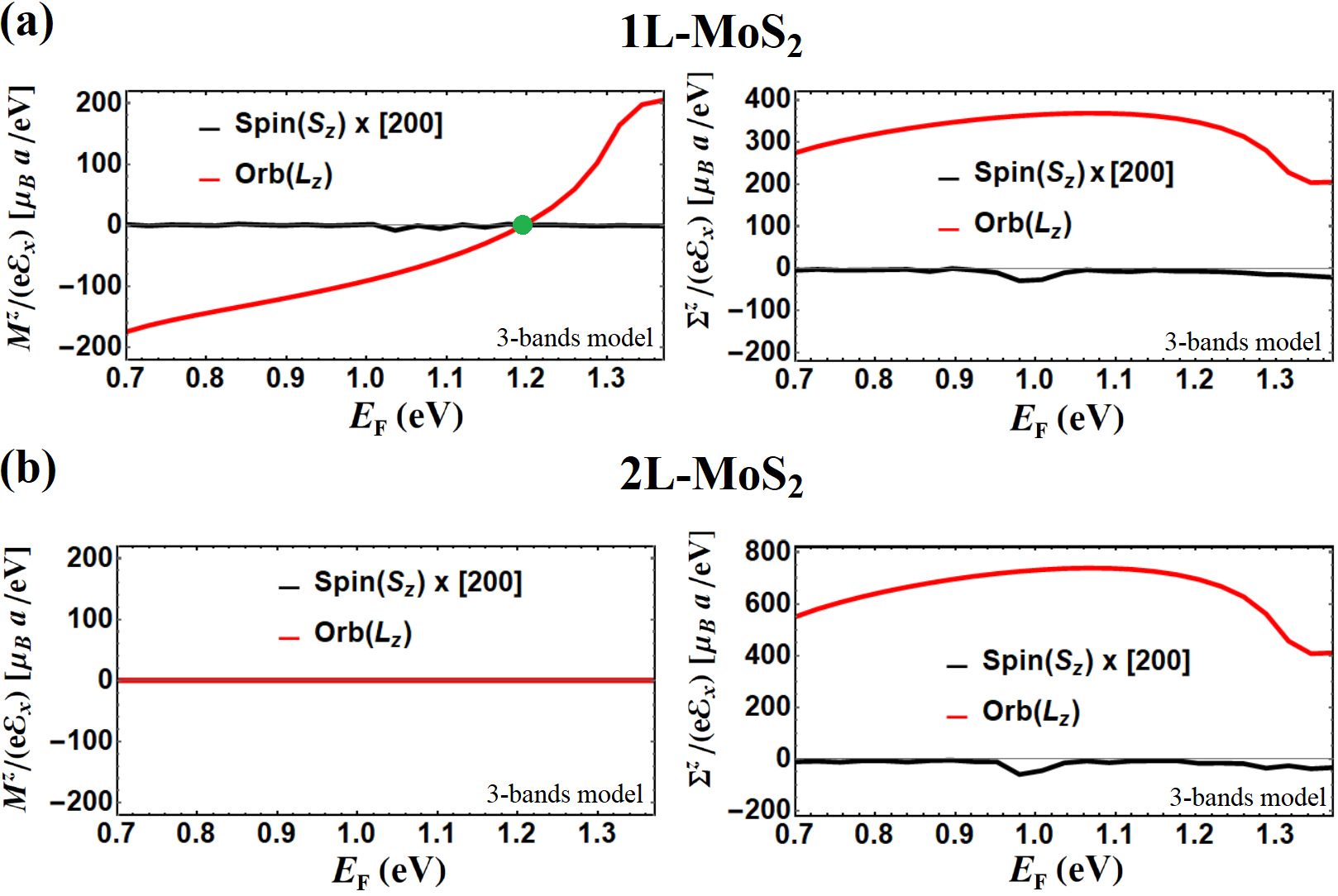}     
	\caption{Current-induced magnetization (left) and angular momentum Hall accumulations (right) for 1L-MoS$_2$ (a) and 2L-MoS$_2$ (b) ZZ-nanoribbons with the inclusion of intrinsic atomic SOC $\lambda=0.073/\hbar^2$ eV. Red lines show the orbital angular momentum ($L_z$) response, and black lines show the spin ($S_z$) response. We multiplied the spin response by a numerical factor of 200 to permit its visualization. Here we set $\Gamma=1$ meV.}
	\label{figSOC}
\end{figure}

In the results presented in the main text, we did not take into account the spin-orbit interaction, which is relatively high in the TMDs, in comparison with other two-dimensional materials. Within the three-band model the spin-orbit interaction may be written as $H_{\text{SOC}}=(\lambda/2) L^z\otimes S^z$, where $\lambda=0.073$ eV$/\hbar^2$ is the intrinsic spin-orbit coupling in the Mo atoms, and $S^z=\hbar \ \text{diag}(1,-1)$ represents electronic spin operator \cite{Xiao-three-band}. We use Eqs. (\ref{Ointra}) and (\ref{Ointer}), substituting $L^z$ by $\frac{1}{2} S^z$, to compute the line-resolved spin response in the ZZ-nanoribbons. The same procedure is also employed with Eqs. (\ref{MZ}) and (\ref{SigmaOHz}) to calculate spin contribution to the current-induced magnetization and to the Hall angular momentum accumulation. The results are depicted in Fig. \ref{figSOC} for 1L-MoS$_2$ and 2L-MoS$_2$ nanoribbons with zigzag edges. We see that the spin contribution is three orders of magnitude lower than the OAM one. This is in line with recent theoretical works on TMDs \cite{Us3-PRL, Cysne-Bhowal-Vignale-Rappoport,Us4-PRB, Bhowal-Satpathy-PhysRevB.101.121112} and justifies disregarding the spin-orbit interaction in our work.

\subsection{The modern theory of orbital angular momentum}

In our calculations we have employed the intra-atomic approximation for the OAM operator. This approximation neglects the inter-site contribution coming from closed trajectories of electrons in periodic systems. The so-called \emph{modern theory of orbital magnetization} (MTOM) provides a more accurate description of the electronic OAM operator in solids \cite{vanderbiltBook}. The accuracy of the intra-atomic approximation highly depends on the system under study, and in some cases may be poor in comparison with the MTOM  \cite{Ceresoli-ModernTheoXIntraAtom-PhysRevB.81.060409,Nikolaev-ModernTheoXIntraAtom-PhysRevB.89.064428,Hanke-ModernTheoXIntraAtom-PhysRevB.94.121114}. The usual formulation of the MTOM expresses the OAM operator of a periodic system as a $k-$space integral of a geometric quantity over the Brillouin zone \cite{ModTheo-Real-Space-PhysRevLett.95.137205, ModTheo-Real-Space-PhysRevB.106.075136}. However, application of the MTOM to nanoribbons is not straightforward. Nanoribbons are systems that have translation symmetry in the longitudinal direction, but open boundary conditions in the transverse one. This imposes some technical difficulties for the application of the MTOM, at least in its early formulations. Very recently only these types of hybrid systems have been considered in the context of the modern theory, even so, for a relatively simple Haldane model \cite{HybridSystem-Resta-PhysRevB.101.165120}. Extending this formulation for multi-orbital systems and quantifying the accuracy of intra-atomic approximation in hybrid systems is an interesting point to be explored in future works.

Nevertheless, one may possibly estimate the importance of the corrections given by the MTOM by considering a strained single layer of MoS$_2$. The strain reduces the point group symmetry of the 2D material from $D_{3h}$ to $C_{2v}$, enabling the occurrence of current-induced orbital magnetization \cite{Son-Kim-2019-PhysRevLett.123.036806,Bhowal-Satpahy-kOME-PhysRevB.102.201403, Lee-2017-NatMaterials-887-9, Faria_Junior_2022}. The presence of translation symmetry in both the $\hat{x}$ and $\hat{y}$ directions allows the customary use of the MTOM. We have performed calculations of the OME in strained 2D-MoS$_2$ using the intra-atomic approximation and the MTOM. Near the top of the valence band, the agreement between the two approaches is fairly good. This is in line with the fact that the intra-atomic approximation provides reasonable results for the orbital Hall conductivity in this energy region, as reported in reference \cite{Cysne-Bhowal-Vignale-Rappoport}. We note that the region around the top of the valence band located at valleys $K$ and $K'$ of 2D Brillouin zone of strained 2D-MoS$_2$ is governed by the states $(d_{xy}\mp i d_{x^2+y^2} )/\sqrt{2}$, which have orbital angular momentum $\langle L_z\rangle=\pm 2 \hbar$, respectively . The same linear combinations dominate the neighborhood of energy region where the edge states cross in the 1L-MoS$_2$ ZZ-nanoribbon, as Fig. \ref{Fig_Lz} illustrates. This indicates that the correction arising from the MTOM in this energy range may be relatively small in our case.

We remark that this is merely an argument of plausibility, since the conducting states near the top of the strained 2D-MoS$_2$ valence band and the edge-states of the ZZ-nanoribbons are of distinct nature and belong to different energy ranges. In fact, to test the accuracy of the intra-atomic approximation for ZZ-nanoribbons, one should directly compare our results with calculations based on the MTOM for hybrid systems \cite{HybridSystem-Resta-PhysRevB.101.165120} or its real space representation for very long supercells \cite{ModTheo-Real-Space-PhysRevLett.95.137205, ModTheo-Real-Space-PhysRevB.106.075136}. This task is numerically challenging for multi-orbital systems but would be much instructive.

\bibliographystyle{apsrev}

\begin{thebibliography}{94}
\expandafter\ifx\csname natexlab\endcsname\relax\def\natexlab#1{#1}\fi
\expandafter\ifx\csname bibnamefont\endcsname\relax
  \def\bibnamefont#1{#1}\fi
\expandafter\ifx\csname bibfnamefont\endcsname\relax
  \def\bibfnamefont#1{#1}\fi
\expandafter\ifx\csname citenamefont\endcsname\relax
  \def\citenamefont#1{#1}\fi
\expandafter\ifx\csname url\endcsname\relax
  \def\url#1{\texttt{#1}}\fi
\expandafter\ifx\csname urlprefix\endcsname\relax\def\urlprefix{URL }\fi
\providecommand{\bibinfo}[2]{#2}
\providecommand{\eprint}[2][]{\url{#2}}

\bibitem[{\citenamefont{Go et~al.}(2021{\natexlab{a}})\citenamefont{Go, Jo,
  Lee, Kl\"{a}ui, and Mokrousov}}]{DGo-EPL-2021-review}
\bibinfo{author}{\bibfnamefont{D.}~\bibnamefont{Go}},
  \bibinfo{author}{\bibfnamefont{D.}~\bibnamefont{Jo}},
  \bibinfo{author}{\bibfnamefont{H.-W.} \bibnamefont{Lee}},
  \bibinfo{author}{\bibfnamefont{M.}~\bibnamefont{Kl\"{a}ui}},
  \bibnamefont{and}
  \bibinfo{author}{\bibfnamefont{Y.}~\bibnamefont{Mokrousov}},
  \bibinfo{journal}{{EPL} (Europhysics Letters)}
  \textbf{\bibinfo{volume}{135}}, \bibinfo{pages}{37001}
  (\bibinfo{year}{2021}{\natexlab{a}}),
  \urlprefix\url{https://doi.org/10.1209/0295-5075/ac2653}.

\bibitem[{\citenamefont{Bernevig et~al.}(2005)\citenamefont{Bernevig, Hughes,
  and Zhang}}]{Bernevig-Hughes-Zhang-PhysRevLett.95.066601}
\bibinfo{author}{\bibfnamefont{B.~A.} \bibnamefont{Bernevig}},
  \bibinfo{author}{\bibfnamefont{T.~L.} \bibnamefont{Hughes}},
  \bibnamefont{and} \bibinfo{author}{\bibfnamefont{S.-C.} \bibnamefont{Zhang}},
  \bibinfo{journal}{Phys. Rev. Lett.} \textbf{\bibinfo{volume}{95}},
  \bibinfo{pages}{066601} (\bibinfo{year}{2005}),
  \urlprefix\url{https://link.aps.org/doi/10.1103/PhysRevLett.95.066601}.

\bibitem[{\citenamefont{Han et~al.}(2022{\natexlab{a}})\citenamefont{Han, Lee,
  and Kim}}]{Han-HyunWooLeePhysRevLett.128.176601}
\bibinfo{author}{\bibfnamefont{S.}~\bibnamefont{Han}},
  \bibinfo{author}{\bibfnamefont{H.-W.} \bibnamefont{Lee}}, \bibnamefont{and}
  \bibinfo{author}{\bibfnamefont{K.-W.} \bibnamefont{Kim}},
  \bibinfo{journal}{Phys. Rev. Lett.} \textbf{\bibinfo{volume}{128}},
  \bibinfo{pages}{176601} (\bibinfo{year}{2022}{\natexlab{a}}),
  \urlprefix\url{https://link.aps.org/doi/10.1103/PhysRevLett.128.176601}.

\bibitem[{\citenamefont{Go et~al.}(2018)\citenamefont{Go, Jo, Kim, and
  Lee}}]{Go-Jo-Kim-Lee-PhysRevLett.121.086602}
\bibinfo{author}{\bibfnamefont{D.}~\bibnamefont{Go}},
  \bibinfo{author}{\bibfnamefont{D.}~\bibnamefont{Jo}},
  \bibinfo{author}{\bibfnamefont{C.}~\bibnamefont{Kim}}, \bibnamefont{and}
  \bibinfo{author}{\bibfnamefont{H.-W.} \bibnamefont{Lee}},
  \bibinfo{journal}{Phys. Rev. Lett.} \textbf{\bibinfo{volume}{121}},
  \bibinfo{pages}{086602} (\bibinfo{year}{2018}),
  \urlprefix\url{https://link.aps.org/doi/10.1103/PhysRevLett.121.086602}.

\bibitem[{\citenamefont{Ding et~al.}(2022)\citenamefont{Ding, Liang, Go, Yun,
  Xue, Liu, Becker, Yang, Du, Wang
  et~al.}}]{Mokrousov-Go-Experiment-PhysRevLett.128.067201}
\bibinfo{author}{\bibfnamefont{S.}~\bibnamefont{Ding}},
  \bibinfo{author}{\bibfnamefont{Z.}~\bibnamefont{Liang}},
  \bibinfo{author}{\bibfnamefont{D.}~\bibnamefont{Go}},
  \bibinfo{author}{\bibfnamefont{C.}~\bibnamefont{Yun}},
  \bibinfo{author}{\bibfnamefont{M.}~\bibnamefont{Xue}},
  \bibinfo{author}{\bibfnamefont{Z.}~\bibnamefont{Liu}},
  \bibinfo{author}{\bibfnamefont{S.}~\bibnamefont{Becker}},
  \bibinfo{author}{\bibfnamefont{W.}~\bibnamefont{Yang}},
  \bibinfo{author}{\bibfnamefont{H.}~\bibnamefont{Du}},
  \bibinfo{author}{\bibfnamefont{C.}~\bibnamefont{Wang}}, \bibnamefont{et~al.},
  \bibinfo{journal}{Phys. Rev. Lett.} \textbf{\bibinfo{volume}{128}},
  \bibinfo{pages}{067201} (\bibinfo{year}{2022}),
  \urlprefix\url{https://link.aps.org/doi/10.1103/PhysRevLett.128.067201}.

\bibitem[{\citenamefont{Chen et~al.}(2018)\citenamefont{Chen, Liu, Yang, Shi,
  Hu, Li, and Zeng}}]{Chen-Zeng-NatCommun2018}
\bibinfo{author}{\bibfnamefont{X.}~\bibnamefont{Chen}},
  \bibinfo{author}{\bibfnamefont{Y.}~\bibnamefont{Liu}},
  \bibinfo{author}{\bibfnamefont{G.}~\bibnamefont{Yang}},
  \bibinfo{author}{\bibfnamefont{H.}~\bibnamefont{Shi}},
  \bibinfo{author}{\bibfnamefont{C.}~\bibnamefont{Hu}},
  \bibinfo{author}{\bibfnamefont{M.}~\bibnamefont{Li}}, \bibnamefont{and}
  \bibinfo{author}{\bibfnamefont{H.}~\bibnamefont{Zeng}},
  \bibinfo{journal}{Nature Communications} \textbf{\bibinfo{volume}{9}},
  \bibinfo{pages}{2569} (\bibinfo{year}{2018}),
  \urlprefix\url{https://doi.org/10.1038/s41467-018-05057-z}.

\bibitem[{\citenamefont{Go et~al.}(2020)\citenamefont{Go, Freimuth, Hanke, Xue,
  Gomonay, Lee, Bl\"ugel, Haney, Lee, and Mokrousov}}]{Go1PRR2.033401}
\bibinfo{author}{\bibfnamefont{D.}~\bibnamefont{Go}},
  \bibinfo{author}{\bibfnamefont{F.}~\bibnamefont{Freimuth}},
  \bibinfo{author}{\bibfnamefont{J.-P.} \bibnamefont{Hanke}},
  \bibinfo{author}{\bibfnamefont{F.}~\bibnamefont{Xue}},
  \bibinfo{author}{\bibfnamefont{O.}~\bibnamefont{Gomonay}},
  \bibinfo{author}{\bibfnamefont{K.-J.} \bibnamefont{Lee}},
  \bibinfo{author}{\bibfnamefont{S.}~\bibnamefont{Bl\"ugel}},
  \bibinfo{author}{\bibfnamefont{P.~M.} \bibnamefont{Haney}},
  \bibinfo{author}{\bibfnamefont{H.-W.} \bibnamefont{Lee}}, \bibnamefont{and}
  \bibinfo{author}{\bibfnamefont{Y.}~\bibnamefont{Mokrousov}},
  \bibinfo{journal}{Phys. Rev. Research} \textbf{\bibinfo{volume}{2}},
  \bibinfo{pages}{033401} (\bibinfo{year}{2020}),
  \urlprefix\url{https://link.aps.org/doi/10.1103/PhysRevResearch.2.033401}.

\bibitem[{\citenamefont{Bose et~al.}(2022)\citenamefont{Bose, Kammerbauer, Go,
  Mokrousov, Jakob, and Klaeui}}]{Arnab-Orbital-Torque-Exp}
\bibinfo{author}{\bibfnamefont{A.}~\bibnamefont{Bose}},
  \bibinfo{author}{\bibfnamefont{F.}~\bibnamefont{Kammerbauer}},
  \bibinfo{author}{\bibfnamefont{D.}~\bibnamefont{Go}},
  \bibinfo{author}{\bibfnamefont{Y.}~\bibnamefont{Mokrousov}},
  \bibinfo{author}{\bibfnamefont{G.}~\bibnamefont{Jakob}}, \bibnamefont{and}
  \bibinfo{author}{\bibfnamefont{M.}~\bibnamefont{Klaeui}},
  \emph{\bibinfo{title}{Detection of long-range orbital-hall torques}}
  (\bibinfo{year}{2022}), \urlprefix\url{https://arxiv.org/abs/2210.02283}.

\bibitem[{\citenamefont{Go and Lee}(2020)}]{Go2PRR2.013177}
\bibinfo{author}{\bibfnamefont{D.}~\bibnamefont{Go}} \bibnamefont{and}
  \bibinfo{author}{\bibfnamefont{H.-W.} \bibnamefont{Lee}},
  \bibinfo{journal}{Phys. Rev. Research} \textbf{\bibinfo{volume}{2}},
  \bibinfo{pages}{013177} (\bibinfo{year}{2020}),
  \urlprefix\url{https://link.aps.org/doi/10.1103/PhysRevResearch.2.013177}.

\bibitem[{\citenamefont{Go et~al.}(2021{\natexlab{b}})\citenamefont{Go, Jo,
  Kim, Lee, Kang, Park, Blügel, Lee, and Mokrousov}}]{Go3_arxiv.2106.07928}
\bibinfo{author}{\bibfnamefont{D.}~\bibnamefont{Go}},
  \bibinfo{author}{\bibfnamefont{D.}~\bibnamefont{Jo}},
  \bibinfo{author}{\bibfnamefont{K.-W.} \bibnamefont{Kim}},
  \bibinfo{author}{\bibfnamefont{S.}~\bibnamefont{Lee}},
  \bibinfo{author}{\bibfnamefont{M.-G.} \bibnamefont{Kang}},
  \bibinfo{author}{\bibfnamefont{B.-G.} \bibnamefont{Park}},
  \bibinfo{author}{\bibfnamefont{S.}~\bibnamefont{Blügel}},
  \bibinfo{author}{\bibfnamefont{H.-W.} \bibnamefont{Lee}}, \bibnamefont{and}
  \bibinfo{author}{\bibfnamefont{Y.}~\bibnamefont{Mokrousov}},
  \emph{\bibinfo{title}{Long-range orbital magnetoelectric torque in
  ferromagnets}} (\bibinfo{year}{2021}{\natexlab{b}}),
  \urlprefix\url{https://arxiv.org/abs/2106.07928}.

\bibitem[{\citenamefont{Hayashi et~al.}(2023)\citenamefont{Hayashi, Jo, Go,
  Gao, Haku, Mokrousov, Lee, and Ando}}]{Go4_arxiv.2202.13896}
\bibinfo{author}{\bibfnamefont{H.}~\bibnamefont{Hayashi}},
  \bibinfo{author}{\bibfnamefont{D.}~\bibnamefont{Jo}},
  \bibinfo{author}{\bibfnamefont{D.}~\bibnamefont{Go}},
  \bibinfo{author}{\bibfnamefont{T.}~\bibnamefont{Gao}},
  \bibinfo{author}{\bibfnamefont{S.}~\bibnamefont{Haku}},
  \bibinfo{author}{\bibfnamefont{Y.}~\bibnamefont{Mokrousov}},
  \bibinfo{author}{\bibfnamefont{H.-W.} \bibnamefont{Lee}}, \bibnamefont{and}
  \bibinfo{author}{\bibfnamefont{K.}~\bibnamefont{Ando}},
  \bibinfo{journal}{Communications Physics} \textbf{\bibinfo{volume}{6}},
  \bibinfo{pages}{32} (\bibinfo{year}{2023}),
  \urlprefix\url{https://doi.org/10.1038/s42005-023-01139-7}.

\bibitem[{\citenamefont{Bhowal and
  Satpathy}(2020{\natexlab{a}})}]{Bhowal&Satpathy3}
\bibinfo{author}{\bibfnamefont{S.}~\bibnamefont{Bhowal}} \bibnamefont{and}
  \bibinfo{author}{\bibfnamefont{S.}~\bibnamefont{Satpathy}},
  \bibinfo{journal}{Phys. Rev. B} \textbf{\bibinfo{volume}{102}},
  \bibinfo{pages}{201403} (\bibinfo{year}{2020}{\natexlab{a}}),
  \urlprefix\url{https://link.aps.org/doi/10.1103/PhysRevB.102.201403}.

\bibitem[{\citenamefont{Yoda et~al.}(2018{\natexlab{a}})\citenamefont{Yoda,
  Yokoyama, and Murakami}}]{Murakami1}
\bibinfo{author}{\bibfnamefont{T.}~\bibnamefont{Yoda}},
  \bibinfo{author}{\bibfnamefont{T.}~\bibnamefont{Yokoyama}}, \bibnamefont{and}
  \bibinfo{author}{\bibfnamefont{S.}~\bibnamefont{Murakami}},
  \bibinfo{journal}{Nano Letters} \textbf{\bibinfo{volume}{18}},
  \bibinfo{pages}{916} (\bibinfo{year}{2018}{\natexlab{a}}),
  \urlprefix\url{https://doi.org/10.1021/acs.nanolett.7b04300}.

\bibitem[{\citenamefont{Osumi et~al.}(2021)\citenamefont{Osumi, Zhang, and
  Murakami}}]{Osumi-Zhang-Murakami-CommunPhys}
\bibinfo{author}{\bibfnamefont{K.}~\bibnamefont{Osumi}},
  \bibinfo{author}{\bibfnamefont{T.}~\bibnamefont{Zhang}}, \bibnamefont{and}
  \bibinfo{author}{\bibfnamefont{S.}~\bibnamefont{Murakami}},
  \bibinfo{journal}{Communications Physics} \textbf{\bibinfo{volume}{4}},
  \bibinfo{pages}{211} (\bibinfo{year}{2021}),
  \urlprefix\url{https://doi.org/10.1038/s42005-021-00702-4}.

\bibitem[{\citenamefont{Johansson et~al.}(2021)\citenamefont{Johansson,
  G\"obel, Henk, Bibes, and
  Mertig}}]{Johansson-Merting-PhysRevResearch.3.013275}
\bibinfo{author}{\bibfnamefont{A.}~\bibnamefont{Johansson}},
  \bibinfo{author}{\bibfnamefont{B.}~\bibnamefont{G\"obel}},
  \bibinfo{author}{\bibfnamefont{J.}~\bibnamefont{Henk}},
  \bibinfo{author}{\bibfnamefont{M.}~\bibnamefont{Bibes}}, \bibnamefont{and}
  \bibinfo{author}{\bibfnamefont{I.}~\bibnamefont{Mertig}},
  \bibinfo{journal}{Phys. Rev. Research} \textbf{\bibinfo{volume}{3}},
  \bibinfo{pages}{013275} (\bibinfo{year}{2021}),
  \urlprefix\url{https://link.aps.org/doi/10.1103/PhysRevResearch.3.013275}.

\bibitem[{\citenamefont{Salemi et~al.}(2021)\citenamefont{Salemi, Berritta, and
  Oppeneer}}]{Salemi-Oppeneer-PRMaterials-2021}
\bibinfo{author}{\bibfnamefont{L.}~\bibnamefont{Salemi}},
  \bibinfo{author}{\bibfnamefont{M.}~\bibnamefont{Berritta}}, \bibnamefont{and}
  \bibinfo{author}{\bibfnamefont{P.~M.} \bibnamefont{Oppeneer}},
  \bibinfo{journal}{Phys. Rev. Materials} \textbf{\bibinfo{volume}{5}},
  \bibinfo{pages}{074407} (\bibinfo{year}{2021}),
  \urlprefix\url{https://link.aps.org/doi/10.1103/PhysRevMaterials.5.074407}.

\bibitem[{\citenamefont{Yoda et~al.}(2018{\natexlab{b}})\citenamefont{Yoda,
  Yokoyama, and Murakami}}]{Yoda-Yokoyama-Murakami-NanoLett-2018}
\bibinfo{author}{\bibfnamefont{T.}~\bibnamefont{Yoda}},
  \bibinfo{author}{\bibfnamefont{T.}~\bibnamefont{Yokoyama}}, \bibnamefont{and}
  \bibinfo{author}{\bibfnamefont{S.}~\bibnamefont{Murakami}},
  \bibinfo{journal}{Nano Letters} \textbf{\bibinfo{volume}{18}},
  \bibinfo{pages}{916} (\bibinfo{year}{2018}{\natexlab{b}}),
  \urlprefix\url{https://doi.org/10.1021/acs.nanolett.7b04300}.

\bibitem[{\citenamefont{Massarelli et~al.}(2019)\citenamefont{Massarelli, Wu,
  and Paramekanti}}]{Massarelli-Wu-Paramekanti-PhysRevB.100.075136}
\bibinfo{author}{\bibfnamefont{G.}~\bibnamefont{Massarelli}},
  \bibinfo{author}{\bibfnamefont{B.}~\bibnamefont{Wu}}, \bibnamefont{and}
  \bibinfo{author}{\bibfnamefont{A.}~\bibnamefont{Paramekanti}},
  \bibinfo{journal}{Phys. Rev. B} \textbf{\bibinfo{volume}{100}},
  \bibinfo{pages}{075136} (\bibinfo{year}{2019}),
  \urlprefix\url{https://link.aps.org/doi/10.1103/PhysRevB.100.075136}.

\bibitem[{\citenamefont{Hayami et~al.}(2018)\citenamefont{Hayami, Yatsushiro,
  Yanagi, and Kusunose}}]{Hayami-Kusunose-PhysRevB.98.165110}
\bibinfo{author}{\bibfnamefont{S.}~\bibnamefont{Hayami}},
  \bibinfo{author}{\bibfnamefont{M.}~\bibnamefont{Yatsushiro}},
  \bibinfo{author}{\bibfnamefont{Y.}~\bibnamefont{Yanagi}}, \bibnamefont{and}
  \bibinfo{author}{\bibfnamefont{H.}~\bibnamefont{Kusunose}},
  \bibinfo{journal}{Phys. Rev. B} \textbf{\bibinfo{volume}{98}},
  \bibinfo{pages}{165110} (\bibinfo{year}{2018}),
  \urlprefix\url{https://link.aps.org/doi/10.1103/PhysRevB.98.165110}.

\bibitem[{\citenamefont{Hayami et~al.}(2016)\citenamefont{Hayami, Kusunose, and
  Motome}}]{Hayami-JPCM-2016}
\bibinfo{author}{\bibfnamefont{S.}~\bibnamefont{Hayami}},
  \bibinfo{author}{\bibfnamefont{H.}~\bibnamefont{Kusunose}}, \bibnamefont{and}
  \bibinfo{author}{\bibfnamefont{Y.}~\bibnamefont{Motome}},
  \bibinfo{journal}{Journal of Physics: Condensed Matter}
  \textbf{\bibinfo{volume}{28}}, \bibinfo{pages}{395601}
  (\bibinfo{year}{2016}),
  \urlprefix\url{https://doi.org/10.1088/0953-8984/28/39/395601}.

\bibitem[{\citenamefont{He and Law}(2020)}]{He-Law-PhysRevResearch.2.012073}
\bibinfo{author}{\bibfnamefont{W.-Y.} \bibnamefont{He}} \bibnamefont{and}
  \bibinfo{author}{\bibfnamefont{K.~T.} \bibnamefont{Law}},
  \bibinfo{journal}{Phys. Rev. Research} \textbf{\bibinfo{volume}{2}},
  \bibinfo{pages}{012073} (\bibinfo{year}{2020}),
  \urlprefix\url{https://link.aps.org/doi/10.1103/PhysRevResearch.2.012073}.

\bibitem[{\citenamefont{Furukawa et~al.}(2021)\citenamefont{Furukawa, Watanabe,
  Ogasawara, Kobayashi, and Itou}}]{Furukawa-Itou-PhysRevResearch.3.023111}
\bibinfo{author}{\bibfnamefont{T.}~\bibnamefont{Furukawa}},
  \bibinfo{author}{\bibfnamefont{Y.}~\bibnamefont{Watanabe}},
  \bibinfo{author}{\bibfnamefont{N.}~\bibnamefont{Ogasawara}},
  \bibinfo{author}{\bibfnamefont{K.}~\bibnamefont{Kobayashi}},
  \bibnamefont{and} \bibinfo{author}{\bibfnamefont{T.}~\bibnamefont{Itou}},
  \bibinfo{journal}{Phys. Rev. Research} \textbf{\bibinfo{volume}{3}},
  \bibinfo{pages}{023111} (\bibinfo{year}{2021}),
  \urlprefix\url{https://link.aps.org/doi/10.1103/PhysRevResearch.3.023111}.

\bibitem[{\citenamefont{Phong et~al.}(2019)\citenamefont{Phong, Addison, Ahn,
  Min, Agarwal, and Mele}}]{Mele-PRL-PhysRevLett.123.236403-2019}
\bibinfo{author}{\bibfnamefont{V.~o.~T.} \bibnamefont{Phong}},
  \bibinfo{author}{\bibfnamefont{Z.}~\bibnamefont{Addison}},
  \bibinfo{author}{\bibfnamefont{S.}~\bibnamefont{Ahn}},
  \bibinfo{author}{\bibfnamefont{H.}~\bibnamefont{Min}},
  \bibinfo{author}{\bibfnamefont{R.}~\bibnamefont{Agarwal}}, \bibnamefont{and}
  \bibinfo{author}{\bibfnamefont{E.~J.} \bibnamefont{Mele}},
  \bibinfo{journal}{Phys. Rev. Lett.} \textbf{\bibinfo{volume}{123}},
  \bibinfo{pages}{236403} (\bibinfo{year}{2019}),
  \urlprefix\url{https://link.aps.org/doi/10.1103/PhysRevLett.123.236403}.

\bibitem[{\citenamefont{Canonico
  et~al.}(2020{\natexlab{a}})\citenamefont{Canonico, Cysne, Rappoport, and
  Muniz}}]{Us1-PRB}
\bibinfo{author}{\bibfnamefont{L.~M.} \bibnamefont{Canonico}},
  \bibinfo{author}{\bibfnamefont{T.~P.} \bibnamefont{Cysne}},
  \bibinfo{author}{\bibfnamefont{T.~G.} \bibnamefont{Rappoport}},
  \bibnamefont{and} \bibinfo{author}{\bibfnamefont{R.~B.} \bibnamefont{Muniz}},
  \bibinfo{journal}{Phys. Rev. B} \textbf{\bibinfo{volume}{101}},
  \bibinfo{pages}{075429} (\bibinfo{year}{2020}{\natexlab{a}}),
  \urlprefix\url{https://link.aps.org/doi/10.1103/PhysRevB.101.075429}.

\bibitem[{\citenamefont{Canonico
  et~al.}(2020{\natexlab{b}})\citenamefont{Canonico, Cysne, Molina-Sanchez,
  Muniz, and Rappoport}}]{Us2-PRBR}
\bibinfo{author}{\bibfnamefont{L.~M.} \bibnamefont{Canonico}},
  \bibinfo{author}{\bibfnamefont{T.~P.} \bibnamefont{Cysne}},
  \bibinfo{author}{\bibfnamefont{A.}~\bibnamefont{Molina-Sanchez}},
  \bibinfo{author}{\bibfnamefont{R.~B.} \bibnamefont{Muniz}}, \bibnamefont{and}
  \bibinfo{author}{\bibfnamefont{T.~G.} \bibnamefont{Rappoport}},
  \bibinfo{journal}{Phys. Rev. B} \textbf{\bibinfo{volume}{101}},
  \bibinfo{pages}{161409} (\bibinfo{year}{2020}{\natexlab{b}}),
  \urlprefix\url{https://link.aps.org/doi/10.1103/PhysRevB.101.161409}.

\bibitem[{\citenamefont{Cysne et~al.}(2021{\natexlab{a}})\citenamefont{Cysne,
  Costa, Canonico, Nardelli, Muniz, and Rappoport}}]{Us3-PRL}
\bibinfo{author}{\bibfnamefont{T.~P.} \bibnamefont{Cysne}},
  \bibinfo{author}{\bibfnamefont{M.}~\bibnamefont{Costa}},
  \bibinfo{author}{\bibfnamefont{L.~M.} \bibnamefont{Canonico}},
  \bibinfo{author}{\bibfnamefont{M.~B.} \bibnamefont{Nardelli}},
  \bibinfo{author}{\bibfnamefont{R.~B.} \bibnamefont{Muniz}}, \bibnamefont{and}
  \bibinfo{author}{\bibfnamefont{T.~G.} \bibnamefont{Rappoport}},
  \bibinfo{journal}{Phys. Rev. Lett.} \textbf{\bibinfo{volume}{126}},
  \bibinfo{pages}{056601} (\bibinfo{year}{2021}{\natexlab{a}}),
  \urlprefix\url{https://link.aps.org/doi/10.1103/PhysRevLett.126.056601}.

\bibitem[{\citenamefont{Cysne et~al.}(2022)\citenamefont{Cysne, Bhowal,
  Vignale, and Rappoport}}]{Cysne-Bhowal-Vignale-Rappoport}
\bibinfo{author}{\bibfnamefont{T.~P.} \bibnamefont{Cysne}},
  \bibinfo{author}{\bibfnamefont{S.}~\bibnamefont{Bhowal}},
  \bibinfo{author}{\bibfnamefont{G.}~\bibnamefont{Vignale}}, \bibnamefont{and}
  \bibinfo{author}{\bibfnamefont{T.~G.} \bibnamefont{Rappoport}},
  \bibinfo{journal}{Phys. Rev. B} \textbf{\bibinfo{volume}{105}},
  \bibinfo{pages}{195421} (\bibinfo{year}{2022}),
  \urlprefix\url{https://link.aps.org/doi/10.1103/PhysRevB.105.195421}.

\bibitem[{\citenamefont{Bhowal and
  Satpathy}(2020{\natexlab{b}})}]{Bhowal-Satpathy-PhysRevB.101.121112}
\bibinfo{author}{\bibfnamefont{S.}~\bibnamefont{Bhowal}} \bibnamefont{and}
  \bibinfo{author}{\bibfnamefont{S.}~\bibnamefont{Satpathy}},
  \bibinfo{journal}{Phys. Rev. B} \textbf{\bibinfo{volume}{101}},
  \bibinfo{pages}{121112} (\bibinfo{year}{2020}{\natexlab{b}}),
  \urlprefix\url{https://link.aps.org/doi/10.1103/PhysRevB.101.121112}.

\bibitem[{\citenamefont{Mu et~al.}(2021)\citenamefont{Mu, Pan, and
  Zhou}}]{Jian-Zhou-NPJCompMat-2021}
\bibinfo{author}{\bibfnamefont{X.}~\bibnamefont{Mu}},
  \bibinfo{author}{\bibfnamefont{Y.}~\bibnamefont{Pan}}, \bibnamefont{and}
  \bibinfo{author}{\bibfnamefont{J.}~\bibnamefont{Zhou}}, \bibinfo{journal}{npj
  Computational Materials} \textbf{\bibinfo{volume}{7}} (\bibinfo{year}{2021}),
  \urlprefix\url{https://doi.org/10.1038/s41524-021-00531-7}.

\bibitem[{\citenamefont{Bhowal and
  Vignale}(2021)}]{Bhowal-Vignale-PhysRevB.103.195309}
\bibinfo{author}{\bibfnamefont{S.}~\bibnamefont{Bhowal}} \bibnamefont{and}
  \bibinfo{author}{\bibfnamefont{G.}~\bibnamefont{Vignale}},
  \bibinfo{journal}{Phys. Rev. B} \textbf{\bibinfo{volume}{103}},
  \bibinfo{pages}{195309} (\bibinfo{year}{2021}),
  \urlprefix\url{https://link.aps.org/doi/10.1103/PhysRevB.103.195309}.

\bibitem[{\citenamefont{He et~al.}(2020)\citenamefont{He, Goldhaber-Gordon, and
  Law}}]{He-2020-tbg-kome-NatCommun}
\bibinfo{author}{\bibfnamefont{W.-Y.} \bibnamefont{He}},
  \bibinfo{author}{\bibfnamefont{D.}~\bibnamefont{Goldhaber-Gordon}},
  \bibnamefont{and} \bibinfo{author}{\bibfnamefont{K.~T.} \bibnamefont{Law}},
  \bibinfo{journal}{Nature Communications} \textbf{\bibinfo{volume}{11}}
  (\bibinfo{year}{2020}),
  \urlprefix\url{https://doi.org/10.1038/s41467-020-15473-9}.

\bibitem[{\citenamefont{Schaefer and
  Nowack}(2021)}]{Schaefer-Katja-PhysRevB.103.224426}
\bibinfo{author}{\bibfnamefont{B.~T.} \bibnamefont{Schaefer}} \bibnamefont{and}
  \bibinfo{author}{\bibfnamefont{K.~C.} \bibnamefont{Nowack}},
  \bibinfo{journal}{Phys. Rev. B} \textbf{\bibinfo{volume}{103}},
  \bibinfo{pages}{224426} (\bibinfo{year}{2021}),
  \urlprefix\url{https://link.aps.org/doi/10.1103/PhysRevB.103.224426}.

\bibitem[{\citenamefont{Sch\"uler et~al.}(2022)\citenamefont{Sch\"uler,
  Pincelli, Dong, Devereaux, Wolf, Rettig, Ernstorfer, and
  Beaulieu}}]{Schuler-Pincelli-Beaulieu-PhysRevX.12.011019}
\bibinfo{author}{\bibfnamefont{M.}~\bibnamefont{Sch\"uler}},
  \bibinfo{author}{\bibfnamefont{T.}~\bibnamefont{Pincelli}},
  \bibinfo{author}{\bibfnamefont{S.}~\bibnamefont{Dong}},
  \bibinfo{author}{\bibfnamefont{T.~P.} \bibnamefont{Devereaux}},
  \bibinfo{author}{\bibfnamefont{M.}~\bibnamefont{Wolf}},
  \bibinfo{author}{\bibfnamefont{L.}~\bibnamefont{Rettig}},
  \bibinfo{author}{\bibfnamefont{R.}~\bibnamefont{Ernstorfer}},
  \bibnamefont{and} \bibinfo{author}{\bibfnamefont{S.}~\bibnamefont{Beaulieu}},
  \bibinfo{journal}{Phys. Rev. X} \textbf{\bibinfo{volume}{12}},
  \bibinfo{pages}{011019} (\bibinfo{year}{2022}),
  \urlprefix\url{https://link.aps.org/doi/10.1103/PhysRevX.12.011019}.

\bibitem[{\citenamefont{Beaulieu et~al.}(2020)\citenamefont{Beaulieu, Schusser,
  Dong, Sch\"uler, Pincelli, Dendzik, Maklar, Neef, Ebert, Hricovini
  et~al.}}]{Beaulieu-Schusser-2020-PhysRevLett.125.216404}
\bibinfo{author}{\bibfnamefont{S.}~\bibnamefont{Beaulieu}},
  \bibinfo{author}{\bibfnamefont{J.}~\bibnamefont{Schusser}},
  \bibinfo{author}{\bibfnamefont{S.}~\bibnamefont{Dong}},
  \bibinfo{author}{\bibfnamefont{M.}~\bibnamefont{Sch\"uler}},
  \bibinfo{author}{\bibfnamefont{T.}~\bibnamefont{Pincelli}},
  \bibinfo{author}{\bibfnamefont{M.}~\bibnamefont{Dendzik}},
  \bibinfo{author}{\bibfnamefont{J.}~\bibnamefont{Maklar}},
  \bibinfo{author}{\bibfnamefont{A.}~\bibnamefont{Neef}},
  \bibinfo{author}{\bibfnamefont{H.}~\bibnamefont{Ebert}},
  \bibinfo{author}{\bibfnamefont{K.}~\bibnamefont{Hricovini}},
  \bibnamefont{et~al.}, \bibinfo{journal}{Phys. Rev. Lett.}
  \textbf{\bibinfo{volume}{125}}, \bibinfo{pages}{216404}
  (\bibinfo{year}{2020}),
  \urlprefix\url{https://link.aps.org/doi/10.1103/PhysRevLett.125.216404}.

\bibitem[{\citenamefont{Han et~al.}(2022{\natexlab{b}})\citenamefont{Han, Lee,
  and Kim}}]{orbitaltexture-microscopicstudy-Seungyun}
\bibinfo{author}{\bibfnamefont{S.}~\bibnamefont{Han}},
  \bibinfo{author}{\bibfnamefont{H.-W.} \bibnamefont{Lee}}, \bibnamefont{and}
  \bibinfo{author}{\bibfnamefont{K.-W.} \bibnamefont{Kim}},
  \emph{\bibinfo{title}{Microscopic study of orbital textures}}
  (\bibinfo{year}{2022}{\natexlab{b}}),
  \urlprefix\url{https://arxiv.org/abs/2210.08876}.

\bibitem[{\citenamefont{Qian et~al.}(2022)\citenamefont{Qian, Liu, Liu, and
  Yao}}]{Quian-Liu-Liu-PhysRevB.105.045417}
\bibinfo{author}{\bibfnamefont{S.}~\bibnamefont{Qian}},
  \bibinfo{author}{\bibfnamefont{G.-B.} \bibnamefont{Liu}},
  \bibinfo{author}{\bibfnamefont{C.-C.} \bibnamefont{Liu}}, \bibnamefont{and}
  \bibinfo{author}{\bibfnamefont{Y.}~\bibnamefont{Yao}},
  \bibinfo{journal}{Phys. Rev. B} \textbf{\bibinfo{volume}{105}},
  \bibinfo{pages}{045417} (\bibinfo{year}{2022}),
  \urlprefix\url{https://link.aps.org/doi/10.1103/PhysRevB.105.045417}.

\bibitem[{\citenamefont{Zeng et~al.}(2021)\citenamefont{Zeng, Liu, Jiang, Sun,
  and Xie}}]{Jiang-Xie-PhysRevB.104.L161108}
\bibinfo{author}{\bibfnamefont{J.}~\bibnamefont{Zeng}},
  \bibinfo{author}{\bibfnamefont{H.}~\bibnamefont{Liu}},
  \bibinfo{author}{\bibfnamefont{H.}~\bibnamefont{Jiang}},
  \bibinfo{author}{\bibfnamefont{Q.-F.} \bibnamefont{Sun}}, \bibnamefont{and}
  \bibinfo{author}{\bibfnamefont{X.~C.} \bibnamefont{Xie}},
  \bibinfo{journal}{Phys. Rev. B} \textbf{\bibinfo{volume}{104}},
  \bibinfo{pages}{L161108} (\bibinfo{year}{2021}),
  \urlprefix\url{https://link.aps.org/doi/10.1103/PhysRevB.104.L161108}.

\bibitem[{\citenamefont{Costa et~al.}(2022)\citenamefont{Costa, Focassio,
  Cysne, Canonico, Schleder, Muniz, Fazzio, and
  Rappoport}}]{Us_Plus_FazzioPeople-Arxiv}
\bibinfo{author}{\bibfnamefont{M.}~\bibnamefont{Costa}},
  \bibinfo{author}{\bibfnamefont{B.}~\bibnamefont{Focassio}},
  \bibinfo{author}{\bibfnamefont{T.~P.} \bibnamefont{Cysne}},
  \bibinfo{author}{\bibfnamefont{L.~M.} \bibnamefont{Canonico}},
  \bibinfo{author}{\bibfnamefont{G.~R.} \bibnamefont{Schleder}},
  \bibinfo{author}{\bibfnamefont{R.~B.} \bibnamefont{Muniz}},
  \bibinfo{author}{\bibfnamefont{A.}~\bibnamefont{Fazzio}}, \bibnamefont{and}
  \bibinfo{author}{\bibfnamefont{T.~G.} \bibnamefont{Rappoport}},
  \emph{\bibinfo{title}{Connecting higher-order topology with the orbital hall
  effect in monolayers of transition metal dichalcogenides}}
  (\bibinfo{year}{2022}), \urlprefix\url{https://arxiv.org/abs/2205.00997}.

\bibitem[{\citenamefont{Xiao et~al.}(2021)\citenamefont{Xiao, Liu, Zhao, Yang,
  and Niu}}]{Xiao-Niu-PhysRevB.103.045401}
\bibinfo{author}{\bibfnamefont{C.}~\bibnamefont{Xiao}},
  \bibinfo{author}{\bibfnamefont{H.}~\bibnamefont{Liu}},
  \bibinfo{author}{\bibfnamefont{J.}~\bibnamefont{Zhao}},
  \bibinfo{author}{\bibfnamefont{S.~A.} \bibnamefont{Yang}}, \bibnamefont{and}
  \bibinfo{author}{\bibfnamefont{Q.}~\bibnamefont{Niu}},
  \bibinfo{journal}{Phys. Rev. B} \textbf{\bibinfo{volume}{103}},
  \bibinfo{pages}{045401} (\bibinfo{year}{2021}),
  \urlprefix\url{https://link.aps.org/doi/10.1103/PhysRevB.103.045401}.

\bibitem[{\citenamefont{Son et~al.}(2019)\citenamefont{Son, Kim, Ahn, Lee, and
  Lee}}]{Son-Kim-2019-PhysRevLett.123.036806}
\bibinfo{author}{\bibfnamefont{J.}~\bibnamefont{Son}},
  \bibinfo{author}{\bibfnamefont{K.-H.} \bibnamefont{Kim}},
  \bibinfo{author}{\bibfnamefont{Y.~H.} \bibnamefont{Ahn}},
  \bibinfo{author}{\bibfnamefont{H.-W.} \bibnamefont{Lee}}, \bibnamefont{and}
  \bibinfo{author}{\bibfnamefont{J.}~\bibnamefont{Lee}},
  \bibinfo{journal}{Phys. Rev. Lett.} \textbf{\bibinfo{volume}{123}},
  \bibinfo{pages}{036806} (\bibinfo{year}{2019}),
  \urlprefix\url{https://link.aps.org/doi/10.1103/PhysRevLett.123.036806}.

\bibitem[{\citenamefont{Bhowal and
  Satpathy}(2020{\natexlab{c}})}]{Bhowal-Satpahy-kOME-PhysRevB.102.201403}
\bibinfo{author}{\bibfnamefont{S.}~\bibnamefont{Bhowal}} \bibnamefont{and}
  \bibinfo{author}{\bibfnamefont{S.}~\bibnamefont{Satpathy}},
  \bibinfo{journal}{Phys. Rev. B} \textbf{\bibinfo{volume}{102}},
  \bibinfo{pages}{201403} (\bibinfo{year}{2020}{\natexlab{c}}),
  \urlprefix\url{https://link.aps.org/doi/10.1103/PhysRevB.102.201403}.

\bibitem[{\citenamefont{Lee et~al.}(2017)\citenamefont{Lee, Wang, Xie, Mak, and
  Shan}}]{Lee-2017-NatMaterials-887-9}
\bibinfo{author}{\bibfnamefont{J.}~\bibnamefont{Lee}},
  \bibinfo{author}{\bibfnamefont{Z.}~\bibnamefont{Wang}},
  \bibinfo{author}{\bibfnamefont{H.}~\bibnamefont{Xie}},
  \bibinfo{author}{\bibfnamefont{K.~F.} \bibnamefont{Mak}}, \bibnamefont{and}
  \bibinfo{author}{\bibfnamefont{J.}~\bibnamefont{Shan}},
  \bibinfo{journal}{Nature Materials} \textbf{\bibinfo{volume}{16}},
  \bibinfo{pages}{887} (\bibinfo{year}{2017}),
  \urlprefix\url{https://doi.org/10.1038/nmat4931}.

\bibitem[{\citenamefont{Junior et~al.}(2022)\citenamefont{Junior, Zollner,
  Wo{\'{z}}niak, Kurpas, Gmitra, and Fabian}}]{Faria_Junior_2022}
\bibinfo{author}{\bibfnamefont{P.~E.~F.} \bibnamefont{Junior}},
  \bibinfo{author}{\bibfnamefont{K.}~\bibnamefont{Zollner}},
  \bibinfo{author}{\bibfnamefont{T.}~\bibnamefont{Wo{\'{z}}niak}},
  \bibinfo{author}{\bibfnamefont{M.}~\bibnamefont{Kurpas}},
  \bibinfo{author}{\bibfnamefont{M.}~\bibnamefont{Gmitra}}, \bibnamefont{and}
  \bibinfo{author}{\bibfnamefont{J.}~\bibnamefont{Fabian}},
  \bibinfo{journal}{New Journal of Physics} \textbf{\bibinfo{volume}{24}},
  \bibinfo{pages}{083004} (\bibinfo{year}{2022}),
  \urlprefix\url{https://doi.org/10.1088/1367-2630/ac7e21}.

\bibitem[{\citenamefont{Cysne et~al.}(2018)\citenamefont{Cysne, Ferreira, and
  Rappoport}}]{CF-PhysRevB.98.045407}
\bibinfo{author}{\bibfnamefont{T.~P.} \bibnamefont{Cysne}},
  \bibinfo{author}{\bibfnamefont{A.}~\bibnamefont{Ferreira}}, \bibnamefont{and}
  \bibinfo{author}{\bibfnamefont{T.~G.} \bibnamefont{Rappoport}},
  \bibinfo{journal}{Phys. Rev. B} \textbf{\bibinfo{volume}{98}},
  \bibinfo{pages}{045407} (\bibinfo{year}{2018}),
  \urlprefix\url{https://link.aps.org/doi/10.1103/PhysRevB.98.045407}.

\bibitem[{\citenamefont{Wang and Wang}(2015)}]{C2v-Suplatt-PhysRevB.92.075419}
\bibinfo{author}{\bibfnamefont{S.~K.} \bibnamefont{Wang}} \bibnamefont{and}
  \bibinfo{author}{\bibfnamefont{J.}~\bibnamefont{Wang}},
  \bibinfo{journal}{Phys. Rev. B} \textbf{\bibinfo{volume}{92}},
  \bibinfo{pages}{075419} (\bibinfo{year}{2015}),
  \urlprefix\url{https://link.aps.org/doi/10.1103/PhysRevB.92.075419}.

\bibitem[{\citenamefont{Liu et~al.}(2013)\citenamefont{Liu, Shan, Yao, Yao, and
  Xiao}}]{Xiao-three-band}
\bibinfo{author}{\bibfnamefont{G.-B.} \bibnamefont{Liu}},
  \bibinfo{author}{\bibfnamefont{W.-Y.} \bibnamefont{Shan}},
  \bibinfo{author}{\bibfnamefont{Y.}~\bibnamefont{Yao}},
  \bibinfo{author}{\bibfnamefont{W.}~\bibnamefont{Yao}}, \bibnamefont{and}
  \bibinfo{author}{\bibfnamefont{D.}~\bibnamefont{Xiao}},
  \bibinfo{journal}{Phys. Rev. B} \textbf{\bibinfo{volume}{88}},
  \bibinfo{pages}{085433} (\bibinfo{year}{2013}),
  \urlprefix\url{https://link.aps.org/doi/10.1103/PhysRevB.88.085433}.

\bibitem[{\citenamefont{Wu et~al.}(2019)\citenamefont{Wu, Zhou, Cai, Cheung,
  Liu, Huang, Lin, Han, An, Wang
  et~al.}}]{Wu-2019-NatComn-10.1038/s41467-019-08629-9}
\bibinfo{author}{\bibfnamefont{Z.}~\bibnamefont{Wu}},
  \bibinfo{author}{\bibfnamefont{B.~T.} \bibnamefont{Zhou}},
  \bibinfo{author}{\bibfnamefont{X.}~\bibnamefont{Cai}},
  \bibinfo{author}{\bibfnamefont{P.}~\bibnamefont{Cheung}},
  \bibinfo{author}{\bibfnamefont{G.-B.} \bibnamefont{Liu}},
  \bibinfo{author}{\bibfnamefont{M.}~\bibnamefont{Huang}},
  \bibinfo{author}{\bibfnamefont{J.}~\bibnamefont{Lin}},
  \bibinfo{author}{\bibfnamefont{T.}~\bibnamefont{Han}},
  \bibinfo{author}{\bibfnamefont{L.}~\bibnamefont{An}},
  \bibinfo{author}{\bibfnamefont{Y.}~\bibnamefont{Wang}}, \bibnamefont{et~al.},
  \bibinfo{journal}{Nature Communications} \textbf{\bibinfo{volume}{10}},
  \bibinfo{pages}{611} (\bibinfo{year}{2019}),
  \urlprefix\url{https://doi.org/10.1038/s41467-019-08629-9}.

\bibitem[{\citenamefont{Cysne et~al.}(2021{\natexlab{b}})\citenamefont{Cysne,
  Guimar\~aes, Canonico, Rappoport, and Muniz}}]{Us4-PRB}
\bibinfo{author}{\bibfnamefont{T.~P.} \bibnamefont{Cysne}},
  \bibinfo{author}{\bibfnamefont{F.~S.~M.} \bibnamefont{Guimar\~aes}},
  \bibinfo{author}{\bibfnamefont{L.~M.} \bibnamefont{Canonico}},
  \bibinfo{author}{\bibfnamefont{T.~G.} \bibnamefont{Rappoport}},
  \bibnamefont{and} \bibinfo{author}{\bibfnamefont{R.~B.} \bibnamefont{Muniz}},
  \bibinfo{journal}{Phys. Rev. B} \textbf{\bibinfo{volume}{104}},
  \bibinfo{pages}{165403} (\bibinfo{year}{2021}{\natexlab{b}}),
  \urlprefix\url{https://link.aps.org/doi/10.1103/PhysRevB.104.165403}.

\bibitem[{\citenamefont{Sinova et~al.}(2015)\citenamefont{Sinova, Valenzuela,
  Wunderlich, Back, and Jungwirth}}]{SHE-Review-RevModPhys.87.1213}
\bibinfo{author}{\bibfnamefont{J.}~\bibnamefont{Sinova}},
  \bibinfo{author}{\bibfnamefont{S.~O.} \bibnamefont{Valenzuela}},
  \bibinfo{author}{\bibfnamefont{J.}~\bibnamefont{Wunderlich}},
  \bibinfo{author}{\bibfnamefont{C.~H.} \bibnamefont{Back}}, \bibnamefont{and}
  \bibinfo{author}{\bibfnamefont{T.}~\bibnamefont{Jungwirth}},
  \bibinfo{journal}{Rev. Mod. Phys.} \textbf{\bibinfo{volume}{87}},
  \bibinfo{pages}{1213} (\bibinfo{year}{2015}),
  \urlprefix\url{https://link.aps.org/doi/10.1103/RevModPhys.87.1213}.

\bibitem[{\citenamefont{Nikoli\ifmmode~\acute{c}\else \'{c}\fi{}
  et~al.}(2005)\citenamefont{Nikoli\ifmmode~\acute{c}\else \'{c}\fi{}, Souma,
  Z\^arbo, and Sinova}}]{2DEG-Rashba-Acc-PhysRevLett.95.046601}
\bibinfo{author}{\bibfnamefont{B.~K.}
  \bibnamefont{Nikoli\ifmmode~\acute{c}\else \'{c}\fi{}}},
  \bibinfo{author}{\bibfnamefont{S.}~\bibnamefont{Souma}},
  \bibinfo{author}{\bibfnamefont{L.~P.} \bibnamefont{Z\^arbo}},
  \bibnamefont{and} \bibinfo{author}{\bibfnamefont{J.}~\bibnamefont{Sinova}},
  \bibinfo{journal}{Phys. Rev. Lett.} \textbf{\bibinfo{volume}{95}},
  \bibinfo{pages}{046601} (\bibinfo{year}{2005}),
  \urlprefix\url{https://link.aps.org/doi/10.1103/PhysRevLett.95.046601}.

\bibitem[{\citenamefont{Nomura et~al.}(2005)\citenamefont{Nomura, Wunderlich,
  Sinova, Kaestner, MacDonald, and
  Jungwirth}}]{2DEG-Rashba-Acc-PhysRevB.72.245330}
\bibinfo{author}{\bibfnamefont{K.}~\bibnamefont{Nomura}},
  \bibinfo{author}{\bibfnamefont{J.}~\bibnamefont{Wunderlich}},
  \bibinfo{author}{\bibfnamefont{J.}~\bibnamefont{Sinova}},
  \bibinfo{author}{\bibfnamefont{B.}~\bibnamefont{Kaestner}},
  \bibinfo{author}{\bibfnamefont{A.~H.} \bibnamefont{MacDonald}},
  \bibnamefont{and}
  \bibinfo{author}{\bibfnamefont{T.}~\bibnamefont{Jungwirth}},
  \bibinfo{journal}{Phys. Rev. B} \textbf{\bibinfo{volume}{72}},
  \bibinfo{pages}{245330} (\bibinfo{year}{2005}),
  \urlprefix\url{https://link.aps.org/doi/10.1103/PhysRevB.72.245330}.

\bibitem[{\citenamefont{Johansson et~al.}(2018)\citenamefont{Johansson, Henk,
  and Mertig}}]{Mertig-PhysRevB.97.085417}
\bibinfo{author}{\bibfnamefont{A.}~\bibnamefont{Johansson}},
  \bibinfo{author}{\bibfnamefont{J.}~\bibnamefont{Henk}}, \bibnamefont{and}
  \bibinfo{author}{\bibfnamefont{I.}~\bibnamefont{Mertig}},
  \bibinfo{journal}{Phys. Rev. B} \textbf{\bibinfo{volume}{97}},
  \bibinfo{pages}{085417} (\bibinfo{year}{2018}),
  \urlprefix\url{https://link.aps.org/doi/10.1103/PhysRevB.97.085417}.

\bibitem[{\citenamefont{Das et~al.}(2013)\citenamefont{Das, Chen, Penumatcha,
  and Appenzeller}}]{ChargeCurrent-MoS2}
\bibinfo{author}{\bibfnamefont{S.}~\bibnamefont{Das}},
  \bibinfo{author}{\bibfnamefont{H.-Y.} \bibnamefont{Chen}},
  \bibinfo{author}{\bibfnamefont{A.~V.} \bibnamefont{Penumatcha}},
  \bibnamefont{and}
  \bibinfo{author}{\bibfnamefont{J.}~\bibnamefont{Appenzeller}},
  \bibinfo{journal}{Nano Letters} \textbf{\bibinfo{volume}{13}},
  \bibinfo{pages}{100} (\bibinfo{year}{2013}), \bibinfo{note}{pMID: 23240655},
  \eprint{https://doi.org/10.1021/nl303583v},
  \urlprefix\url{https://doi.org/10.1021/nl303583v}.

\bibitem[{\citenamefont{McClellan et~al.}(2021)\citenamefont{McClellan, Yalon,
  Smithe, Suryavanshi, and Pop}}]{McClellan-doi:10.1021/acsnano.0c09078}
\bibinfo{author}{\bibfnamefont{C.~J.} \bibnamefont{McClellan}},
  \bibinfo{author}{\bibfnamefont{E.}~\bibnamefont{Yalon}},
  \bibinfo{author}{\bibfnamefont{K.~K.~H.} \bibnamefont{Smithe}},
  \bibinfo{author}{\bibfnamefont{S.~V.} \bibnamefont{Suryavanshi}},
  \bibnamefont{and} \bibinfo{author}{\bibfnamefont{E.}~\bibnamefont{Pop}},
  \bibinfo{journal}{ACS Nano} \textbf{\bibinfo{volume}{15}},
  \bibinfo{pages}{1587} (\bibinfo{year}{2021}), \bibinfo{note}{pMID: 33405894},
  \eprint{https://doi.org/10.1021/acsnano.0c09078},
  \urlprefix\url{https://doi.org/10.1021/acsnano.0c09078}.

\bibitem[{\citenamefont{Agapito et~al.}(2013)\citenamefont{Agapito, Ferretti,
  Calzolari, Curtarolo, and Buongiorno~Nardelli}}]{PAO1}
\bibinfo{author}{\bibfnamefont{L.~A.} \bibnamefont{Agapito}},
  \bibinfo{author}{\bibfnamefont{A.}~\bibnamefont{Ferretti}},
  \bibinfo{author}{\bibfnamefont{A.}~\bibnamefont{Calzolari}},
  \bibinfo{author}{\bibfnamefont{S.}~\bibnamefont{Curtarolo}},
  \bibnamefont{and}
  \bibinfo{author}{\bibfnamefont{M.}~\bibnamefont{Buongiorno~Nardelli}},
  \bibinfo{journal}{Phys. Rev. B} \textbf{\bibinfo{volume}{88}},
  \bibinfo{pages}{165127} (\bibinfo{year}{2013}),
  \urlprefix\url{https://link.aps.org/doi/10.1103/PhysRevB.88.165127}.

\bibitem[{\citenamefont{Cerasoli et~al.}(2021)\citenamefont{Cerasoli, Supka,
  Jayaraj, Costa, Siloi, Sławińska, Curtarolo, Fornari, Ceresoli, and
  {Buongiorno Nardelli}}}]{PAO6}
\bibinfo{author}{\bibfnamefont{F.~T.} \bibnamefont{Cerasoli}},
  \bibinfo{author}{\bibfnamefont{A.~R.} \bibnamefont{Supka}},
  \bibinfo{author}{\bibfnamefont{A.}~\bibnamefont{Jayaraj}},
  \bibinfo{author}{\bibfnamefont{M.}~\bibnamefont{Costa}},
  \bibinfo{author}{\bibfnamefont{I.}~\bibnamefont{Siloi}},
  \bibinfo{author}{\bibfnamefont{J.}~\bibnamefont{Sławińska}},
  \bibinfo{author}{\bibfnamefont{S.}~\bibnamefont{Curtarolo}},
  \bibinfo{author}{\bibfnamefont{M.}~\bibnamefont{Fornari}},
  \bibinfo{author}{\bibfnamefont{D.}~\bibnamefont{Ceresoli}}, \bibnamefont{and}
  \bibinfo{author}{\bibfnamefont{M.}~\bibnamefont{{Buongiorno Nardelli}}},
  \bibinfo{journal}{Computational Materials Science}
  \textbf{\bibinfo{volume}{200}}, \bibinfo{pages}{110828}
  (\bibinfo{year}{2021}), ISSN \bibinfo{issn}{0927-0256},
  \urlprefix\url{https://www.sciencedirect.com/science/article/pii/S0927025621005486}.

\bibitem[{\citenamefont{Han et~al.}(2007)\citenamefont{Han, \"Ozyilmaz, Zhang,
  and Kim}}]{PKim-PhysRevLett.98.206805}
\bibinfo{author}{\bibfnamefont{M.~Y.} \bibnamefont{Han}},
  \bibinfo{author}{\bibfnamefont{B.}~\bibnamefont{\"Ozyilmaz}},
  \bibinfo{author}{\bibfnamefont{Y.}~\bibnamefont{Zhang}}, \bibnamefont{and}
  \bibinfo{author}{\bibfnamefont{P.}~\bibnamefont{Kim}},
  \bibinfo{journal}{Phys. Rev. Lett.} \textbf{\bibinfo{volume}{98}},
  \bibinfo{pages}{206805} (\bibinfo{year}{2007}),
  \urlprefix\url{https://link.aps.org/doi/10.1103/PhysRevLett.98.206805}.

\bibitem[{\citenamefont{Ridolfi et~al.}(2017)\citenamefont{Ridolfi, Lima,
  Mucciolo, and Lewenkopf}}]{Emilia-Caio-PhysRevB.95.035430}
\bibinfo{author}{\bibfnamefont{E.}~\bibnamefont{Ridolfi}},
  \bibinfo{author}{\bibfnamefont{L.~R.~F.} \bibnamefont{Lima}},
  \bibinfo{author}{\bibfnamefont{E.~R.} \bibnamefont{Mucciolo}},
  \bibnamefont{and} \bibinfo{author}{\bibfnamefont{C.~H.}
  \bibnamefont{Lewenkopf}}, \bibinfo{journal}{Phys. Rev. B}
  \textbf{\bibinfo{volume}{95}}, \bibinfo{pages}{035430}
  (\bibinfo{year}{2017}),
  \urlprefix\url{https://link.aps.org/doi/10.1103/PhysRevB.95.035430}.

\bibitem[{\citenamefont{Salemi et~al.}(2019)\citenamefont{Salemi, Berritta,
  Nandy, and Oppeneer}}]{NatComn-10.1038-Salemi2019}
\bibinfo{author}{\bibfnamefont{L.}~\bibnamefont{Salemi}},
  \bibinfo{author}{\bibfnamefont{M.}~\bibnamefont{Berritta}},
  \bibinfo{author}{\bibfnamefont{A.~K.} \bibnamefont{Nandy}}, \bibnamefont{and}
  \bibinfo{author}{\bibfnamefont{P.~M.} \bibnamefont{Oppeneer}},
  \bibinfo{journal}{Nature Communications} \textbf{\bibinfo{volume}{10}},
  \bibinfo{pages}{5381} (\bibinfo{year}{2019}),
  \urlprefix\url{https://doi.org/10.1038/s41467-019-13367-z}.

\bibitem[{\citenamefont{Chen et~al.}(2022)\citenamefont{Chen, Gu, Li, Wang, and
  Liu}}]{Layer-resolved-polarization}
\bibinfo{author}{\bibfnamefont{W.}~\bibnamefont{Chen}},
  \bibinfo{author}{\bibfnamefont{M.}~\bibnamefont{Gu}},
  \bibinfo{author}{\bibfnamefont{J.}~\bibnamefont{Li}},
  \bibinfo{author}{\bibfnamefont{P.}~\bibnamefont{Wang}}, \bibnamefont{and}
  \bibinfo{author}{\bibfnamefont{Q.}~\bibnamefont{Liu}},
  \bibinfo{journal}{Phys. Rev. Lett.} \textbf{\bibinfo{volume}{129}},
  \bibinfo{pages}{276601} (\bibinfo{year}{2022}),
  \urlprefix\url{https://link.aps.org/doi/10.1103/PhysRevLett.129.276601}.

\bibitem[{\citenamefont{Baugher et~al.}(2013)\citenamefont{Baugher, Churchill,
  Yang, and Jarillo-Herrero}}]{ElectronicDensity-baugher2013intrinsic}
\bibinfo{author}{\bibfnamefont{B.~W.~H.} \bibnamefont{Baugher}},
  \bibinfo{author}{\bibfnamefont{H.~O.~H.} \bibnamefont{Churchill}},
  \bibinfo{author}{\bibfnamefont{Y.}~\bibnamefont{Yang}}, \bibnamefont{and}
  \bibinfo{author}{\bibfnamefont{P.}~\bibnamefont{Jarillo-Herrero}},
  \bibinfo{journal}{Nano Letters} \textbf{\bibinfo{volume}{13}},
  \bibinfo{pages}{4212} (\bibinfo{year}{2013}), \bibinfo{note}{pMID: 23930826},
  \eprint{https://doi.org/10.1021/nl401916s},
  \urlprefix\url{https://doi.org/10.1021/nl401916s}.

\bibitem[{\citenamefont{Pezo et~al.}(2019)\citenamefont{Pezo, Lima, Costa, and
  Fazzio}}]{Pezo2019-FermiLevel-NR}
\bibinfo{author}{\bibfnamefont{A.}~\bibnamefont{Pezo}},
  \bibinfo{author}{\bibfnamefont{M.~P.} \bibnamefont{Lima}},
  \bibinfo{author}{\bibfnamefont{M.}~\bibnamefont{Costa}}, \bibnamefont{and}
  \bibinfo{author}{\bibfnamefont{A.}~\bibnamefont{Fazzio}},
  \bibinfo{journal}{Physical Chemistry Chemical Physics}
  \textbf{\bibinfo{volume}{21}}, \bibinfo{pages}{11359} (\bibinfo{year}{2019}),
  \urlprefix\url{https://doi.org/10.1039/c9cp01590f}.

\bibitem[{\citenamefont{Davelou et~al.}(2017)\citenamefont{Davelou, Kopidakis,
  Kaxiras, and Remediakis}}]{FermiLevel-NR-PhysRevB.96.165436}
\bibinfo{author}{\bibfnamefont{D.}~\bibnamefont{Davelou}},
  \bibinfo{author}{\bibfnamefont{G.}~\bibnamefont{Kopidakis}},
  \bibinfo{author}{\bibfnamefont{E.}~\bibnamefont{Kaxiras}}, \bibnamefont{and}
  \bibinfo{author}{\bibfnamefont{I.~N.} \bibnamefont{Remediakis}},
  \bibinfo{journal}{Phys. Rev. B} \textbf{\bibinfo{volume}{96}},
  \bibinfo{pages}{165436} (\bibinfo{year}{2017}),
  \urlprefix\url{https://link.aps.org/doi/10.1103/PhysRevB.96.165436}.

\bibitem[{\citenamefont{Ivchenko and Pikus}(1978)}]{Ivchenko-Pikus-Chiral-JETP}
\bibinfo{author}{\bibfnamefont{E.}~\bibnamefont{Ivchenko}} \bibnamefont{and}
  \bibinfo{author}{\bibfnamefont{G.}~\bibnamefont{Pikus}},
  \bibinfo{journal}{ZhETF Pisma Redaktsiiu} \textbf{\bibinfo{volume}{27}},
  \bibinfo{pages}{640} (\bibinfo{year}{1978}).

\bibitem[{\citenamefont{Vorob'ev et~al.}(1979)\citenamefont{Vorob'ev, Ivchenko,
  Pikus, Farbshteǐn, Shalygin, and Shturbin}}]{Vorobev-JETP}
\bibinfo{author}{\bibfnamefont{L.}~\bibnamefont{Vorob'ev}},
  \bibinfo{author}{\bibfnamefont{E.}~\bibnamefont{Ivchenko}},
  \bibinfo{author}{\bibfnamefont{G.}~\bibnamefont{Pikus}},
  \bibinfo{author}{\bibfnamefont{I.}~\bibnamefont{Farbshteǐn}},
  \bibinfo{author}{\bibfnamefont{V.}~\bibnamefont{Shalygin}}, \bibnamefont{and}
  \bibinfo{author}{\bibfnamefont{A.}~\bibnamefont{Shturbin}},
  \bibinfo{journal}{Soviet Journal of Experimental and Theoretical Physics
  Letters} \textbf{\bibinfo{volume}{29}}, \bibinfo{pages}{441}
  (\bibinfo{year}{1979}).

\bibitem[{\citenamefont{Edelstein}(1990)}]{Edelstein1990}
\bibinfo{author}{\bibfnamefont{V.}~\bibnamefont{Edelstein}},
  \bibinfo{journal}{Solid State Communications} \textbf{\bibinfo{volume}{73}},
  \bibinfo{pages}{233} (\bibinfo{year}{1990}),
  \urlprefix\url{https://doi.org/10.1016/0038-1098(90)90963-c}.

\bibitem[{\citenamefont{Aljarb et~al.}(2020)\citenamefont{Aljarb, Fu, Hsu,
  Chuu, Wan, Hakami, Naphade, Yengel, Lee, Brems
  et~al.}}]{MOS2Nanoribbons-Syntetization-NatMaterials}
\bibinfo{author}{\bibfnamefont{A.}~\bibnamefont{Aljarb}},
  \bibinfo{author}{\bibfnamefont{J.-H.} \bibnamefont{Fu}},
  \bibinfo{author}{\bibfnamefont{C.-C.} \bibnamefont{Hsu}},
  \bibinfo{author}{\bibfnamefont{C.-P.} \bibnamefont{Chuu}},
  \bibinfo{author}{\bibfnamefont{Y.}~\bibnamefont{Wan}},
  \bibinfo{author}{\bibfnamefont{M.}~\bibnamefont{Hakami}},
  \bibinfo{author}{\bibfnamefont{D.~R.} \bibnamefont{Naphade}},
  \bibinfo{author}{\bibfnamefont{E.}~\bibnamefont{Yengel}},
  \bibinfo{author}{\bibfnamefont{C.-J.} \bibnamefont{Lee}},
  \bibinfo{author}{\bibfnamefont{S.}~\bibnamefont{Brems}},
  \bibnamefont{et~al.}, \bibinfo{journal}{Nature Materials}
  \textbf{\bibinfo{volume}{19}}, \bibinfo{pages}{1300} (\bibinfo{year}{2020}),
  \urlprefix\url{https://doi.org/10.1038/s41563-020-0795-4}.

\bibitem[{\citenamefont{{J\"{u}lich Supercomputing Centre}}(2018)}]{jureca}
\bibinfo{author}{\bibnamefont{{J\"{u}lich Supercomputing Centre}}},
  \bibinfo{journal}{Journal of large-scale research facilities}
  \textbf{\bibinfo{volume}{4}} (\bibinfo{year}{2018}),
  \urlprefix\url{http://dx.doi.org/10.17815/jlsrf-4-121-1}.

\bibitem[{\citenamefont{Vanderbilt}(2018)}]{vanderbiltBook}
\bibinfo{author}{\bibfnamefont{D.}~\bibnamefont{Vanderbilt}},
  \emph{\bibinfo{title}{Berry Phases in Electronic Structure Theory: Electric
  Polarization, Orbital Magnetization and Topological Insulators}}
  (\bibinfo{publisher}{Cambridge University Press}, \bibinfo{year}{2018}),
  \bibinfo{note}{\url{www.cambridge.org/9781107157651}}.

\bibitem[{\citenamefont{Gong et~al.}(2013)\citenamefont{Gong, Liu, Yu, Xiao,
  Cui, Xu, and Yao}}]{Gong_2013}
\bibinfo{author}{\bibfnamefont{Z.}~\bibnamefont{Gong}},
  \bibinfo{author}{\bibfnamefont{G.-B.} \bibnamefont{Liu}},
  \bibinfo{author}{\bibfnamefont{H.}~\bibnamefont{Yu}},
  \bibinfo{author}{\bibfnamefont{D.}~\bibnamefont{Xiao}},
  \bibinfo{author}{\bibfnamefont{X.}~\bibnamefont{Cui}},
  \bibinfo{author}{\bibfnamefont{X.}~\bibnamefont{Xu}}, \bibnamefont{and}
  \bibinfo{author}{\bibfnamefont{W.}~\bibnamefont{Yao}},
  \bibinfo{journal}{Nature Communications} \textbf{\bibinfo{volume}{4}}
  (\bibinfo{year}{2013}), ISSN \bibinfo{issn}{2041-1723},
  \urlprefix\url{http://dx.doi.org/10.1038/ncomms3053}.

\bibitem[{\citenamefont{Hohenberg and Kohn}(1964)}]{DFT1}
\bibinfo{author}{\bibfnamefont{P.}~\bibnamefont{Hohenberg}} \bibnamefont{and}
  \bibinfo{author}{\bibfnamefont{W.}~\bibnamefont{Kohn}},
  \bibinfo{journal}{Phys. Rev.} \textbf{\bibinfo{volume}{136}},
  \bibinfo{pages}{B864} (\bibinfo{year}{1964}),
  \urlprefix\url{https://link.aps.org/doi/10.1103/PhysRev.136.B864}.

\bibitem[{\citenamefont{Kohn and Sham}(1965)}]{DFT2}
\bibinfo{author}{\bibfnamefont{W.}~\bibnamefont{Kohn}} \bibnamefont{and}
  \bibinfo{author}{\bibfnamefont{L.~J.} \bibnamefont{Sham}},
  \bibinfo{journal}{Phys. Rev.} \textbf{\bibinfo{volume}{140}},
  \bibinfo{pages}{A1133} (\bibinfo{year}{1965}),
  \urlprefix\url{https://link.aps.org/doi/10.1103/PhysRev.140.A1133}.

\bibitem[{\citenamefont{Giannozzi et~al.}(2017)\citenamefont{Giannozzi,
  Andreussi, Brumme, Bunau, Buongiorno~Nardelli, Calandra, Car, Cavazzoni,
  Ceresoli, Cococcioni et~al.}}]{QE-2017}
\bibinfo{author}{\bibfnamefont{P.}~\bibnamefont{Giannozzi}},
  \bibinfo{author}{\bibfnamefont{O.}~\bibnamefont{Andreussi}},
  \bibinfo{author}{\bibfnamefont{T.}~\bibnamefont{Brumme}},
  \bibinfo{author}{\bibfnamefont{O.}~\bibnamefont{Bunau}},
  \bibinfo{author}{\bibfnamefont{M.}~\bibnamefont{Buongiorno~Nardelli}},
  \bibinfo{author}{\bibfnamefont{M.}~\bibnamefont{Calandra}},
  \bibinfo{author}{\bibfnamefont{R.}~\bibnamefont{Car}},
  \bibinfo{author}{\bibfnamefont{C.}~\bibnamefont{Cavazzoni}},
  \bibinfo{author}{\bibfnamefont{D.}~\bibnamefont{Ceresoli}},
  \bibinfo{author}{\bibfnamefont{M.}~\bibnamefont{Cococcioni}},
  \bibnamefont{et~al.}, \bibinfo{journal}{Journal of Physics: Condensed Matter}
  \textbf{\bibinfo{volume}{29}}, \bibinfo{pages}{465901}
  (\bibinfo{year}{2017}),
  \urlprefix\url{http://stacks.iop.org/0953-8984/29/i=46/a=465901}.

\bibitem[{\citenamefont{Perdew et~al.}(1996)\citenamefont{Perdew, Burke, and
  Ernzerhof}}]{PBE}
\bibinfo{author}{\bibfnamefont{J.~P.} \bibnamefont{Perdew}},
  \bibinfo{author}{\bibfnamefont{K.}~\bibnamefont{Burke}}, \bibnamefont{and}
  \bibinfo{author}{\bibfnamefont{M.}~\bibnamefont{Ernzerhof}},
  \bibinfo{journal}{Phys. Rev. Lett.} \textbf{\bibinfo{volume}{77}},
  \bibinfo{pages}{3865} (\bibinfo{year}{1996}),
  \urlprefix\url{https://link.aps.org/doi/10.1103/PhysRevLett.77.3865}.

\bibitem[{\citenamefont{Kresse and Joubert}(1999)}]{PAW}
\bibinfo{author}{\bibfnamefont{G.}~\bibnamefont{Kresse}} \bibnamefont{and}
  \bibinfo{author}{\bibfnamefont{D.}~\bibnamefont{Joubert}},
  \bibinfo{journal}{Phys. Rev. B} \textbf{\bibinfo{volume}{59}},
  \bibinfo{pages}{1758} (\bibinfo{year}{1999}),
  \urlprefix\url{https://link.aps.org/doi/10.1103/PhysRevB.59.1758}.

\bibitem[{\citenamefont{Grimme}(2006)}]{DFT-D3}
\bibinfo{author}{\bibfnamefont{S.}~\bibnamefont{Grimme}},
  \bibinfo{journal}{Journal of Computational Chemistry}
  \textbf{\bibinfo{volume}{27}}, \bibinfo{pages}{1787} (\bibinfo{year}{2006}),
  \eprint{https://onlinelibrary.wiley.com/doi/pdf/10.1002/jcc.20495},
  \urlprefix\url{https://onlinelibrary.wiley.com/doi/abs/10.1002/jcc.20495}.

\bibitem[{\citenamefont{Agapito et~al.}(2016)\citenamefont{Agapito, Fornari,
  Ceresoli, Ferretti, Curtarolo, and Buongiorno~Nardelli}}]{PAO3}
\bibinfo{author}{\bibfnamefont{L.~A.} \bibnamefont{Agapito}},
  \bibinfo{author}{\bibfnamefont{M.}~\bibnamefont{Fornari}},
  \bibinfo{author}{\bibfnamefont{D.}~\bibnamefont{Ceresoli}},
  \bibinfo{author}{\bibfnamefont{A.}~\bibnamefont{Ferretti}},
  \bibinfo{author}{\bibfnamefont{S.}~\bibnamefont{Curtarolo}},
  \bibnamefont{and}
  \bibinfo{author}{\bibfnamefont{M.}~\bibnamefont{Buongiorno~Nardelli}},
  \bibinfo{journal}{Phys. Rev. B} \textbf{\bibinfo{volume}{93}},
  \bibinfo{pages}{125137} (\bibinfo{year}{2016}),
  \urlprefix\url{https://link.aps.org/doi/10.1103/PhysRevB.93.125137}.

\bibitem[{\citenamefont{Buongiorno~Nardelli
  et~al.}(2018)\citenamefont{Buongiorno~Nardelli, Cerasoli, Costa, Curtarolo,
  Gennaro, Fornari, Liyanage, Supka, and Wang}}]{PAO5}
\bibinfo{author}{\bibfnamefont{M.}~\bibnamefont{Buongiorno~Nardelli}},
  \bibinfo{author}{\bibfnamefont{F.~T.} \bibnamefont{Cerasoli}},
  \bibinfo{author}{\bibfnamefont{M.}~\bibnamefont{Costa}},
  \bibinfo{author}{\bibfnamefont{S.}~\bibnamefont{Curtarolo}},
  \bibinfo{author}{\bibfnamefont{R.~D.} \bibnamefont{Gennaro}},
  \bibinfo{author}{\bibfnamefont{M.}~\bibnamefont{Fornari}},
  \bibinfo{author}{\bibfnamefont{L.}~\bibnamefont{Liyanage}},
  \bibinfo{author}{\bibfnamefont{A.~R.} \bibnamefont{Supka}}, \bibnamefont{and}
  \bibinfo{author}{\bibfnamefont{H.}~\bibnamefont{Wang}},
  \bibinfo{journal}{Computational Materials Science}
  \textbf{\bibinfo{volume}{143}}, \bibinfo{pages}{462 } (\bibinfo{year}{2018}),
  ISSN \bibinfo{issn}{0927-0256},
  \urlprefix\url{http://www.sciencedirect.com/science/article/pii/S0927025617306651}.

\bibitem[{\citenamefont{Costa et~al.}(2019)\citenamefont{Costa, Schleder,
  Buongiorno~Nardelli, Lewenkopf, and Fazzio}}]{Costa2019}
\bibinfo{author}{\bibfnamefont{M.}~\bibnamefont{Costa}},
  \bibinfo{author}{\bibfnamefont{G.~R.} \bibnamefont{Schleder}},
  \bibinfo{author}{\bibfnamefont{M.}~\bibnamefont{Buongiorno~Nardelli}},
  \bibinfo{author}{\bibfnamefont{C.}~\bibnamefont{Lewenkopf}},
  \bibnamefont{and} \bibinfo{author}{\bibfnamefont{A.}~\bibnamefont{Fazzio}},
  \bibinfo{journal}{Nano Letters} \textbf{\bibinfo{volume}{19}},
  \bibinfo{pages}{8941} (\bibinfo{year}{2019}),
  \urlprefix\url{https://doi.org/10.1021/acs.nanolett.9b03881}.

\bibitem[{\citenamefont{Costa et~al.}(2018)\citenamefont{Costa, Costa, Freitas,
  Schmidt, Buongiorno~Nardelli, and Fazzio}}]{Costa2018}
\bibinfo{author}{\bibfnamefont{M.}~\bibnamefont{Costa}},
  \bibinfo{author}{\bibfnamefont{A.~T.} \bibnamefont{Costa}},
  \bibinfo{author}{\bibfnamefont{W.~A.} \bibnamefont{Freitas}},
  \bibinfo{author}{\bibfnamefont{T.~M.} \bibnamefont{Schmidt}},
  \bibinfo{author}{\bibfnamefont{M.}~\bibnamefont{Buongiorno~Nardelli}},
  \bibnamefont{and} \bibinfo{author}{\bibfnamefont{A.}~\bibnamefont{Fazzio}},
  \bibinfo{journal}{ACS Omega} \textbf{\bibinfo{volume}{3}},
  \bibinfo{pages}{15900} (\bibinfo{year}{2018}),
  \urlprefix\url{https://doi.org/10.1021/acsomega.8b01836}.

\bibitem[{\citenamefont{Costa et~al.}(2020)\citenamefont{Costa, Peres,
  Fern\'andez-Rossier, and Costa}}]{fegete}
\bibinfo{author}{\bibfnamefont{M.}~\bibnamefont{Costa}},
  \bibinfo{author}{\bibfnamefont{N.~M.~R.} \bibnamefont{Peres}},
  \bibinfo{author}{\bibfnamefont{J.}~\bibnamefont{Fern\'andez-Rossier}},
  \bibnamefont{and} \bibinfo{author}{\bibfnamefont{A.~T.} \bibnamefont{Costa}},
  \bibinfo{journal}{Phys. Rev. B} \textbf{\bibinfo{volume}{102}},
  \bibinfo{pages}{014450} (\bibinfo{year}{2020}),
  \urlprefix\url{https://link.aps.org/doi/10.1103/PhysRevB.102.014450}.

\bibitem[{\citenamefont{Costa et~al.}(2021)\citenamefont{Costa, Schleder,
  Acosta, Padilha, Cerasoli, Nardelli, and Fazzio}}]{hoti}
\bibinfo{author}{\bibfnamefont{M.}~\bibnamefont{Costa}},
  \bibinfo{author}{\bibfnamefont{G.~R.} \bibnamefont{Schleder}},
  \bibinfo{author}{\bibfnamefont{C.~M.} \bibnamefont{Acosta}},
  \bibinfo{author}{\bibfnamefont{A.~C.~M.} \bibnamefont{Padilha}},
  \bibinfo{author}{\bibfnamefont{F.}~\bibnamefont{Cerasoli}},
  \bibinfo{author}{\bibfnamefont{M.~B.} \bibnamefont{Nardelli}},
  \bibnamefont{and} \bibinfo{author}{\bibfnamefont{A.}~\bibnamefont{Fazzio}},
  \bibinfo{journal}{npj Computational Materials} \textbf{\bibinfo{volume}{7}},
  \bibinfo{pages}{49} (\bibinfo{year}{2021}),
  \urlprefix\url{https://doi.org/10.1038/s41524-021-00518-4}.

\bibitem[{\citenamefont{Guimar\~aes et~al.}(2015)\citenamefont{Guimar\~aes,
  Lounis, Costa, and Muniz}}]{PhysRevB.92.220410}
\bibinfo{author}{\bibfnamefont{F.~S.~M.} \bibnamefont{Guimar\~aes}},
  \bibinfo{author}{\bibfnamefont{S.}~\bibnamefont{Lounis}},
  \bibinfo{author}{\bibfnamefont{A.~T.} \bibnamefont{Costa}}, \bibnamefont{and}
  \bibinfo{author}{\bibfnamefont{R.~B.} \bibnamefont{Muniz}},
  \bibinfo{journal}{Phys. Rev. B} \textbf{\bibinfo{volume}{92}},
  \bibinfo{pages}{220410} (\bibinfo{year}{2015}),
  \urlprefix\url{https://link.aps.org/doi/10.1103/PhysRevB.92.220410}.

\bibitem[{\citenamefont{Guimarães et~al.}(2017)\citenamefont{Guimarães, dos
  Santos~Dias, Bouaziz, Costa, Muniz, and Lounis}}]{Guimaraes2017}
\bibinfo{author}{\bibfnamefont{F.~S.~M.} \bibnamefont{Guimarães}},
  \bibinfo{author}{\bibfnamefont{M.}~\bibnamefont{dos Santos~Dias}},
  \bibinfo{author}{\bibfnamefont{J.}~\bibnamefont{Bouaziz}},
  \bibinfo{author}{\bibfnamefont{A.~T.} \bibnamefont{Costa}},
  \bibinfo{author}{\bibfnamefont{R.~B.} \bibnamefont{Muniz}}, \bibnamefont{and}
  \bibinfo{author}{\bibfnamefont{S.}~\bibnamefont{Lounis}},
  \bibinfo{journal}{Scientific Reports} \textbf{\bibinfo{volume}{7}}
  (\bibinfo{year}{2017}), ISSN \bibinfo{issn}{2045-2322},
  \urlprefix\url{http://dx.doi.org/10.1038/s41598-017-03924-1}.

\bibitem[{\citenamefont{Bonbien and
  Manchon}(2020)}]{Bobien-Manchon-PhysRevB.102.085113}
\bibinfo{author}{\bibfnamefont{V.}~\bibnamefont{Bonbien}} \bibnamefont{and}
  \bibinfo{author}{\bibfnamefont{A.}~\bibnamefont{Manchon}},
  \bibinfo{journal}{Phys. Rev. B} \textbf{\bibinfo{volume}{102}},
  \bibinfo{pages}{085113} (\bibinfo{year}{2020}),
  \urlprefix\url{https://link.aps.org/doi/10.1103/PhysRevB.102.085113}.

\bibitem[{\citenamefont{Watanabe and
  Yanase}(2017)}]{Hikaru-Youichi-PhysRevB.96.064432}
\bibinfo{author}{\bibfnamefont{H.}~\bibnamefont{Watanabe}} \bibnamefont{and}
  \bibinfo{author}{\bibfnamefont{Y.}~\bibnamefont{Yanase}},
  \bibinfo{journal}{Phys. Rev. B} \textbf{\bibinfo{volume}{96}},
  \bibinfo{pages}{064432} (\bibinfo{year}{2017}),
  \urlprefix\url{https://link.aps.org/doi/10.1103/PhysRevB.96.064432}.

\bibitem[{\citenamefont{\ifmmode~\check{Z}\else \v{Z}\fi{}elezn\'y
  et~al.}(2017)\citenamefont{\ifmmode~\check{Z}\else \v{Z}\fi{}elezn\'y, Gao,
  Manchon, Freimuth, Mokrousov, Zemen, Ma\ifmmode~\check{s}\else \v{s}\fi{}ek,
  Sinova, and Jungwirth}}]{PhysRevB.95.014403}
\bibinfo{author}{\bibfnamefont{J.}~\bibnamefont{\ifmmode~\check{Z}\else
  \v{Z}\fi{}elezn\'y}}, \bibinfo{author}{\bibfnamefont{H.}~\bibnamefont{Gao}},
  \bibinfo{author}{\bibfnamefont{A.}~\bibnamefont{Manchon}},
  \bibinfo{author}{\bibfnamefont{F.}~\bibnamefont{Freimuth}},
  \bibinfo{author}{\bibfnamefont{Y.}~\bibnamefont{Mokrousov}},
  \bibinfo{author}{\bibfnamefont{J.}~\bibnamefont{Zemen}},
  \bibinfo{author}{\bibfnamefont{J.}~\bibnamefont{Ma\ifmmode~\check{s}\else
  \v{s}\fi{}ek}}, \bibinfo{author}{\bibfnamefont{J.}~\bibnamefont{Sinova}},
  \bibnamefont{and}
  \bibinfo{author}{\bibfnamefont{T.}~\bibnamefont{Jungwirth}},
  \bibinfo{journal}{Phys. Rev. B} \textbf{\bibinfo{volume}{95}},
  \bibinfo{pages}{014403} (\bibinfo{year}{2017}),
  \urlprefix\url{https://link.aps.org/doi/10.1103/PhysRevB.95.014403}.

\bibitem[{\citenamefont{Shao et~al.}(2016)\citenamefont{Shao, Yu, Lan, Shi, Li,
  Zheng, Zhu, Li, Amiri, and Wang}}]{Spin-Orbit-Torques-MoS2-nanolett}
\bibinfo{author}{\bibfnamefont{Q.}~\bibnamefont{Shao}},
  \bibinfo{author}{\bibfnamefont{G.}~\bibnamefont{Yu}},
  \bibinfo{author}{\bibfnamefont{Y.-W.} \bibnamefont{Lan}},
  \bibinfo{author}{\bibfnamefont{Y.}~\bibnamefont{Shi}},
  \bibinfo{author}{\bibfnamefont{M.-Y.} \bibnamefont{Li}},
  \bibinfo{author}{\bibfnamefont{C.}~\bibnamefont{Zheng}},
  \bibinfo{author}{\bibfnamefont{X.}~\bibnamefont{Zhu}},
  \bibinfo{author}{\bibfnamefont{L.-J.} \bibnamefont{Li}},
  \bibinfo{author}{\bibfnamefont{P.~K.} \bibnamefont{Amiri}}, \bibnamefont{and}
  \bibinfo{author}{\bibfnamefont{K.~L.} \bibnamefont{Wang}},
  \bibinfo{journal}{Nano Letters} \textbf{\bibinfo{volume}{16}},
  \bibinfo{pages}{7514} (\bibinfo{year}{2016}), \bibinfo{note}{pMID: 27960524},
  \eprint{https://doi.org/10.1021/acs.nanolett.6b03300},
  \urlprefix\url{https://doi.org/10.1021/acs.nanolett.6b03300}.

\bibitem[{\citenamefont{Ceresoli et~al.}(2010)\citenamefont{Ceresoli,
  Gerstmann, Seitsonen, and
  Mauri}}]{Ceresoli-ModernTheoXIntraAtom-PhysRevB.81.060409}
\bibinfo{author}{\bibfnamefont{D.}~\bibnamefont{Ceresoli}},
  \bibinfo{author}{\bibfnamefont{U.}~\bibnamefont{Gerstmann}},
  \bibinfo{author}{\bibfnamefont{A.~P.} \bibnamefont{Seitsonen}},
  \bibnamefont{and} \bibinfo{author}{\bibfnamefont{F.}~\bibnamefont{Mauri}},
  \bibinfo{journal}{Phys. Rev. B} \textbf{\bibinfo{volume}{81}},
  \bibinfo{pages}{060409} (\bibinfo{year}{2010}),
  \urlprefix\url{https://link.aps.org/doi/10.1103/PhysRevB.81.060409}.

\bibitem[{\citenamefont{Nikolaev and
  Solovyev}(2014)}]{Nikolaev-ModernTheoXIntraAtom-PhysRevB.89.064428}
\bibinfo{author}{\bibfnamefont{S.~A.} \bibnamefont{Nikolaev}} \bibnamefont{and}
  \bibinfo{author}{\bibfnamefont{I.~V.} \bibnamefont{Solovyev}},
  \bibinfo{journal}{Phys. Rev. B} \textbf{\bibinfo{volume}{89}},
  \bibinfo{pages}{064428} (\bibinfo{year}{2014}),
  \urlprefix\url{https://link.aps.org/doi/10.1103/PhysRevB.89.064428}.

\bibitem[{\citenamefont{Hanke et~al.}(2016)\citenamefont{Hanke, Freimuth,
  Nandy, Zhang, Bl\"ugel, and
  Mokrousov}}]{Hanke-ModernTheoXIntraAtom-PhysRevB.94.121114}
\bibinfo{author}{\bibfnamefont{J.-P.} \bibnamefont{Hanke}},
  \bibinfo{author}{\bibfnamefont{F.}~\bibnamefont{Freimuth}},
  \bibinfo{author}{\bibfnamefont{A.~K.} \bibnamefont{Nandy}},
  \bibinfo{author}{\bibfnamefont{H.}~\bibnamefont{Zhang}},
  \bibinfo{author}{\bibfnamefont{S.}~\bibnamefont{Bl\"ugel}}, \bibnamefont{and}
  \bibinfo{author}{\bibfnamefont{Y.}~\bibnamefont{Mokrousov}},
  \bibinfo{journal}{Phys. Rev. B} \textbf{\bibinfo{volume}{94}},
  \bibinfo{pages}{121114} (\bibinfo{year}{2016}),
  \urlprefix\url{https://link.aps.org/doi/10.1103/PhysRevB.94.121114}.

\bibitem[{\citenamefont{Thonhauser et~al.}(2005)\citenamefont{Thonhauser,
  Ceresoli, Vanderbilt, and Resta}}]{ModTheo-Real-Space-PhysRevLett.95.137205}
\bibinfo{author}{\bibfnamefont{T.}~\bibnamefont{Thonhauser}},
  \bibinfo{author}{\bibfnamefont{D.}~\bibnamefont{Ceresoli}},
  \bibinfo{author}{\bibfnamefont{D.}~\bibnamefont{Vanderbilt}},
  \bibnamefont{and} \bibinfo{author}{\bibfnamefont{R.}~\bibnamefont{Resta}},
  \bibinfo{journal}{Phys. Rev. Lett.} \textbf{\bibinfo{volume}{95}},
  \bibinfo{pages}{137205} (\bibinfo{year}{2005}),
  \urlprefix\url{https://link.aps.org/doi/10.1103/PhysRevLett.95.137205}.

\bibitem[{\citenamefont{Wang et~al.}(2022)\citenamefont{Wang, Yu, Guan, Dai,
  Wang, and Zhang}}]{ModTheo-Real-Space-PhysRevB.106.075136}
\bibinfo{author}{\bibfnamefont{S.-S.} \bibnamefont{Wang}},
  \bibinfo{author}{\bibfnamefont{Y.}~\bibnamefont{Yu}},
  \bibinfo{author}{\bibfnamefont{J.-H.} \bibnamefont{Guan}},
  \bibinfo{author}{\bibfnamefont{Y.-M.} \bibnamefont{Dai}},
  \bibinfo{author}{\bibfnamefont{H.-H.} \bibnamefont{Wang}}, \bibnamefont{and}
  \bibinfo{author}{\bibfnamefont{Y.-Y.} \bibnamefont{Zhang}},
  \bibinfo{journal}{Phys. Rev. B} \textbf{\bibinfo{volume}{106}},
  \bibinfo{pages}{075136} (\bibinfo{year}{2022}),
  \urlprefix\url{https://link.aps.org/doi/10.1103/PhysRevB.106.075136}.

\bibitem[{\citenamefont{Drigo and
  Resta}(2020)}]{HybridSystem-Resta-PhysRevB.101.165120}
\bibinfo{author}{\bibfnamefont{E.}~\bibnamefont{Drigo}} \bibnamefont{and}
  \bibinfo{author}{\bibfnamefont{R.}~\bibnamefont{Resta}},
  \bibinfo{journal}{Phys. Rev. B} \textbf{\bibinfo{volume}{101}},
  \bibinfo{pages}{165120} (\bibinfo{year}{2020}),
  \urlprefix\url{https://link.aps.org/doi/10.1103/PhysRevB.101.165120}.

\end{thebibliography}

\end{document}